%% file: main.tex
\documentclass[tradiabstract]{aa}

\usepackage{graphicx}
\usepackage{txfonts}
\usepackage[dvipsnames]{xcolor}
\usepackage{ulem}

\usepackage{subcaption}
\usepackage{hyperref}
\usepackage{placeins}

\begin{document}

   \title{Simultaneous X-ray and optical variability of M dwarfs observed with eROSITA and TESS}


   \author{W. M. Joseph
          \inst{1}
          \and
          B. Stelzer
          \inst{1}
          \and
          E. Magaudda
          \inst{1}
          \and
          T. Vičánek Martínez
          \inst{2}}

   \institute{Institut für Astronomie \& Astrophysik, Eberhard Karls Universität Tübingen, Sand 1, 72076 Tübingen, Germany\
   \and Hamburger Sternwarte, Universit\"at Hamburg, Gojenbergsweg 112, 21029 Hamburg
             }

   \date{}

 
  \abstract
   {
   M dwarf stars are the most numerous stars in the Galaxy, and are highly magnetically active. 
   They exhibit 
   bursts of radiation and matter, called flares and Coronal Mass Ejections which have the potential to strongly affect the habitability of their planets.
   }
   {We 
   investigate variability through simultaneous optical and X-ray observations for the first time in a statistical sample of M dwarfs.
   Such simultaneous observations
   at different wavelengths, which correspond to 
   emissions from different layers of the stellar atmosphere, are required to constrain the flare frequency and energetics and to understand the physics of flares. 
   }
   {We 
   use light curves from the extended ROentgen Survey with an Imaging Telescope Array (eROSITA) on board the Russian Spektrum-Roentgen-Gamma mission (SRG) and the Transiting Exoplanet Survey Satellite (TESS) for a sample of M dwarfs observed simultaneously with both instruments. 
   Specifically, we identified 
   256 
   M dwarfs in the TESS Southern Continuous Viewing Zone (SCVZ) 
   with 
   simultaneous TESS and 
   eROSITA detection.
   For this work, we selected the $25$ 
   X-ray brightest or most X-ray variable stars.
   We used photometric data from {\it Gaia} and 2MASS to obtain stellar parameters such as 
   distances, colours, masses, radii, and bolometric luminosities. X-ray fluxes and luminosities were determined from observed eROSITA count rates using appropriate rate-to-flux conversion factors. We defined and examined various variability diagnostics in both wavebands 
   and how these parameters are related to each other.}
   {
   Our stars are nearby (mostly within $\sim 100$\,pc ), rotating fast ($P_{\rm rot} < 9$\,d), and displaying a high optical flare frequency, as expected from the selection of particularly X-ray active objects. 
   The X-ray and optical duty cycles -- defined as the fraction of observing time in which the stars were in a high activity state -- 
   are positively correlated, and there is a trend of faster rotators tending to have higher 
   X-ray and optical variability. 
   For stars with many X-ray flaring events, 
   chances of these events to be found together with optical flares are high. 
   A quantitative  variability study of individual flares in the X-ray light curves is severely affected by data gaps due to the low (4\,h) cadence during the eROSITA all-sky survey.
   To mitigate this, we make use of the optical flares observed with TESS combined with knowledge accumulated from solar flares to put additional constraints on the peak flux and timing of X-ray events. 
  With this method we could perform an exponential fit to $21$ X-ray light curves in the aftermath of an optical flare, and we find that the energies for these X-ray flares are well-correlated with the corresponding optical flare energy.}
   {Despite the substantial uncertainties associated to our analysis, which are mostly related to the poor sampling of the eROSITA light curves, our results showcase in an exemplary way the relevance of simultaneous all-sky surveys in different wavebands for obtaining  unprecedented quantitative information on stellar variability.}


   \maketitle
%

\section{Introduction}
\label{sec:introduction}
M dwarfs, the most numerous stars in our galaxy \citep{Chabrier2001}, have gained renewed interest due to the discovery of numerous exoplanets around them \citep{Howard2012}, some of which may be potentially habitable \citep{Tarter2007}. For M dwarfs the habitable zones are at small distance from the star,
making the planets particularly sensitive to the 
stellar radiation. The high-energy photons from the upper atmospheres of the stars (UV and X-ray wavelengths, commonly termed XUV radiation) and the accompanying particle flux are responsible for the strongest effects, driving chemistry in planetary atmospheres or eroding them \citep{Tilley2019}.
The notorious variability that is naturally associated with magnetically heated stellar atmospheres is the main driver of photo-induced processes in planet atmospheres. 

M dwarfs constitute the historical group of `flare stars'. Stellar flares are drastic short-term enhancements of the multi-wavelength radiation that go along with particle acceleration, both a consequence of magnetic reconnection events in the corona. 
Flares are well documented at all wavelengths for the Sun and have been observed on M dwarfs since the 1950s when the first events were seen on objects like UV\,Cet \citep{JoyHumason1949} which today is one of the  prototypical flaring M~dwarfs. Due to the high contrast between these energetic events and the fainter quiescent emission as compared to earlier-type stars, magnetic activity can be most efficiently studied in M dwarfs.

In the standard (solar) flare scenario
magnetic reconnection spurs successive emissions from all layers of the stellar atmosphere. The accelerated charged particles generate gyrosynchrotron emission in the hard X-ray and radio bands as they follow magnetic field lines. Upon hitting the chromosphere, these particles heat it, leading to additional hard X-rays through bremsstrahlung emission and white-light continuum emission at the footpoints. The resulting plasma, now in a high-pressure region, ascends along the magnetic field loop, filling the top where dense and hot thermal emission in soft X-rays occurs. While the hard X-ray and white-light emissions share a similar temporal signature, the soft X-rays from the loop's top exhibit a temporal lag due to the sequential nature of the flare mechanism. Time lags between the peak of white-light and X-ray emissions have been observed to be a few minutes both on the solar flares \citep{CastellanosDuran2020} and on the rare examples of contemporaneous optical and X-ray observations of flares on active stars \citep{Stelzer2006,Fuhrmeister2011,Stelzer2022}.

In this scenario  optical and X-ray flares are separated temporally and spatially, with soft X-rays tracing flaring activity and heating in the corona and the optical variability tracing flaring in the photosphere.
To investigate the  activity occurring across various layers of the stellar atmosphere, it is essential to observe it in different wavelengths corresponding to these distinct layers. However, due to the stochastic nature of flares, systematically tracking their multi-wavelength evolution in stars other than our Sun has been unfeasible up to now. In this paper we study for the first time ever light curves of active stars from two instruments that provide simultaneous (nearly) all-sky surveys in the optical and X-ray wavebands: The {\it Transiting Exoplanet Survey Satellite} (TESS) \citep{Ricker2014} in the optical band and the {\it extended Roentgen Survey with an Imaging Telescope Array} (eROSITA) \citep{Predehl2021} in X-rays.

Since flares are unpredictable, next to the simultaneous multi-wavelength coverage, 
long exposures are required to collect statistical samples  of events.
 Both criteria are met by eROSITA and TESS for the sky region around the ecliptic poles where both surveys  provide the highest temporal coverage. 
In this paper, we therefore focus on the so called TESS Southern Continuous Viewing Zone (SCVZ) (see Sect.~\ref{sec:sample}), and we study the simultaneous optical and X-ray variability of selected M dwarfs.


In Sect.~\ref{sec:surveys}, we describe the TESS and eROSITA instruments and their survey strategies with focus on the aspects relevant for our work. In Sect.~\ref{sec:sample} we outline our sample selection. In Sect.~\ref{sec:gaia-2mass-params}, we detail the stellar parameters obtained from the {\it Gaia}, 2MASS, and eRASS catalogues. In Sect.~\ref{sec:erass-lc} and Sect.~\ref{sec:TESS-lc}, we explain the variability diagnostics derived from the eRASS and TESS light curves, respectively. In Sect.~\ref{sec:results} we discuss our results. Finally, Sect.~\ref{sec:conclusions} summarizes our findings. 

\section{All-sky surveys suitable for flare studies}
\label{sec:surveys}
\subsection{TESS}
\label{subsect:TESS}

TESS is a an MIT-led NASA survey mission that is designed to detect and observe transiting exoplanets around bright and nearby stars \citep{Ricker2014}. TESS has four wide-field, red-sensitive (\(\sim\) 600-1000 nm) cameras with a collective field of view of \(24^{\circ}\times96^{\circ}\), with which it scans over 90\(\%\) of the entire sky using various scanning strategies during each year of its mission. 

This paper uses data from Year 3 of the mission during which TESS used the scanning method where it tiles the entire sky in 26 
segments with 13 segments per hemisphere (northern and southern). Each segment extends from the ecliptic pole to a \(6^{\circ}\) ecliptic latitude. TESS continuously observes each segment for 27.4 days or two satellite orbits before pivoting \(\sim27^{\circ}\) eastwards around the ecliptic pole at the orbit perigee and observing the next segment. During each segment observation, the camera's boresight is pointed nearly antisolar. There is significant overlap between the observed sectors, with the regions within \(\sim12^{\circ}\) of the ecliptic poles being observed in all sectors per hemisphere. This region is known as the continuous viewing zone (CVZ) of TESS and has a continuous observational period of a year, since each hemisphere takes one year to complete and an all-sky survey takes two years. 

The TESS data products are stored and then downlinked at every visit of the satellite to the orbit perigee during the Low-Altitude Housekeeping Operations (LAHO) phases of the orbits \citep{tess-inst-handbook}. 
The TESS input catalogue (TIC) is a catalogue of stellar parameters of every optically persistent and stationary object in the sky \citep{Stassun2018}. In the TIC and its subsets, each star has a unique identifier called the TIC ID. Data is taken at a two-minute cadence for stars in the Candidate Target List (CTL), which is a subset of the TIC. The stars from each TESS sector that have such light curves available are listed in the TESS two-minute target lists, which can be accessed through the MIT website (\texttt{\url{https://tess.mit.edu/observations/target-lists/}}). A separate two-minute target list comprising about $20000$ stars is available for each TESS sector. 
All listed stars have their associated pipeline-processed data products, including light curves, available on the Barbara A. Mikulski Archive for Space Telescopes (MAST) portal (\texttt{\url{https://mast.stsci.edu/portal/Mashup/Clients/Mast/Portal.html}}).

\subsection{eROSITA}
\label{subsect:eROSITA}

eROSITA 
is a wide-field X-ray telescope developed under the leadership of the Max-Planck Institut f\"ur extraterrestrische Physik (MPE) in Garching, Germany. It is the primary instrument on-board the Russian-German ``Spectrum-Roentgen-Gamma" (SRG) mission \citep{Sunyaev2021}. eROSITA is designed to take all-sky surveys in the energy band $0.2-10$\,keV, and similar to its predecessor, the X-ray telescope ROSAT \citep{Truemper1982}, eROSITA also provides all-sky X-ray data on active stars, but at a higher sensitivity. The criteria adopted to compile the eROSITA source catalogs will be published with the first data release of eROSITA within the article by \citet{Merloni23.0} on eRASS1. 

The primary observational mode of SRG/eROSITA is the survey mode which was used in the all-sky survey phase \citep{Predehl2021}. In this mode, the spacecraft constantly rotates about an axis and scans great circles on the sky. In order to perform an eROSITA all-sky survey (eRASS) the following parameters are fixed: The `scan rate' which defines the spacecraft's rotation, the `survey rate' which defines the rate at which the scanned great circles progress and the `survey pole' which specifies the plane in which the rotation axis moves. The `scan rate' here is an angular velocity of \(0.025\,{\rm deg/s}\), which corresponds to a revolution of 4 hr (called one `eROday') and a field-of-view (FOV) passage time of $\sim 40$\,s. The `survey rate' used is \(\approx 1 {\rm deg/d}\) which is the spacecraft's angular velocity around the Sun. This separates the scans by \(\approx\)$10^\prime$ and ensures that any sky position
is observed at least six times in a day after which it moves out of the field-of-view until the next all-sky survey. The survey pole used for the eRASS was the ecliptic pole. Each eRASS takes six months to complete and four full sky surveys have been completed by August 2022. 

A ground software system called the eROSITA Science Analysis Software System (eSASS) \citep{Brunner22.0} is used in order to get calibrated science data products as well as perform data analysis tasks. The eROSITA data can then be processed by the  eSASS pipeline in order to get calibrated data products such as event lists, images, exposure maps, background maps, X-ray source catalogues and source-specific data products such as light curves. The all-sky data products are available as 4700 overlapping 3.6° × 3.6° sky tiles which can be accessed by authorized users through a web interface. Access to the eRASS data and associated software used in this paper is restricted and is available only to the members of the eROSITA\_DE consortium.

\section{Sample selection}
\label{sec:sample}

For our goal to study simultaneous optical and X-ray variability of active stars we require (1) long light curves and (2) simultaneous coverage with eROSITA and TESS. The most suitable sky region for this work is, therefore, the ecliptic poles.

As described in Sect.~\ref{subsect:eROSITA}, during the 
eRASS
eROSITA scans the entire sky by rotating on its axis with an orbital period of 4 hours, equivalent to 1 eRODay \citep{Predehl2021}. This scanning strategy yields an exposure time of up to 4000\,s at the ecliptic poles (versus 200 s at low latitudes), resulting in their stars having the longest light curves. Of the two ecliptic poles, only data for the southern one is accessible to us through the eROSITA\_DE consortium.

TESS observed the southern hemisphere in Year~1 and Year~3 of its mission \citep{Ricker2014}. Temporal overlap with the part of eRASS accessible to us was during Year~3 of its mission. 
The 
CVZ at the ecliptic poles overlap all segments in a hemisphere, such that the poles receive the most observation time. The Southern CVZ (SCVZ) of TESS, therefore, is the most favorable location for our project.
TESS continuously observed its SCVZ during Year\,1 (Sectors\,1 - 13) and Year\,3 (Sectors\,27 - 39) of its mission. The eROSITA all-sky surveys took place from 8 December 2019 to 26 February 2022. TESS Year~3 (4 July 2020 to 24 June 2021) corresponds almost exactly to the time of the second and third eROSITA surveys, eRASS2 and eRASS3. The minor temporal overlap with eRASS4 is neglected in this pilot project.


We started our sample selection by making use of the pipeline-produced TESS light curves obtained in two-minute cadence (see Sect.~\ref{subsect:TESS}). The subsequent downselection of a sample of M dwarfs suitable for a joint optical/X-ray variability study involved {\it Gaia} and eRASS data, and is explained in the remainder of this Section.

\subsection{Retrieving {\it Gaia} data}
We obtain accurate coordinates, distances, and magnitudes from  {\it Gaia} DR2  \citep{Gaia2016,Gaia2018}, the latest data release available when we started the project. The retrieval of {\it Gaia} data began with the determination of the {\it Gaia} DR2 {\sc source\_ID} for each star from the TESS two-minute target lists of Sectors $27-39$ that overlap with eRASS. The {\it Gaia} {\sc source\_ID}s are housed within the TIC and were obtained by cross-referencing the two-minute target lists with the TESS Input Catalog - v8.0 \citep{Stassun2019} using the TIC ID. Subsequently, we used the {\it Gaia} {\sc source\_ID}s to retrieve the corresponding {\it Gaia} DR2 data from the {\it Gaia} archive (available at \texttt{\url{https://gea.esac.esa.int/archive/}}).

\subsection{Selection of M dwarfs in the TESS SCVZ}
\label{sec:mdwarf-selection}

We used  {\it Gaia} DR2 coordinates and colours and employed the following criteria to identify M dwarfs within the TESS SCVZ:
\begin{enumerate}
  \item Colour selection:
  We identified M dwarfs on the basis of $G_{\rm BP} - G_{\rm RP}$  color,
  obtained from the integrated {\it Gaia} mean apparent magnitudes measured with the blue and red photometer, $BP$ and $RP$, respectively \footnote{\(BP\) and \(RP\) represent the integrated fluxes
  from the {\it Gaia} BP and RP spectrophotometers, which were available with {\it Gaia} DR2 data \citep{Riello2018}}. We made our colour selection using the table ``A Modern Mean Dwarf Stellar Color and Effective Temperature Sequence" by E.Mamajek (\texttt{\url{https://www.pas.rochester.edu/~emamajek/EEM_dwarf_UBVIJHK_colors_Teff.txt}}), according to which dwarfs of SpT M0V and later have $BP-RP\geq1.815$\,mag.

  \item Exclusion of giant stars:
  Following \citep{Magaudda2022}, we used the absolute {\it Gaia} magnitude, $M_{\rm G}$, with the criterion \(M_{\rm G}\geq5.0\)\,mag to exclude suspected giant stars from our selection.

  \item SCVZ stars:
  We specifically selected stars in the SCVZ limiting the ecliptic latitudes ($b$) to within \(12^{\circ}\) of the Southern ecliptic pole, that is $b \leq -78^\circ$. 
\end{enumerate}

\begin{figure} 
\centering
\includegraphics[width=\columnwidth]{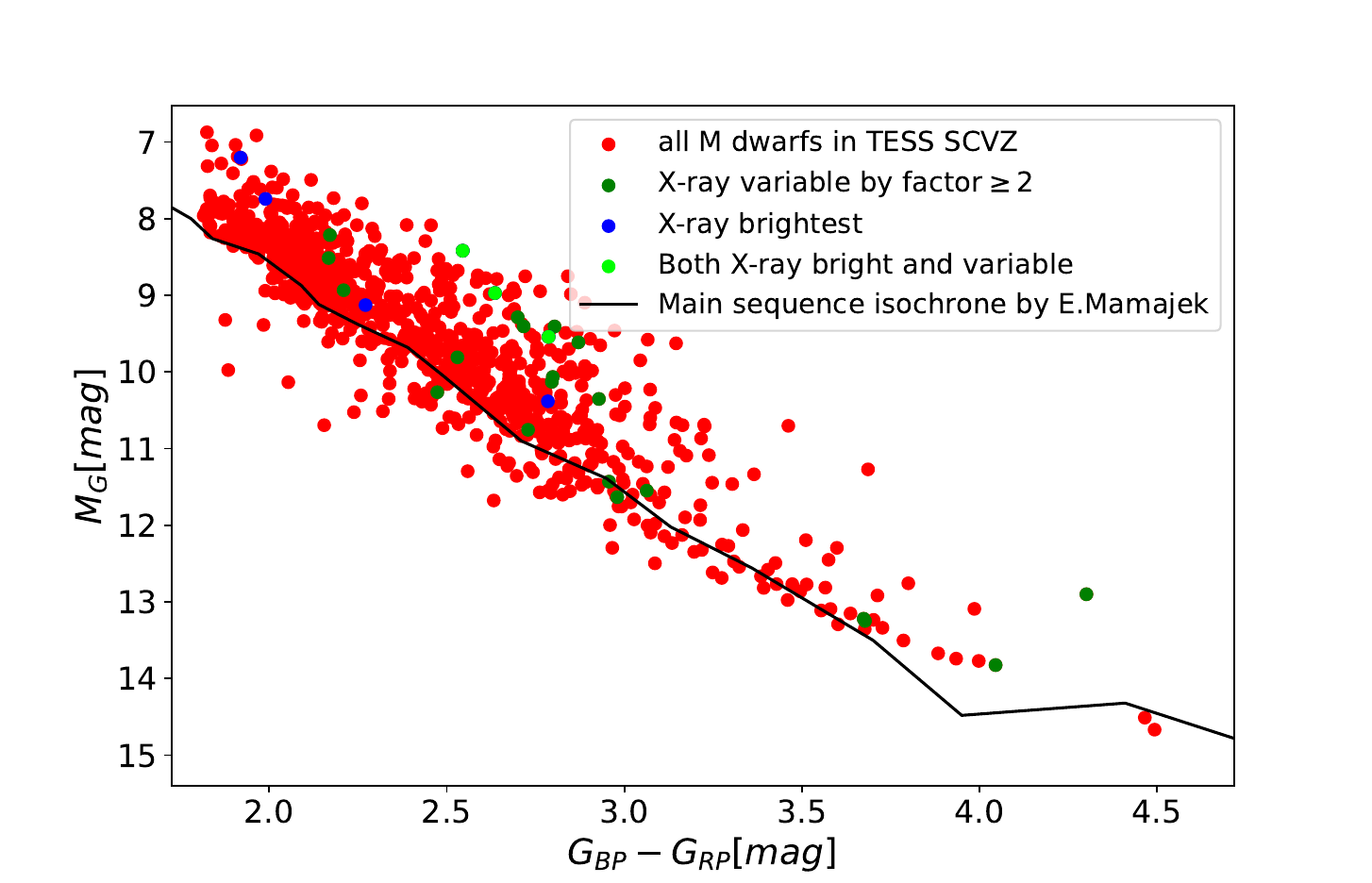}
\caption{Colour-magnitude diagram for M dwarfs in the TESS SCVZ (red) with the two samples defined in Sect.~\ref{sec:sample-selection} highlighted in blue (count rate $\geq$ 1\,cts/s) and green (count rate ratio between eRASS2 and eRASS3 $\geq 2$\,cts/s).  Overplotted is the main sequence isochrone (black) from the 
table ”A Modern Mean Dwarf Stellar Color and Effective Temperature Sequence” by E.Mamajek (\texttt{\url{https://www.pas.rochester.edu/~emamajek/EEM_dwarf_UBVIJHK_colors_Teff.txt}}) }
\label{fig:cmd-subsample}
\end{figure}

\begin{figure} 
\centering
\includegraphics[width=\columnwidth]{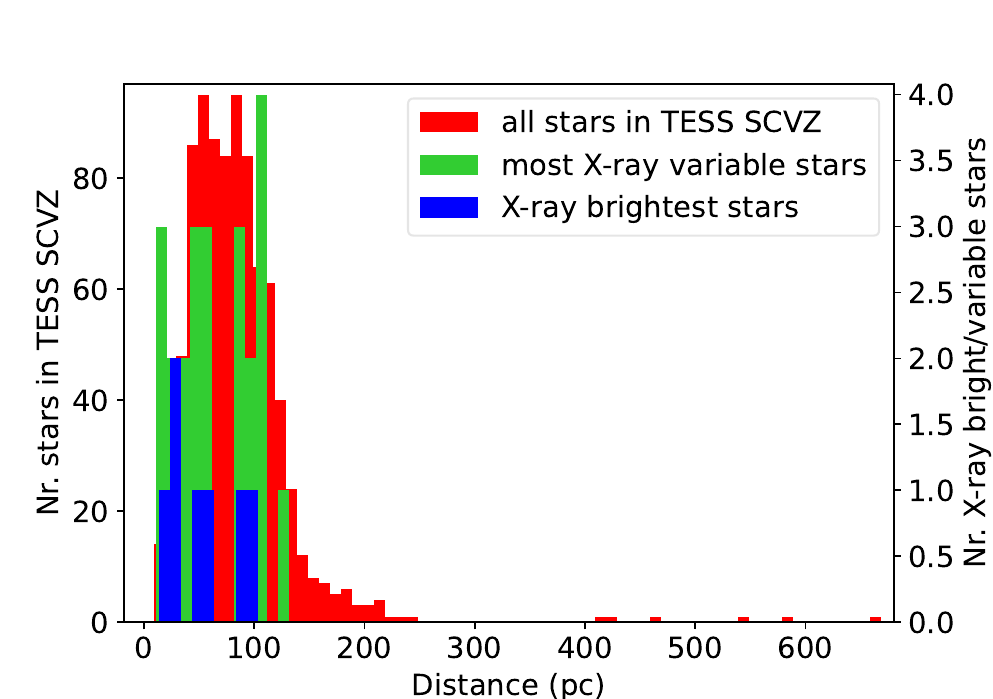}
\caption{Distribution of distances for the two subsamples of X-ray brightest and most X-ray variable defined in Sect.~\ref{sec:sample-selection} as well as the entire M dwarf sample in the TESS SCVZ described in Sect.~\ref{sec:mdwarf-selection}}.
\label{fig:distance-hist-subsample}
\end{figure}

A total of 874 stars met the aforementioned criteria. Figure ~\ref{fig:cmd-subsample} displays a {\it Gaia} color-magnitude diagram (CMD) for these stars and Fig.~\ref{fig:distance-hist-subsample} shows how they are distributed by their distances. In these figures we show the sample selected from TESS and {\it Gaia} as described above, and the subsample of it studied in this work. This latter one resulted from further downselection on the basis of eRASS data (see next section).

\subsection{Search for eRASS X-ray  counterparts}
\label{subsect:select_eRASS}

To identify those M dwarfs from the sample defined in Sect.~\ref{sec:mdwarf-selection} that have an eRASS X-ray detection we performed a match between {\it Gaia} coordinates and eRASS coordinates from the preliminary version of eRASS2 and eRASS3 catalogues produced by the eROSITA\_DE consortium (all\_e2\_201221\_poscorr\_mpe\_clean.fits and all\_e3\_210707\_poscorr\_mpe\_clean.fits, respectively). 
The catalogues we used are the cleaned and position corrected catalogues \citep{Brunner22.0} processed using the eSASS pipeline version c946. These catalogues contain the merged source catalogues of each sky tile, with duplicates from overlapping sky tiles removed (`cleaned') and the coordinates of the sources boresight corrected ({\sc ra\_corr} and {\sc dec\_corr} in the official catalogues). However, the stellar positions used for this match required correction for proper motion because most of the stars are located in proximity
(as shown in Fig.~\ref{fig:distance-hist-subsample}), and the {\it Gaia}-DR2 coordinates pertain to J2015.5, approximately 5 years prior to the eRASS surveys. 

We extrapolated the optical positions of all stars to the mean observing date of both eRASS2 and eRASS3.
The mean observing date epochs of eRASS2 and eRASS3 were 12 September 2020 and 16 March 2021, respectively, which corresponds to a time difference of 1961 and 2146 days between the middle of each of the two eRASS survey epochs and the {\it Gaia}-DR2 epoch. After proper motion correction to these two dates our target list was cross-referenced with the eRASS2 and eRASS3 catalogues. Among the 874 M dwarfs, we find that 303 stars are listed in the eRASS2 catalogue, 293 stars are present in the eRASS3 catalogue, and 256 stars have detections in both the eRASS2 and eRASS3 catalogues. 

\subsection{X-ray brightest and most variable M dwarfs}
\label{sec:sample-selection}

In this work we focus on a sub-sample with high or strongly changing eRASS count rate. This warrants the best signal for our pilot study to quantify the relation between  optical and X-ray variability. We defer the discussion of the X-ray and optical activity signatures of the full sample of M dwarfs in the TESS SCVZ that have 
TESS and eRASS data to a future work.

\begin{figure} 
\centering
\parbox{\textwidth}{
\parbox{0.5\columnwidth}{
\includegraphics[width=0.5\columnwidth]{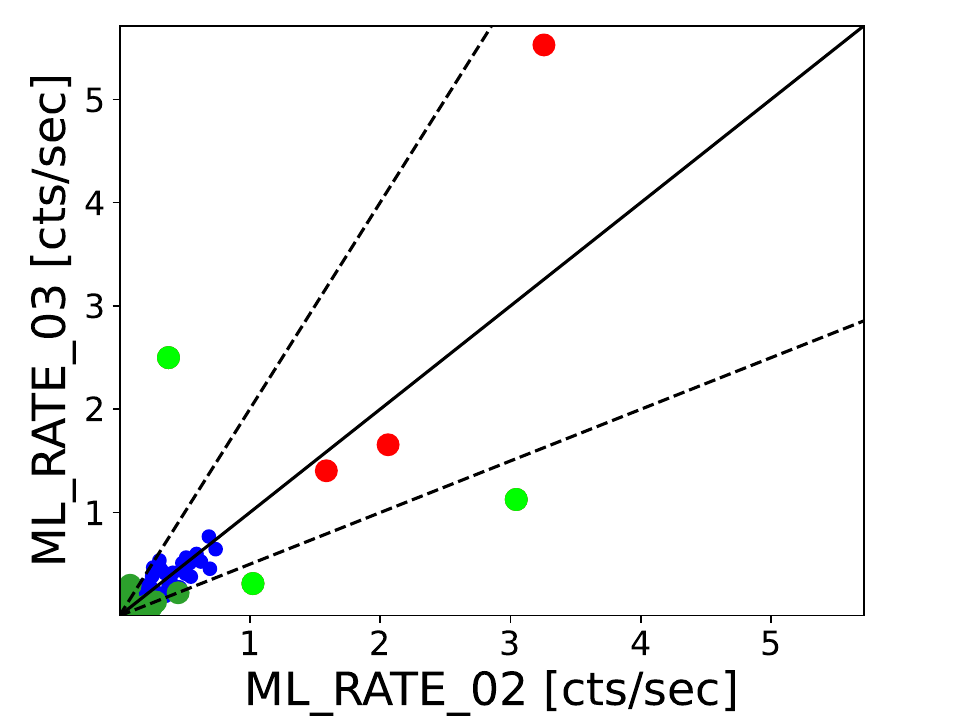}
}
\parbox{0.5\columnwidth}{\includegraphics[width=0.5\columnwidth]{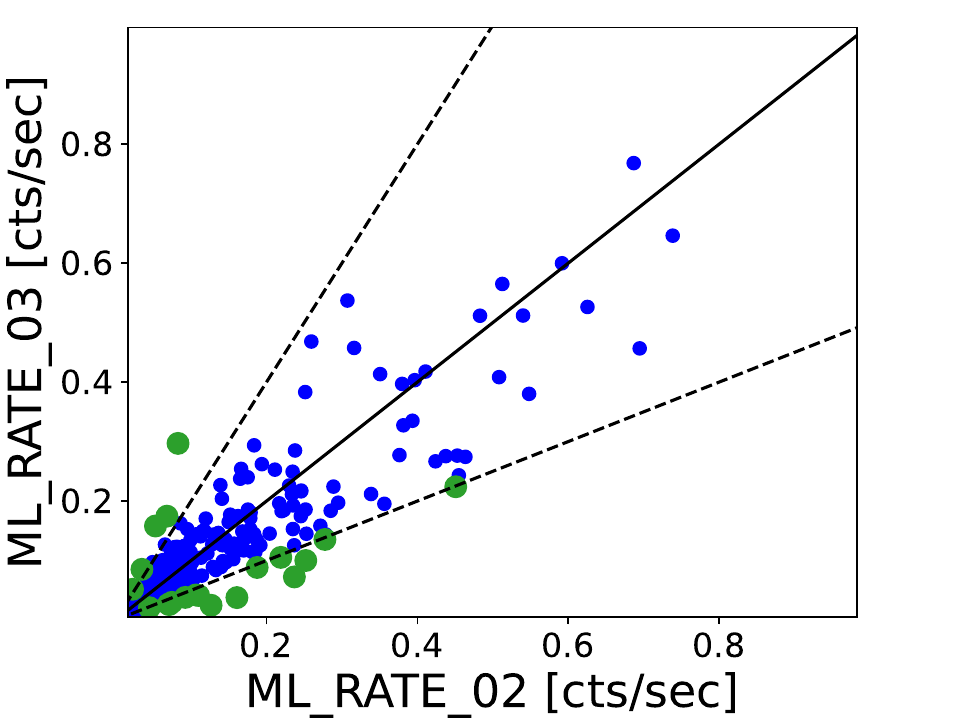}
}
}
\caption{Variation of X-ray count rate in the $0.2-5.0$\,keV band between eRASS2 and eRASS3 for the M dwarfs in the TESS SCVZ. left panel - all stars, right panel - zoom into the low count rate region. The lines denote equal count rates in both surveys (solid) and differences of a factor two (dashed). The X-ray brightest stars (Eq.~1) are high-lighted in red, the most X-ray variable (Eq.~2) in green and stars that fulfil both critera in lime. See Sect.~\ref{sec:sample-selection} for details.
}
\label{fig:brightness-change-erass2and3}
\end{figure}

For our further down-selection to define the X-ray brightest and most variable M dwarfs from the list of 256 stars with both eRASS2 and eRASS3 detections 
we applied the following criteria:
\begin{enumerate}
  \item Our selection of the stars with the highest photon statistics is based on the eRASS count rate in the $0.2-5.0$\,keV energy band (column {\sc ML\_RATE\_0} in the eRASS catalogs). We choose the stars with
  \begin{equation}
  \label{eq:brightest}
    {\sc ML\_RATE\_0}_{\rm i} \geq 1.0 \:cts/s
  \end{equation}
 in at least one of the two eRASS surveys ($i=2,3$ for eRASS2 and eRASS3).
 This criterion identifies a total of seven stars as the X-ray brightest.
  \item In order to find the stars that show the strongest variability between eRASS2 and eRASS3, we choose the stars with
  \begin{equation}
  \label{eq:most-var}
    \frac{{\sc ML\_RATE\_0}_{\rm i}}{{\sc ML\_RATE\_0}_{\rm j}}\geq2.0 \:cts/s 
  \end{equation}
  with $i = 2,3$ and $j = 3,2$.
This condition identifies $23$ stars as the most X-ray variable between eRASS2 and eRASS3. 
\end{enumerate}
The combination of the two criteria leads to the selection of 27 stars. Fig.~\ref{fig:brightness-change-erass2and3} displays {\sc ML\_RATE\_0} from eRASS2 vs. eRASS3 in which the selection criteria can be observed. All the M dwarfs in the TESS SCVZ are marked in blue, the X-ray brightest stars from Eq.~\ref{eq:brightest} in red, the most X-ray variable stars from Eq.~\ref{eq:most-var} in green and stars that fulfil both criteria in lime. The diagonal where stars with no change in brightness between the two eRASS surveys lie is drawn in black and the dashed lines indicate a factor 2 change in X-ray brightness between the two surveys.

In the {\it Gaia} CMD of Fig.~\ref{fig:cmd-subsample} we overlaid the final sample on the full sample of M dwarfs in the SCVZ from our master table defined in Sect.~\ref{sec:mdwarf-selection}. It  illustrates that the stars are evenly distributed across the M dwarf spectral sequence, and in particular among the M dwarfs in the TESS SCVZ. 
The distance distribution of the X-ray brightest and most variable stars together with all M dwarfs in the TESS SCVZ is shown in Fig.~\ref{fig:distance-hist-subsample}, showing that the stars tend to be nearby, mostly within $100$\,pc. This is because predominantly bright and nearby M dwarfs were selected for the TESS two-minute target lists, and an additional preference for stars at short distance from our eRASS selection criteria.

Each of the $27$ X-ray selected stars has, by definition of the sample, one light curve each for eRASS2 and eRASS3.  This results in a total of 54 eRASS light curves. However, not all of the TESS light curves corresponding to the eRASS observations were available in the MAST database. 
Moreover, some of the light curves, both from TESS and eRASS, lacked simultaneous coverage due to 
the TESS LAHO gaps (see Sect.~\ref{subsect:TESS}) and in some cases discarded data within the TESS light curve because of failed quality checks. 
Ultimately, there are $42$ eRASS light curves from $25$ stars that meet the criteria of being the brightest or most variable M dwarfs in X-rays and also having simultaneous TESS two-minute data. This database serves as the basis for the remainder of this paper. 

\section{Stellar parameters from {\it Gaia}, eRASS  and 2\,MASS catalogues}
\label{sec:gaia-2mass-params}
We obtained stellar parameters for the $25$ stars of our final sample using data from the {\it Gaia}, 2\,MASS and eRASS  catalogues. We retrieved the 2\,MASS data for our stars directly from the {\it Gaia} archive using the {\it Gaia}-DR2 {\sc source\_id}, as the cross match between 2\,MASS and {\it Gaia}-DR2 catalogues has been pre-computed by the {\it Gaia} DPAC (Data Processing and Analysis Consortium).  We also obtained the Johnson–Kron–Cousin $V$ from {\it Gaia} magnitudes $G_{\rm BP}$ and $G_{\rm RP}$ using the relation from \citep{Jao2018}.
We used the empirical relations from \citep{Mann2015} to calculate the stellar radii and masses from the absolute $K_{\rm s}$-band magnitude, $M_{\rm Ks}$, the effective temperatures ($T_{\rm eff}$) as an average of the values found individually using $Bp-Rp$ with $J-H$ and $V$ with $J-H$, and bolometric corrections to {\it Gaia} magnitudes, which we then used to calculate bolometric luminosities, $L_{\rm bol}$.
We calculated the time-averaged observed X-ray flux ($\langle F_{\rm x,obs} \rangle$)  from the mean count rate (ML\_RATE\_0) provided in the eRASS catalogues using a conversion factor of $7.81 \cdot 10^{-13}\,{\rm erg\,cts^{-1}\,cm^{-2}}$ derived by \citep{Magaudda2022} from eROSITA spectra of M dwarfs. We used the stellar radii and the distances from ${\it Gaia}$ parallaxes  to calculate the average X-ray luminosity ($\langle L_{\rm x} \rangle$) and X-ray surface flux  ($\langle F_{\rm x,surf} \rangle$) of the given eRASS.
The values for all these parameters for the 25 stars are available through online catalogues. The columns of these catalogues are described in Appendix~\ref{tab:gaia-2mass-tess} and~\ref{tab:erass}.

\section{Analysis of eRASS light curves}
\label{sec:erass-lc}

\subsection{eRASS data extraction} 

We extracted the eRASS light curves using the eSASSusers\_211214 software release \citep{Brunner22.0} for our sources in such a way that they were binned regularly with a bin size of 1 eRODay and any empty bins were discarded (i.e., bins having zero fractional exposure). 
In addition, only the bins with fractional exposure \texttt{fracexp} \(\geq\) 0.00005 are included as bins with low fractional exposure have large error values. The above cutoff value for \texttt{fracexp} was chosen as it eliminated bins with a significantly smaller fractional exposure than the rest of the data points in the light curve. 
As described in Sect.~\ref{subsect:eROSITA}, as the eROSITA scans of the great circles progress, a given source moves in and out of the telescopes' FOV. This means that the \texttt{fracexp} tends to be lowest when the source is at the edge of the FOV, which corresponds to the beginning and end of the light curve.

\subsection{Definition of eRASS quiescent level}
\label{sec:erass-qui-level}

In this paper, variability in the eRASS light curves is quantified by first defining a quiescent count rate, \(R_{\rm qui}\), and then marking the light curve bins that are significantly above it, which we interpret as being part of flares. To do this, the eRASS count rates are divided into equally sized bins and bin size is chosen as the maximum of two optimum bin size estimators,  the Freedman Diaconis Estimator \citep{Freedman1981} and the Sturges estimator (e.g., \citep{SturgesEstimator}) implemented using the Numpy package in Python \citep{harris2020array}. 
The X-ray quiescent level (purple in the example shown in Fig.~\ref{fig:erass-qui-level}) is defined using the 
two criteria described in the following. 

Firstly, the quiescent level should be the upper boundary of the lowest count rate bin with more than two data points. If no such bin exists, we use the lowest count rate bin.  We prioritize bins with more than two data  points because the lowest count rates are sometimes negative and have large error values associated with them. This occurs for stars that are very faint, comparable to, or even fainter than the background.
Secondly, it is required that the quiescent level has its upper boundary above zero cts/sec. If this is not met by the bin selected with the first criterion,
we use the upper boundary of the next highest count rate bin (i.e., the bin above it) as $R_{\rm qui}$ value.
Similar to the first criterion this is necessary as for some light curves the lowest count rate bin has only data points with $ML\_RATE\_0 < 0$\,cts/s, and a negative quiescent level is unphysical.
The quiescent count rates we obtained from each eRASS light curve of our targets, \(R_{\rm qui,eRASS2}\) and \(R_{\rm qui,eRASS3}\), along with the mean of the two values \(R_{\rm qui,mean}\) and the difference between them \(R_{\rm qui,std.dev}\) are presented in Table~\ref{table:tic-rqui-erassdc}.

\input{tables/tic_rqui_erassdc}

\subsection{eRASS duty cycle of variability}
\label{sec:var-diognostics-erass}

\begin{figure} 
\centering
\includegraphics[width=\columnwidth]{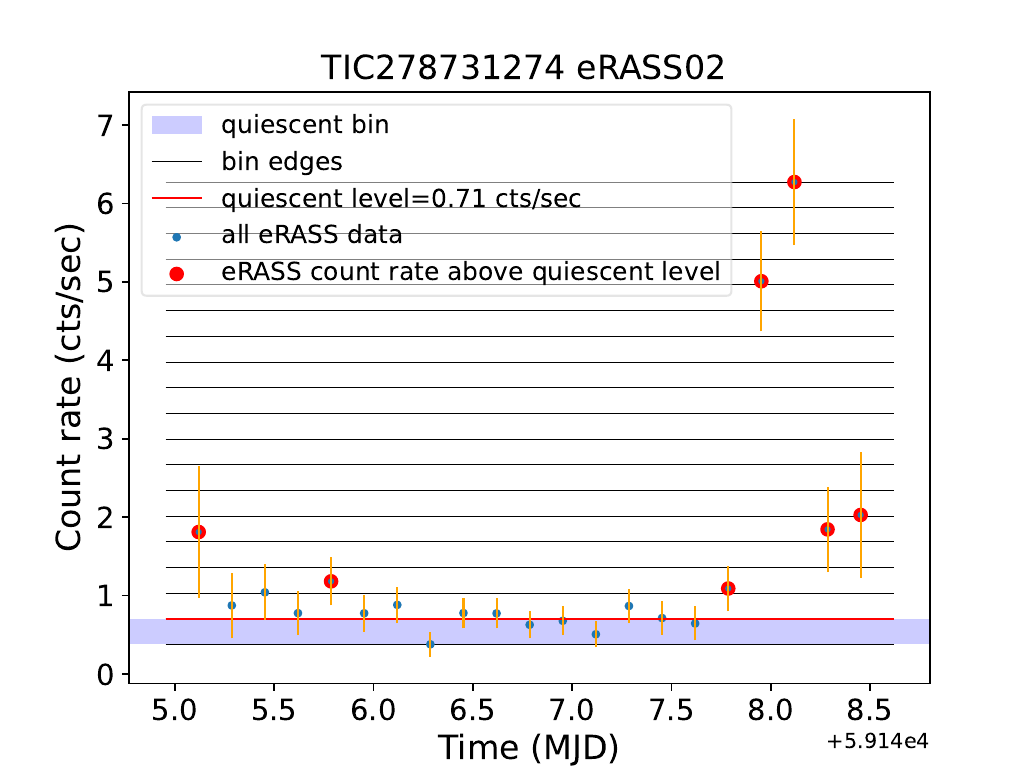}
\caption{Example of an eRASS light curve (the M dwarf TIC278731274 observed in eRASS2)
demonstrating the definition of the quiescent count rate level, \(R_{\rm qui}\), and the selection of variable light curve bins (namely eROdays significantly above $R_{\rm qui})$; see text in Sect.~\ref{sec:erass-qui-level} and Sect.~\ref{sec:var-diognostics-erass}.}
\label{fig:erass-qui-level}
\end{figure}

The bins which are located significantly above the quiescent level are considered to be ascribed to intrinsic variability of the star. These bins are marked in red in Fig.~\ref{fig:erass-qui-level}. 
We define {\it significantly above the quiescent level} as an upwards excursion of at least $1\,\sigma$. That is the lower bound of the $1\,\sigma$ uncertainty on the count rate must lie above the quiescent level, $R_{\rm qui}$.  

Flare counting is difficult in eRASS light curves because the gaps between subsequent data-taking intervals (eROdays) of $4$\,h  are on the order of the duration of or even longer than an X-ray flare while the individual visits to the source are on the order of $100$\,s, thus capturing only a snapshot in the evolution of a flare. To quantify the overall level of variability in eRASS light curves, we define a duty cycle (\(\nu_{\rm x,DC}\)) for the X-ray variability 
as the ratio of the number of bins corresponding to eROdays significantly above the quiescent level (\(N_{\rm x,var}\)) to the total number of bins in the light curve (\(N_{\rm x, total}\)),

\begin{equation}\label{eq:duty-cycle-erass}
\nu_{\rm x,DC}=\frac{N_{\rm x,var}}{N_{\rm x, total}}.   
\end{equation} 

We have computed the eRASS duty cycle individually for the light curves within the time intervals that have TESS observations. The eRASS duty cycle of the stars obtained from the light curves of the two surveys individually, ($\nu_{\rm x,DC,eRASS2}$ and $\nu_{\rm x,DC,eRASS3}$), along with the mean of the two values ($\nu_{\rm x,DC,mean}$) and their standard deviation ($\nu_{\rm x,DC,stddev}$) are provided for all stars in Table~\ref{table:tic-rqui-erassdc}.

\section{Analysis of TESS light curves}
\label{sec:TESS-lc}

The two-minute TESS light curves are of the form Pre-search Data Conditioning Simple Aperture Photometry (PDCSAP) flux vs. TIME. PDCSAP flux is used as long-term trends have been removed from the data and it tends to be cleaner and have fewer systematic errors than Simple Aperture Photometry (SAP) flux. TIME is in units of Barycentric Julian Date (BJD)-2457000.0. The light curves in this thesis have been normalized by us for analysis by dividing the values of all PDCSAP flux data points by their mean flux value. 
We apply our rotation and flare detection code to the normalized TESS 
light curves. Specifically, in this work we use the Python version which mainly consists in a translation of the IDL code used in our previous works \citep{Stelzer2016,Raetz2020,Stelzer2022HabitableZones}. 

\subsection{Detection of rotation and flares}
\label{sec:rot-flare-analysis}

The rotation period of the star can be obtained by analyzing the periodic modulation of optical light curves that arises from the presence of photospheric star spots. The amplitude of the variability provides information about the amount and distribution of spots on the surface. 
Our rotation and flare detection code uses the Lomb-Scargle periodogram from the Python package Astropy \citep{astropy:2013, astropy:2018, astropy:2022} for the search of the  rotation signal. The Lomb-Scargle periodogram is named after N.R.Lomb \citep{1976Ap&SS..39..447L} and J.D.Scargle \citep{1982ApJ...263..835S} (See also \citep{2018ApJS..236...16V}) and is a tool to find the periodicity of (unevenly) spaced observations. 
We calculated the Lomb-Scargle periodogram separately for each of the TESS light curves from the full light curves of the sectors that overlap with the times of eRASS2 and eRASS3 observations of our targets.

As a baseline we adopt the highest peak in the periodogram as $P_{\rm rot}$. To validate these periods we require that (i) a clear signal is seen in the phase-folded light curve and (ii) the highest peak has at least two times as much power as the second highest peak. 
We kept two period values that did not fulfil criterion (ii) because the second highest peak is at one-half the period of the highest peak and the light curve appears `double-dipped' (TIC 33866201 in eRASS2 epoch, TIC 220473309 in eRASS2 epoch).
These cases are usually ascribed to a spot pattern that changed from spots dominating one hemisphere to a more even distribution in terms of stellar longitude \citep{McQuillan2013}. In one case (TIC\,278735797) a clear periodic signal is present at both epochs with the value for $P_{\rm rot}$ differing by about a factor two. Here, visual  inspection showed that the light curve with the detection of the shorter period shows a more complex modulation.  Since  it is not essential for our study whether we have detected the rotation period in one or two epochs, we ignore the shorter $P_{\rm rot}$ value from the complex light curve at eRASS3 epoch for this star. 


To summarize, we were able to determine $P_{\rm rot}$ for $18$ of $25$ stars. All  periods, referred to as $P_{\rm rot,e2}$ and $P_{\rm rot,e3}$ respectively, are presented in Table~\ref{table:rot-tessdc-tessfr}. With the exception of TIC\,278735797 discussed above, if two periods were found for a given star from the two epochs the values are consistent with each other. 


 

The light curve with rotational modulation signal subtracted, obtained by applying boxcar smoothing and subsequently subtracting the smoothed curve, provides a "flattened" light curve from which the upward outliers can be identified as white light flare signatures or 'flare candidates'. From this flare signal, the rise and decay time of white light emission can be found as well as the flare energy emitted in white light by integrating over the flare light curve. The above procedure is explained in detail in \citep{Stelzer2016},\citep{Raetz2020} and \citep{Stelzer2022HabitableZones}. 
The validation of the flare candidates includes the use of an empirical flare template from 
\citep{Davenport2014}. Valid flares must satisfy the following criteria, as described by \citep{Raetz2020}, which we reiterate here:
\begin{enumerate}
    \item The flare shall  not be close to a data gap. 
    \item The ratio of the maxima (or peak) of the flare to the minima of the flare \(f(peak):f(minima)\) must be at least 2:1
    \item The rise time must be shorter than the decay time 
    \item The peak of the flare must not be the first or last point of the light curve
    \item The flare model 
    must fit better than a linear model 
\end{enumerate}

We calculated the TESS flare energies in ergs (also detailed in e.g., \citep{Stelzer2016},\citep{Raetz2020} and \citep{Stelzer2022HabitableZones}, which we reiterate here) by multiplying the equivalent duration ($ED$), which is defined as the area under the flare light curve (e.g.,\citep{Hunt-Walker2012}), with the quiescent TESS luminosity $L_{\rm qui,T}$. The $ED$ has units of seconds as we are integrating over the dimensionless normalized light curve. $L_{\rm qui,T}$ is determined from the quiescent flux in the TESS band ($F_{\rm qui,T}$) and the {\it Gaia} distances. $F_{\rm qui,T}$ is calculated from the zeropoint $ZP_{\rm TESS}=1.34\times10^{-9}\,{\rm erg\,cm^{-2}\,s^{-1}\,Å^{-1}}$, the TESS magnitude ($T$) and the filter bandwidth $W_{\rm eff} = 3898.68 Å$ as $F_{\rm qui,T}=ZP_{\rm TESS}\cdot10^{-0.4} \cdot T \cdot W_{\rm eff}$. $ZP_{\rm TESS}$ and $W_{\rm eff}$ were obtained from the website of the Spanish Virtual Observatory \footnote{\texttt{\url{http://svo2.cab.inta-csic.es/theory/fps/}}}.



%

\subsection{TESS variability diagnostics}
\label{sec:var-diognostics-tess}

For the sake of comparison with the X-ray data, in our study of optical variability we consider only that portion of the TESS light curves that corresponds to the epoch and duration of the eRASS visits to the respective star. We define two  parameters that describe the frequency of flares in TESS light curves. First, the TESS flare rate ($\nu_{\rm o,f}$) for each light curve is the number of validated flares that are detected in the light curve ($N_{\rm o,f}$) normalized by the length of the light curve in days ($\Delta t_{\rm o}$) excluding data gaps,

\begin{equation}\label{eq:flare-rate-tess}
\nu_{\rm o,f}\,[\rm d^{-1}] = \frac{N_{\rm o,f}}{\Delta t_{\rm o}}.
\end{equation}

\noindent
This parameter is the standard way of defining flare rates, see e.g. \citet{Stelzer2022HabitableZones}.

Secondly, a duty cycle can be defined for TESS light curves, $\nu_{\rm o,DC}$, analogously to the eRASS duty cycle as the ratio of the total number of flare bins in the light curve ($N_{\rm f\, pts}$) to the total number of bins in the TESS light curve ($N_{\rm all\, pts}$).

\begin{equation}\label{eq:duty-cycle-tess}
\nu_{\rm o,DC}=\frac{N_{\rm f\, pts}}{N_{\rm all\, pts}}. 
\end{equation}

As for the X-ray duty cycle, we have computed the optical duty cycle individually for the time intervals in the light curves that are covered by the two eRASS surveys ($\nu_{\rm o,DC,eRASS2}$ and $\nu_{\rm o,DC,eRASS3}$).
In some of our analyses, we use the mean of the optical (TESS) duty cycle $\nu_{\rm o,DC,mean}$, which is the average of $\nu_{\rm o,DC,eRASS2}$ and $\nu_{\rm o,DC,eRASS3}$. As a measure for the uncertainty of $\nu_{\rm o,DC}$ we use the standard deviation built from the two values. The flare rates and duty cycles are presented in Table.~\ref{table:rot-tessdc-tessfr}

\input{tables/tic_rot_tessfr_tessdc}

\section{Results}
\label{sec:results}
\subsection{Rotation period and X-ray activity}

\begin{figure}[h]

\centering
\includegraphics[width=\columnwidth]{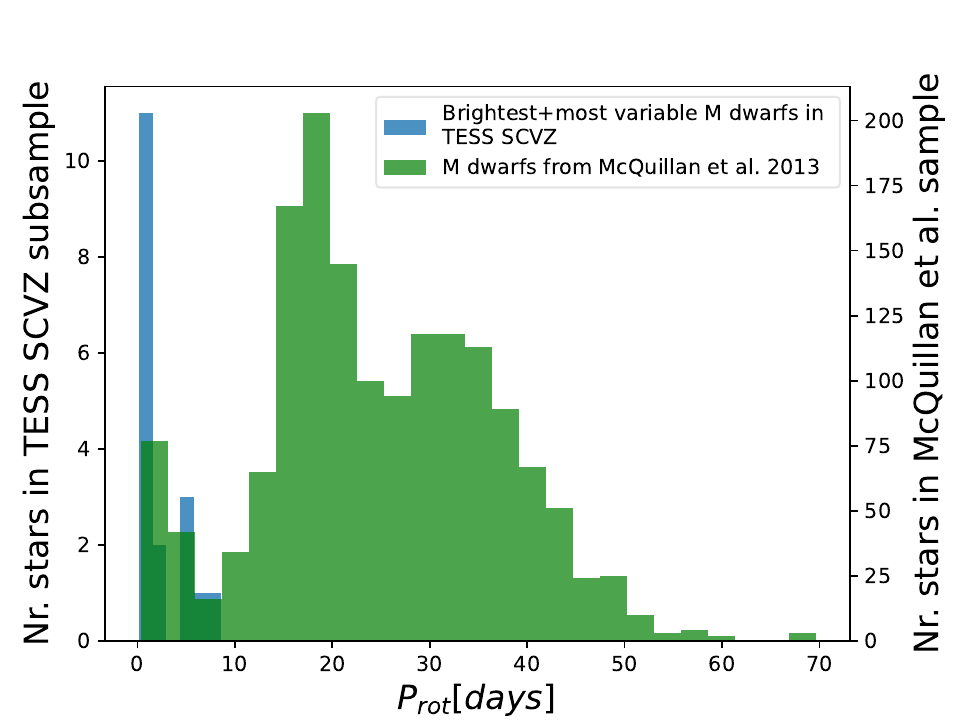}
\caption{ Distribution of rotation periods of our final M dwarf sample of the X-ray brightest or most X-ray variable stars (blue) and of the sample of all M dwarfs detected in a 10-month observational period of the 4-year {\it Kepler} mission from \citep{McQuillan2013} (green). 
}
\label{fig:rotation-period-dist}
\end{figure}

The distribution of $P_{\rm rot}$ values in our sample of $25$ M dwarfs is shown in Fig.~\ref{fig:rotation-period-dist} together with a large sample of M dwarfs observed with the {\it Kepler} mission (see below). 
All stars in the sample for which a period could be detected are fast rotators with the highest value being $P_{\rm rot} \sim 8.6$\,d. 
This can be explained by the fact that active stars have shorter rotation period, and the sub-sample of M dwarfs used in the analysis was specifically chosen as highly X-ray active (either by high count rate or strong variability in eRASS), as specified in Sect.~\ref{sec:sample-selection}. This sample is therefore biased towards active and fast rotating stars in comparison to, e.g., M dwarfs detected by the {\it Kepler} space mission whose properties were studied amongst others by  \citep{McQuillan2013}. 
Fig.~\ref{fig:rotation-period-dist}
illustrates the two  period distributions, from which the bias in our sample towards fast rotators becomes evident. 
On the other hand, periods longer than $15$\,d (roughly half the time-span of a  TESS sector light curve) can not be confidently measured, introducing another bias towards fast-rotating stars into our sample. 

\begin{figure} 
\centering
\includegraphics[width=\columnwidth]{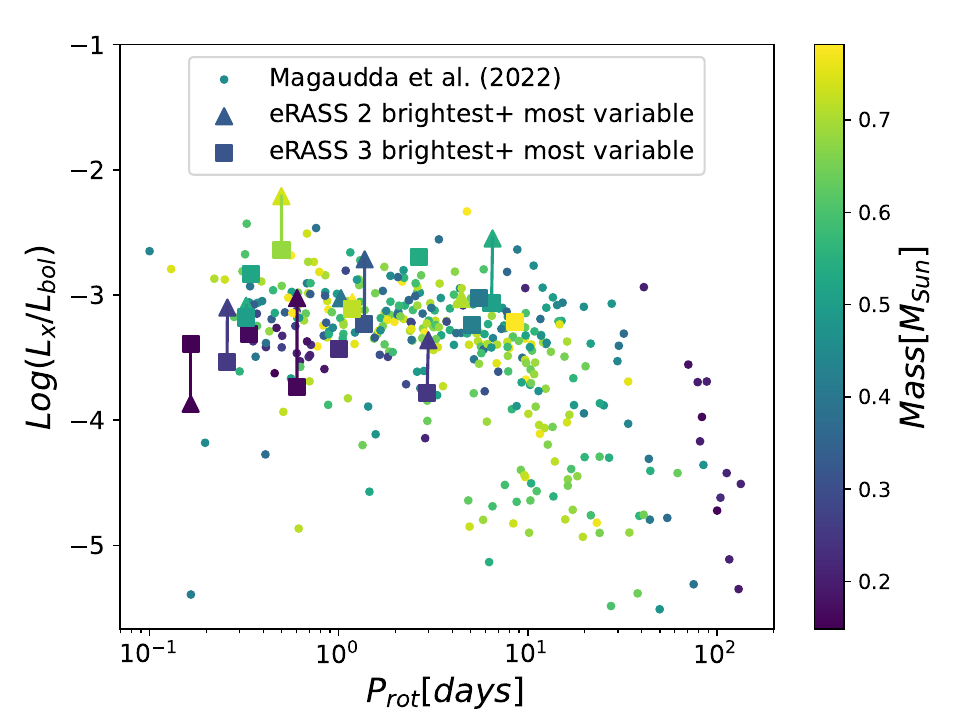}
\caption{Relation between X-ray activity and rotation of our sample (large triangles and squares for eRASS2 and eRASS3 respectively) compared to the M dwarf samples from \citep{Magaudda2020} and \citep{Magaudda2022} (dots). The stars are color-coded by mass and lines connect the two epochs for a given star from our sample.
}
\label{fig:rot-period-magaudda}
\end{figure}

With their short periods our stars lie in the saturated regime of the rotation-activity relation, as can be seen from Fig.~\ref{fig:rot-period-magaudda}. 
The range of $L_{\rm x}/L_{\rm bol}$ covered by our sample 
is in agreement with the spread of the saturation level observed in the so  far largest sample of M dwarfs examined for the rotation-activity connection \citep{Magaudda2020,Magaudda2022}. 
As compared to representations based on $L_{\rm x}$, the use of the $L_{\rm x}/L_{\rm bol}$ ratio removes most of the mass-dependence of activity \citep[except for a possible residual effect; see e.g.][]{Reiners2014,Magaudda2022AN}. The origin of the remaining spread of the saturated level of about one dex is not fully understood. Our data suggests that it is  
determined by variability together with a residual mass-dependence. The mean $\log{(L_{\rm x}/L_{\rm bol})}$ value of our targets shown in Fig.~\ref{fig:rot-period-magaudda} is $-3.13$ with a  standard deviation of $0.37$, and similar values are obtained for our full sample including the stars without rotation period measurement.


\subsection{Optical flare frequency distribution}
\label{subsect:optFFD}

\begin{figure} 
\centering
\includegraphics[width=\columnwidth]{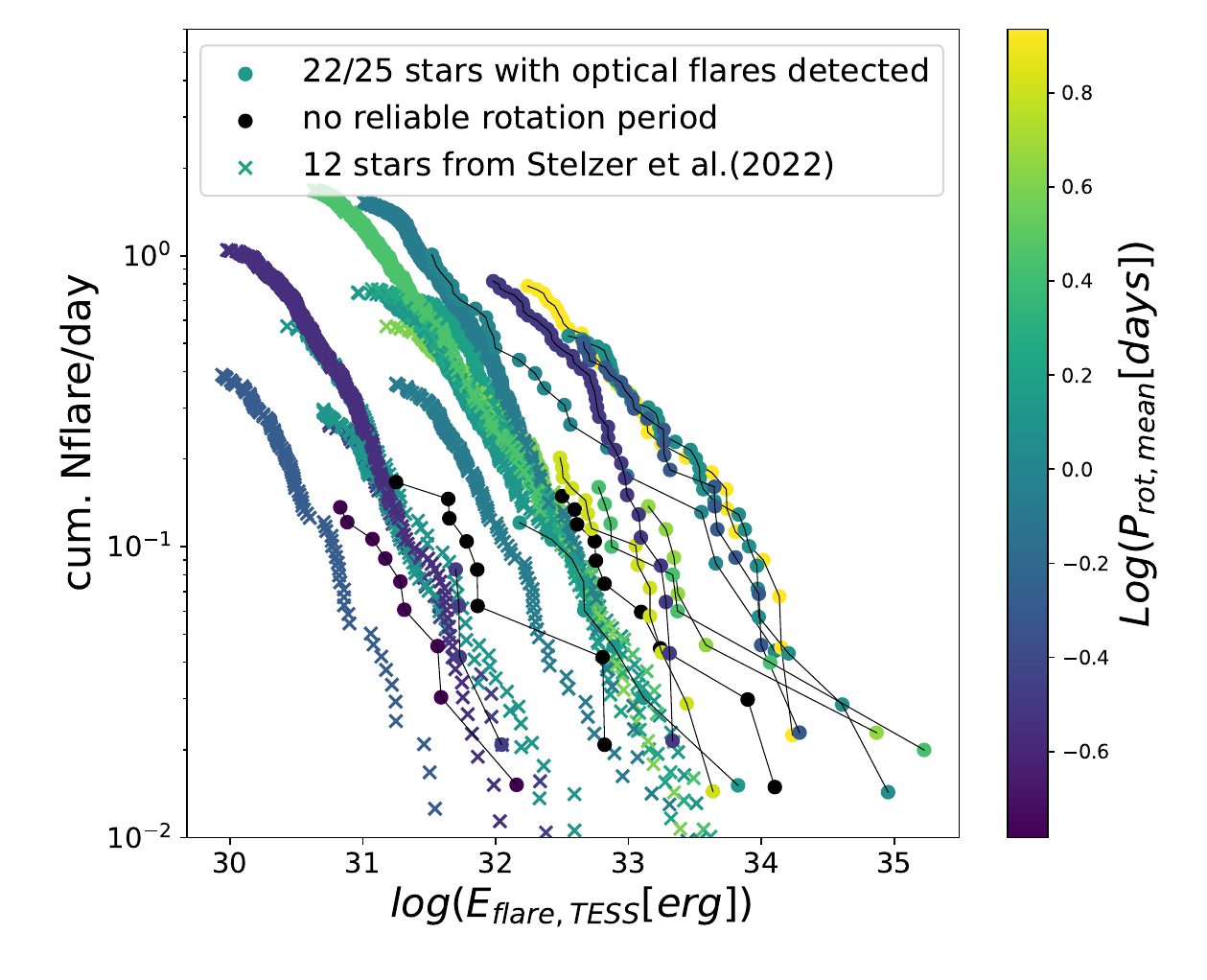}
\caption{
Optical flare frequency distributions of the $22$ targets with flares detected in the TESS light curves of Sectors overlapping with eRASS (circles). For comparison the FFDs for the fast rotating M dwarfs from the catalog by \citep{2019ApJ...874L...8K} that were studied by \citep{Stelzer2022HabitableZones} are shown (crosses).}
\label{fig:optical-ffd}
\end{figure}

We have computed for each of the $22$ stars from our sample that displayed optical flares its cumulative flare frequency distribution (FFD) from the TESS light curves of the two sectors that overlap with eRASS. The FFDs were calculated by combining the flare energies obtained from both sector's light curve into one FFD per star. The FFDs of our stars are shown in Fig.~\ref{fig:optical-ffd}, from which it is evident that our sample spans a large range in flare activity.

The most important parameter of FFDs is the slope of a power law fit, which is crucial for an assessment of the importance of nano-flare heating \citep[e.g.,][]{1991SoPh..133..357H,2002A&A...382.1070V}. 
At low energies FFDs are known to flatten  because small-amplitude flares are not detectable. The minimum detectable flare energy, which depends on the noise level of the light curve, can be determined by simulations. A  quantitative analysis of the TESS FFDs of our targets is not the scope of this paper as systematic studies of FFDs have been carried out on larger samples of M dwarfs 
\citep[e.g.][]{2020AJ....159...60G,Raetz2020}.  
Therefore, we have not computed the slopes and completeness thresholds of our FFDs.

In Fig.~\ref{fig:optical-ffd} we merely display the FFDs of our sample in comparison to a sample of nearby M dwarfs studied by \citep{Stelzer2022HabitableZones}. These latter ones are the $12$ stars from the catalog of \citep{2019ApJ...874L...8K} for which \citep{Stelzer2022HabitableZones} could determine the rotation period from the TESS light curves of all sectors in which the star was observed. Their $P_{\rm rot}$ values range between $0.3$ and $4$\,d, comparable to the periods of the stars from our combined eRASS/TESS sample. Yet, as can be seen from Fig.~\ref{fig:optical-ffd}, our sample is on average characterized by  higher optical flare frequencies. This confirms the bias towards strongly active stars introduced by our X-ray selection criteria.

\subsection{Global X-ray and optical variability diagnostics}

To compare the different information content of the TESS and eRASS light curves that derives from the very different observing cadence of the two surveys, we have defined in Sects.~\ref{sec:var-diognostics-erass} and~\ref{sec:var-diognostics-tess} global variability diagnostics. These represent the overall level of variability of a given light curve as a single number. Here we examine relations between these different variability diagnostics.

\begin{figure} 
\centering
\includegraphics[width=0.95\columnwidth]{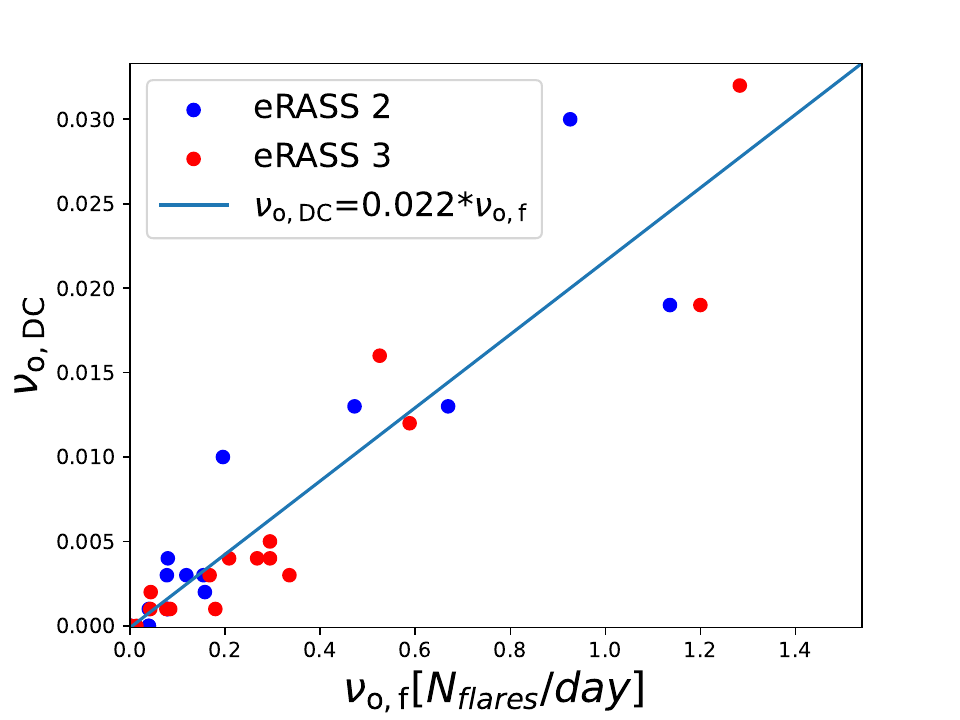}
\caption{TESS duty cycle \(\nu_{\rm o,DC}\) vs. TESS flare rate \(\nu_{\rm o,f}\), along with a linear fit to the data (blue line). For each star the TESS results from the data overlapping in time with the  eRASS2 and eRASS3 light curve, respectively, are distinguished by different color.}
\label{fig:tessdc_vs_tessfr}
\end{figure}

In particular, for both eRASS and TESS we have defined the duty cycle of variability in the same way, as fraction of time the star was in a `high-state'. 
We compare the optical (TESS) duty cycle ($\nu_{\rm o,DC}$) and optical flare rate ($\nu_{\rm o,f}$) to each other. Hereby, we evaluate the parameters for the two timespans corresponding to the eRASS2 and eRASS3 observation of the stars individually.
Fig.~\ref{fig:tessdc_vs_tessfr} demonstrates that the optical duty cycle correlates very well with the optical flare frequency. 
The Pearson's correlation coefficient between \(\nu_{\rm o,DC}\) and \(\nu_{\rm o,f}\) is $0.934$ with a p-value of $1.22 \cdot 10^{-17}$. We also perform a linear fit between \(\nu_{\rm o,DC}\) and \(\nu_{\rm o,f}\) and find that the line \(\nu_{\rm o,DC}\)=0.022*\(\nu_{\rm o,f}\) fits the data the best. 
This high level of correlation is not unexpected, as more flares in the light curve means that the overall time the star has spent at elevated level of brightness is higher. More precisely, the key difference between the TESS flare frequency and the TESS duty cycle is that the latter is function of both the flare rate as well as the duration of the flares. Since the flare frequency scales with the flare energy (Sect.~\ref{subsect:optFFD}), and the flare energy is determined by both the amplitude and the duration of the flare, the flare frequency is higher for events that last longer. Hence, the duty cycle increases as the frequency of flares increases. In the following, we use the TESS duty cycle as a proxy for the optical variability because this parameter is formally equivalent to the X-ray duty cycle that we have defined for eRASS light curves.

\begin{figure} 
\centering
\includegraphics[width=0.95\columnwidth]{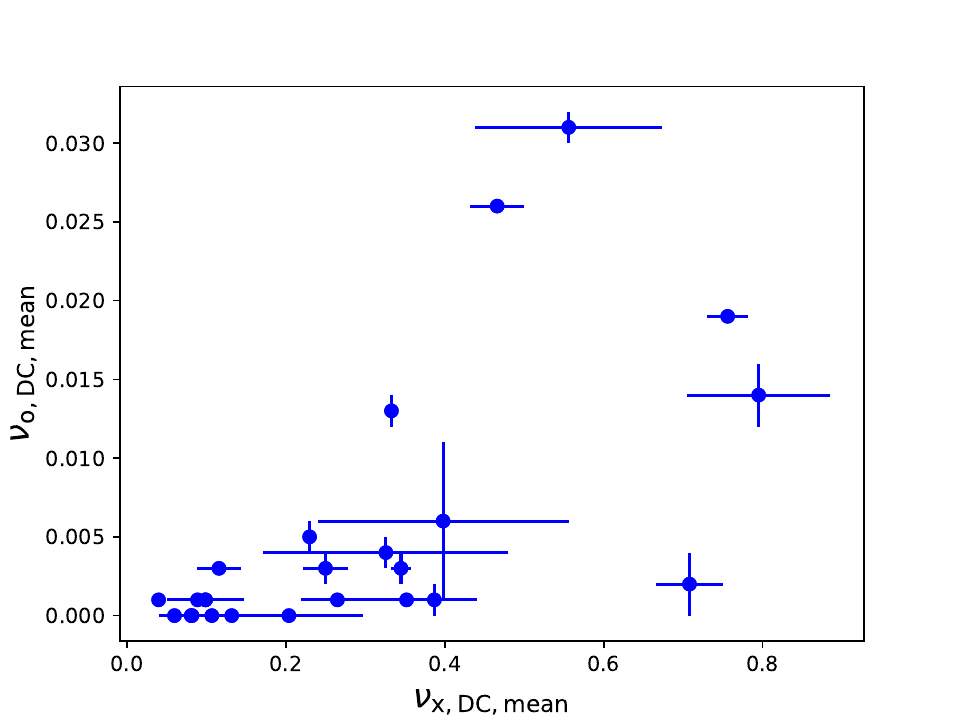}
\caption{TESS mean duty cycle vs. eRASS mean duty cycle, with the standard deviations between the values from the two eRASS epochs represented as uncertainties.}
\label{fig:flare-rate-vs-duty-cycle}
\end{figure}

The mean TESS duty cycle $\nu_{\rm o,DC,mean}$ 
is plotted against the mean eRASS duty cycle $\nu_{\rm x,DC,mean}$ 
in Fig.~\ref{fig:flare-rate-vs-duty-cycle}.
The Pearson's coefficient is $0.578$ with a p-value of $0.01$,  and the Spearman's rank correlation coefficient is $0.638$ with a p-value of $0.003$. The optical and the X-ray duty cycles, thus, show a significant correlation.
The still rather significant scatter in the relation might indicate that a range of multi-wavelength energetics is associated with flares. 
 The values for the TESS duty cycles are much lower than those of the eRASS duty cycles, which is likely a combination of the different data cadences of the two surveys and the different intrinsic duration of optical and X-ray events, as mentioned above.  The eRASS duty cycle 
may not directly be interpreted as a flare rate, because (a) flares may be missed in the $4$\,h gaps between two eROdays and (b) X-ray flares can last several hours such that the duty cycle is not equivalent to flare counting. 

\begin{figure} 
\centering
\parbox{\textwidth}{
\parbox{0.5\columnwidth}{\includegraphics[width=0.5\columnwidth]{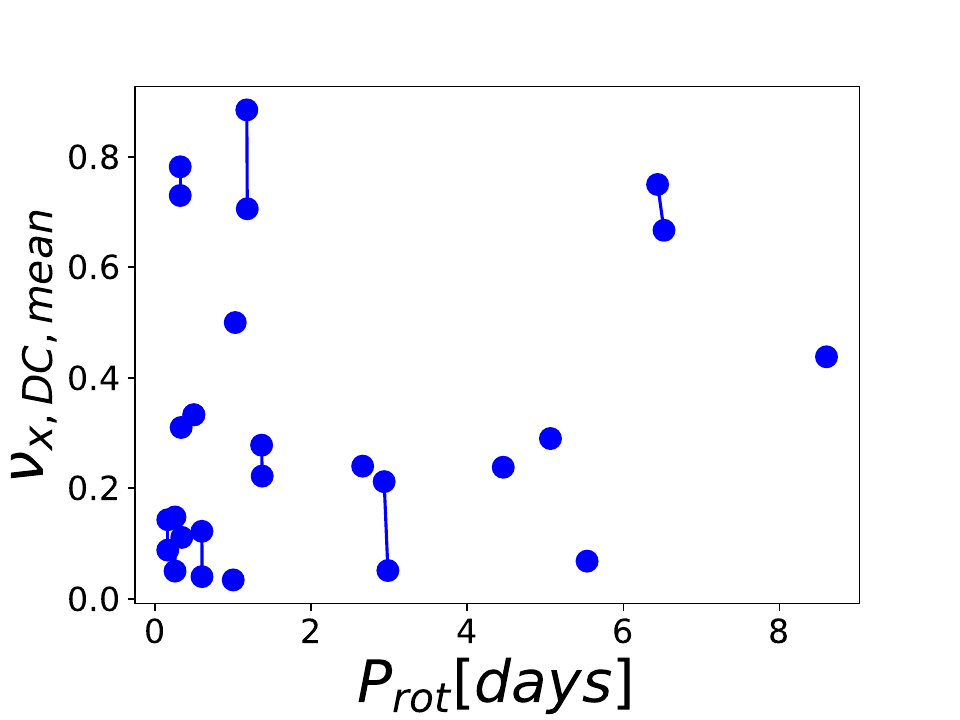}}
\parbox{0.5\columnwidth}{\includegraphics[width=0.5\columnwidth]{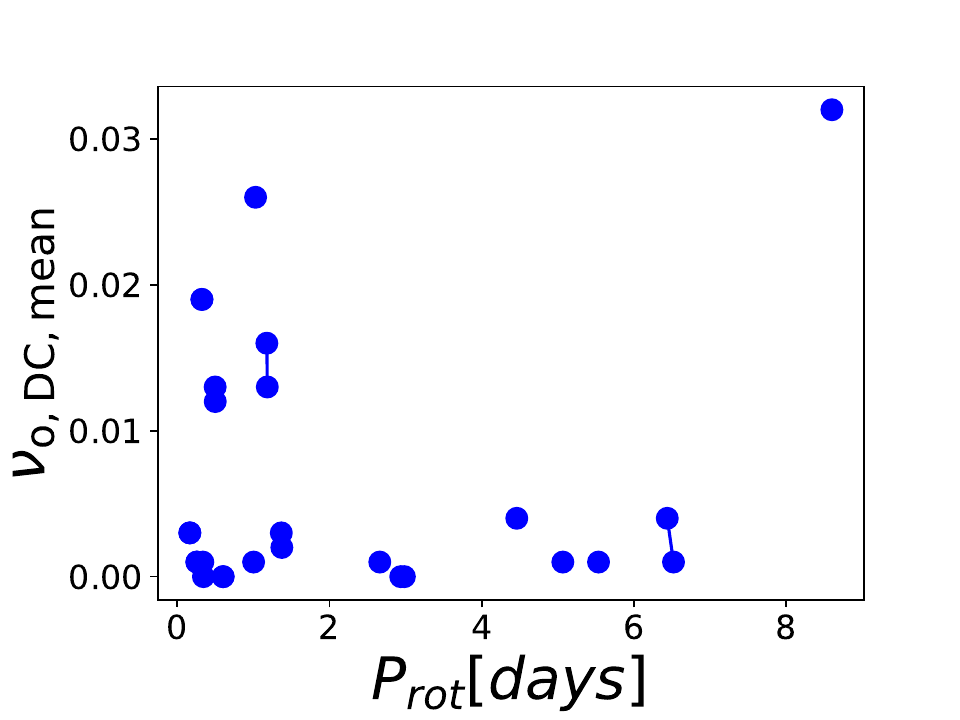}}
}
\caption{X-ray duty cycle \(\nu_{x,DC,mean}\) vs. rotation period 
(left) and optical duty cycle \(\nu_{\rm o,DC,mean}\) vs. rotation period 
(right). Lines connect the two epochs for a given star.}
\label{fig:duty-cycle-rot-period}
\end{figure}

We examine the relation between the activity duty cycles and the rotation period in Fig.~\ref{fig:duty-cycle-rot-period}. Here each star is represented by the black line connecting the duty cycles and rotation periods from the time interval of eRASS2 and eRASS3 observations. Recall that the TESS duty cycle was calculated only for the time of overlap between TESS and eRASS observations.
There is trend of high duty cycles, both in X-rays and in optical, occurring predominantly on fast rotators ($P_{\rm rot} \lesssim 2$\,d). However, the star with the highest value of $\nu_{\rm o,DC}$ has the longest period. This star has a very prominent double-humped TESS light curve, which means it is strongly spotted. A sample with a broader range of $P_{\rm rot}$ would be needed to establish an activity-rotation relation for the duty cycles, as other studies have shown a bimodal distribution separating active from inactive stars at a value of $P_{\rm rot} \sim 10$\,d (e.g., \citep{2011ApJ...743...48W,Raetz2020,Magaudda2022}).

\subsection{Association of optical flares to X-ray flares}
\label{sec:lc-erass-flare-tess-flare}

The association of the eRASS X-ray variability with the distinct optical flares seen with TESS is difficult as a result of the long data gaps (of $4$\,h) between individual X-ray exposures during the eRASS. To identify joint optical and X-ray variability we examined for each
`X-ray flare data point' (XFDP), defined as light curve bin significantly above the quiescent level (red in Fig.~\ref{fig:erass-qui-level}; see Sect.~\ref{sec:erass-qui-level}), whether there is an optical flare nearby in time. 

To overcome the problem  of the low cadence of the eRASS light curves we resort to our knowledge about the typical duration of stellar flares and the timing between the  optical and X-ray component of a flare. In the standard flare scenario \citep{cargill83.0,2017LRSP...14....2B} a time lag is expected between the optical and X-ray flare peak. In fact, the time difference between the peak of the optical and X-ray brightness in a giant flare on the M dwarf AD\,Leo was \(\approx300\,sec\) \citep{Stelzer2022}, and similar delays have been observed in other stellar flares (e.g.,\citep{Katsova2002, Stelzer2006}) as well as solar flares \citep{CastellanosDuran2020}.
The typical optical/X-ray time lag of a flare is, thus, much shorter than one eROday, the cadence between subsequent bins in the eRASS light curves. Therefore, optical flares are expected to be found  around some XFDPs in a time window that is a fraction of 1 eROday.
Other XFDPs are expected to have no optical counterpart within an eROday as the X-ray flare is part of the `gradual' phase of flares \citep{hudson11.0} 
and its duration is far longer than that of the optical flare which belongs to the `impulsive' phase. This solar flare morphology has been confirmed for events on other stars, see e.g.,\citep{Stelzer2022}. 

\begin{figure} 
\centering
\includegraphics[width=\columnwidth]{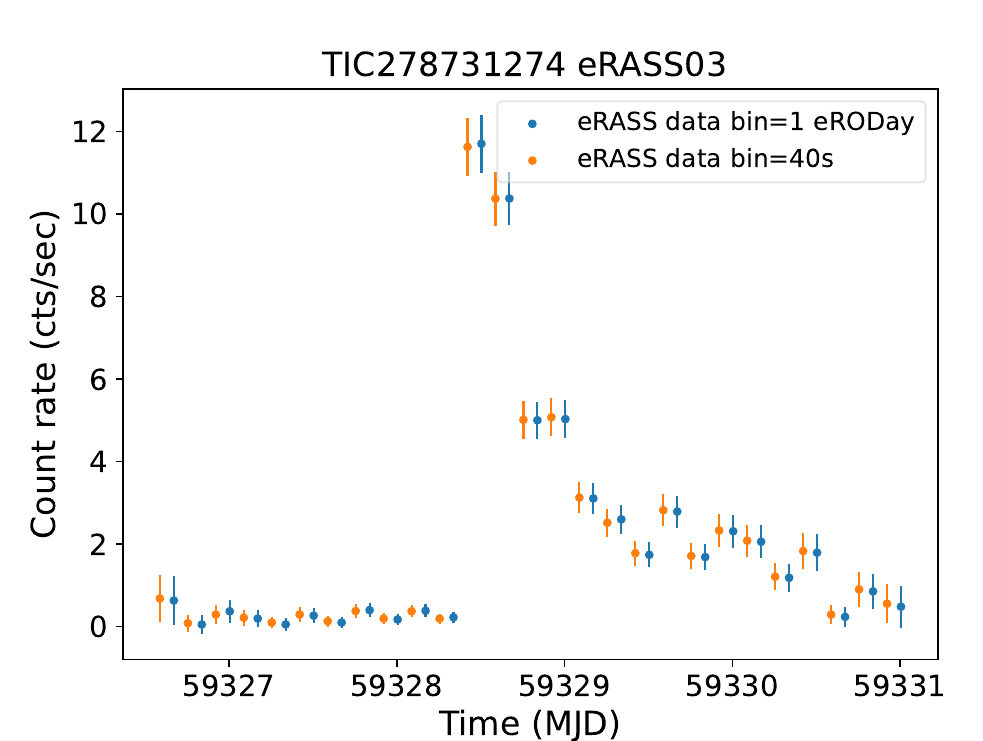}

\caption{Example of an eRASS light curved binned at $40$\,s, with bins close in time combined (orange) and the same light curve binned at 1 eRODay. For this example the time-shift between the two light curves is roughly 2 h or 1/2 eRODay. 
}
\label{fig:erass-timing-diff}
\end{figure}

To make sure that we do not miss joint optical/X-ray flare events we searched all XFDPs for TESS flares in time intervals of $\pm 1/2$\,eROday centered on the time within that eROday when eROSITA observed the star. 
This required a different binning of the eRASS light curves as compared to the one used above in the standard analysis. The eRASS light curves used above were binned with a bin size of $4$\,h which is the cadence of eROSITA visits to a given sky position, as explained in Sect.~\ref{sec:erass-lc}. However, the actual exposure on the source during this $4$\,h timespan is only on the order of $40$\,s, meaning that most of an $4$\,h-bin holds no data. We do not know at which time within the 4\,h interval eROSITA actually took data of the source. In other words, in a light curve binned with the $4$\,h (alias 1 eROday) cadence every data point has a timing uncertainty of $4$\,h. 
Given the typical duration of optical flares of $\lesssim 10$ min \citep{Hawley2014, Namekata17}  and the typical time lapse between optical and X-ray flare peaks of $\sim 5$\,min (see above) such a high uncertainty in the timing prohibits an association between X-ray and optical flares. 
To recover X-ray timing information the binsize must be on the order of the net on-source time within an eRODay, but this information is not trivial to extract.
We, therefore, produced a second set of eRASS light curves, binned into $40$\,s intervals. Evidently, this produces many empty bins but it improves the time-resolution by a factor $360$. An example of the improved accuracy in timing is illustrated in Fig.~\ref{fig:erass-timing-diff} which shows the same light curve binned with the method just described together with the light curve binned at intervals of one eRODay. A systematic shift of the two light curves by roughly 1/2 eRODay is seen in this 
case. Generally, time shifts between the two binnings range from 1/8 eRODays to 1/2 eRODay for our stars which means that the light curve with the coarse (1\,eRODay) binning is far from providing the best possible estimate for the peak time of the flare. The drawback of eRASS light curves with small bin size is that it happens that the short on-target time during one eRASS visit of the source is split onto two $40$\,s bins. This way of binning the eRASS data, thus, may produce some bins with extremely short exposure that have huge errors on the count rate. Since these bins hold only a minor fraction of the counts, excluding them from the analysis does not lead to a major loss of data. For the representation in Fig.~\ref{fig:erass-timing-diff} as well as for the further analysis, we, thus, removed bins with fractional exposure {\sc fracexp} $< 0.01$ and we combined, in the remaining light curve, the bins which were very close in time (i.e. belonging to a single eRODay) by averaging their time, rate and errors.

\begin{figure} 
\centering
\includegraphics[width=\columnwidth]{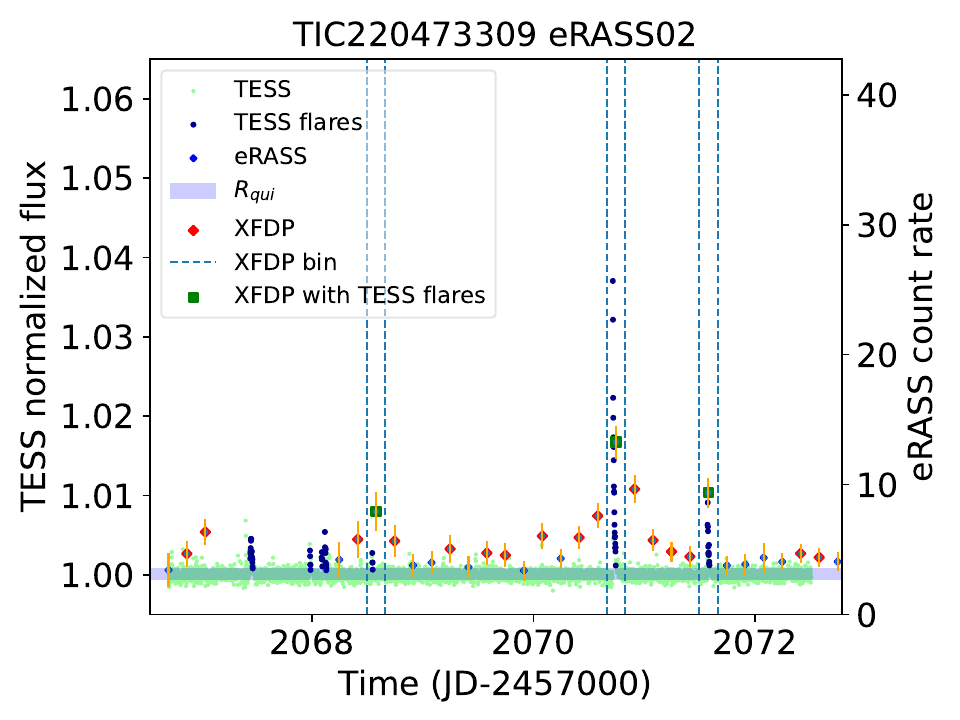}
\caption{
Example of a flattened TESS light curve and eRASS light curve for TIC\,220473309 observed during the eRASS2 epoch of the star. The eRASS data points significantly above the quiescent level (purple-shaded area), also named as the ``X-ray flare data points", are marked in red. The TESS flare data points are marked in dark blue. X-ray flare data points around which TESS flares were found are marked as green squares, and the vertical bins of width 1\,eROday in which they lie are marked with dashed blue lines.}
\label{fig:lc-erass-flare-tess-flare}
\end{figure}

With this procedure we have again one bin per eROday in the light curve, as before in the standard analysis. However, this bin is now positioned at the actual time when eROSITA observed the star.
This means, that in these new eRASS light curves we now know the time of observation with $40$\,s precision, and we can proceed towards our goal of searching for the optical counterpart of X-ray flares. In Fig.~\ref{fig:lc-erass-flare-tess-flare} we show an example of the outcome of this search: The XFDPs are marked in red like in Fig.~\ref{fig:erass-qui-level}. 
The time windows of  $\pm 1/2$\,eRODay around eROSITA X-ray flare bins that comprise/include TESS flare bins are  marked with blue dashed lines, and these XFDPs are highlighted as green squares. The respective TESS flares occurring up to half an eROday before and after an X-ray flare data point are marked in dark blue.  

To scientifically evaluate our search for joint optical/X-ray variability we examine what fraction of all XFDPs in a given eRASS light curve have a TESS flare preceding it  within $1/2$ eRODay {before the XFDP. 
The number of XFDPs with a corresponding optical flare found in this interval is plotted against the total number of XFDPs in the eRASS light curve in Fig.~\ref{fig:xray-flare-bin-vs-tot-bin}. 
First, it is noticed that a high number of X-ray flare bins have no optical event associated with them. This might occur for the following reasons: (i) The XFDP is part of the advanced decay phase of the X-ray flare. 
As discussed above, X-ray flares last hours to days, whereas optical flares decay on timescales of a few minutes. Fig.~\ref{fig:lc-erass-flare-tess-flare} shows several such events in which the X-ray flare probably lasted for several eRODays.  (ii) There is a true absence of an optical counterpart to the X-ray event. There have, indeed, been instances from the solar literature where X-ray flares have been observed with no optical counterpart \citep{Namekata_2020}.
Overall, 
both numbers appear to be correlated. 
The Pearson’s coefficient for the correlation of the two parameters is  0.771 with a p-value of 3.75e-11. 
This means that for light curves with many X-ray flares occurring, chances to find them occurring together with optical flares are high. 

\begin{figure}
\centering
\includegraphics[width=\columnwidth]{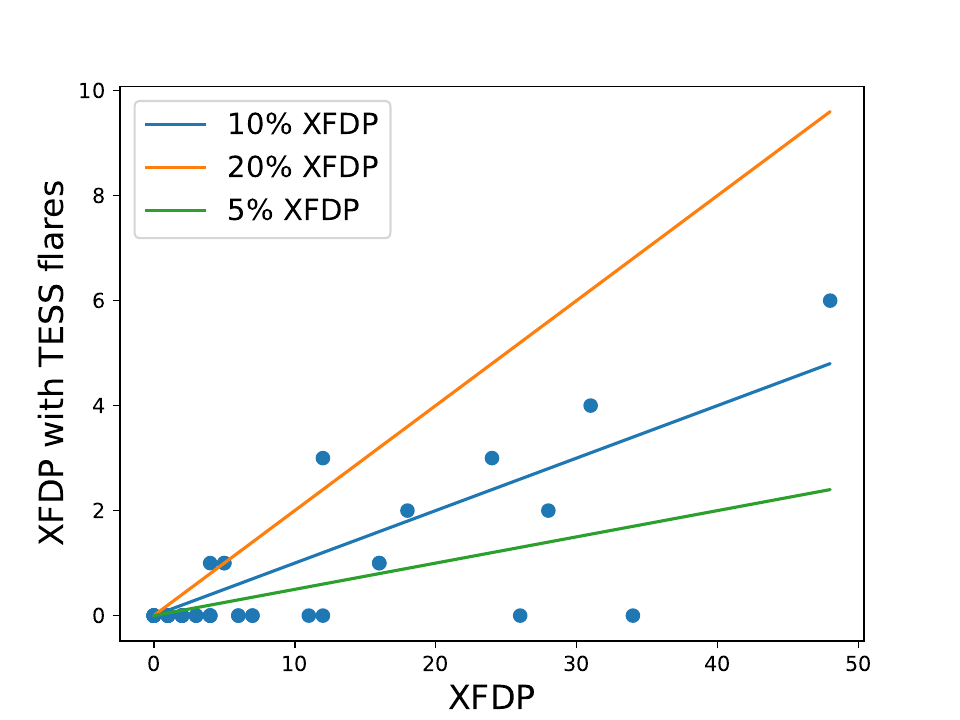}
\caption{XFDPs that are preceded by TESS flares vs. all XFDPs per eRASS light curve. The orange, blue and green lines denote 20\%, 10\% and 5\% of XFDPs that are preceded by TESS flares, respectively. }
\label{fig:xray-flare-bin-vs-tot-bin}
\end{figure}

\subsection{X-ray flare energetics}
\label{subsect:fit_xflare}

As discussed before, the low eRASS data cadence makes it difficult to quantify X-ray flares. Given a typical X-ray flare duration of one to a few hours, plausibly even a single individual XFDP which has `quiescent'  bins on both of its sides may represent a full-fledged flare, for which however both rise and decay phase can not be constrained. To provide an (albeit coarse) estimate of the X-ray flare energies from the poorly sampled eRASS light curves we make inferences from solar data combined with the observed TESS flares to constrain the timing and flux of the peaks of the X-ray flares. Then, we perform an exponential fit starting with the estimated X-ray flare peak. Finally we derive the X-ray flare energy  by integrating under the fit curve. We caution, that this analysis presents many severe approximations and assumptions as we detail in the now following description of the procedure. Yet, as will be shown below it produces reasonable results.

\subsubsection{Estimate of the X-ray flare peaks}
\label{subsubsect:xrayflare_peakestimate}

We estimated the eRASS count rate of the (unobserved) X-ray flare peak from the energy of its corresponding TESS flare with help of the empirical relation from \citep{Namekata17} for solar flares between the GOES soft X-ray flux ($F_{\rm peak,GOES}$) at the peak and the energy of the corresponding WLF ($E_{\rm WLF}$).
We used the updated fit from \citep{Stelzer2022} which includes data from the Nov 2022 Great Flare on the M dwarf AD\,Leo, and spans a much broader range of GOES peak flux and WLF energy than the solar flares alone. To be able to apply the $E_{\rm WLF} - F_{\rm peak,GOES}$ relation, we first calculated the WLF energy of the TESS flare. 
Hereby we assumed the optical flare to be represented by blackbody emission at a constant temperature of $T_{\rm f} = 9000$\,K \citep[as is typically observed for solar WLFs,][]{kretzschmar11.0}. 
Then we followed \citet{Shibayama2013} to calculate the emitting area $A_{\rm F}(t)$ (assumed to be variable along the event), and then we integrated the bolometric flare luminosity $L_{\rm WLF}(t) = \sigma T_{\rm f}^4 \cdot A_{\rm F}(t)$ over the flare light curve to find $E_{\rm WLF}$. The WLF energy results a factor $5$ higher than the flare energy in the TESS band.
Next, we obtained $F_{\rm peak,GOES}$ from the above mentioned relation with $E_{\rm WLF}$. The GOES flux
is the flux of the flare peak 
observed at $1$\,AU from the source (namely the distance of the GOES satellites  from the Sun). This must be translated into the peak flux observed from the star at Earth, by scaling with the star's distance calculated from the {\it Gaia} parallax. The resulting value still refers to the GOES band ($1.5-12.4$\,keV).
The last step was to convert the flux from the GOES band into the eRASS band used for our light curves ($0.2-5.0$\,keV) and obtain the corresponding count rate. To achieve this we used the Great X-ray flare on AD\,Leo \citep{Stelzer2022} for which the peak flux was measured both in the eRASS and in the GOES band using {\it XMM-Newton} data, yielding a flux ratio of 
$1.8$ (priv comm., M. Caramazza). By applying this conversion factor to our data we assume that the relative flux distribution of the flare peak in different X-ray sub-bands is universal for all X-ray flares. Finally, we converted the flux in the eRASS band to count rate for use in our eRASS light curves. Here we applied the  conversion factor $1.1 \cdot 10^{-12}\,{\rm erg/cts/cm^2}$ calculated from eRASS data of Prox Cen which is to our knowledge the best representative of eROSITA flare data that has been analysed so far (Magaudda et al., in prep.).  

\begin{figure} 
\centering
\includegraphics[width=\columnwidth]{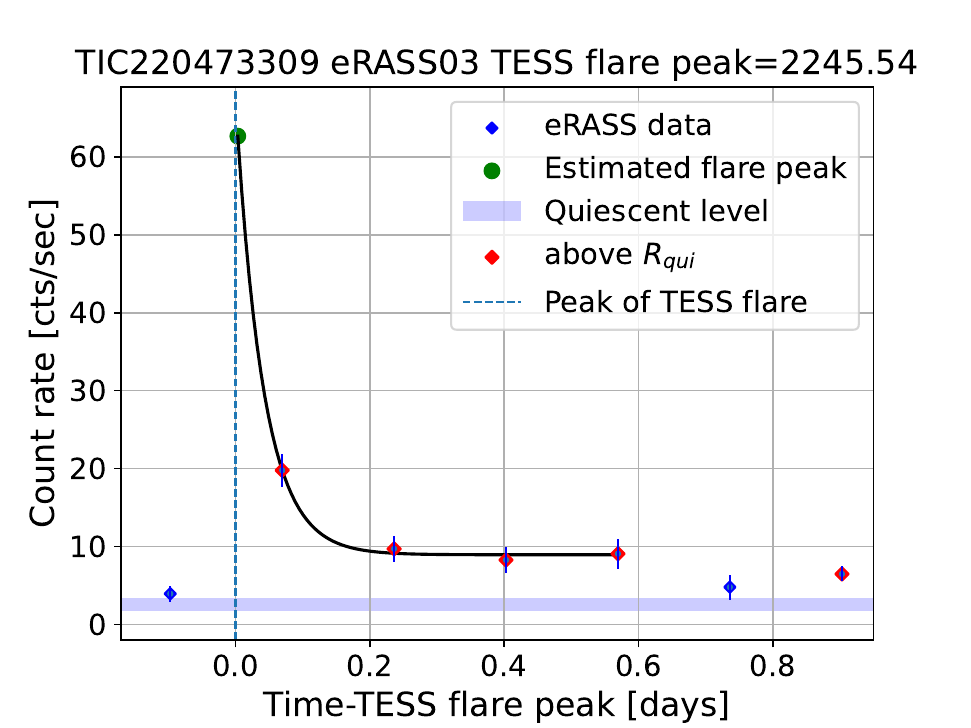}
\caption{eRASS3 light curve of TIC\,220473309 illustrating our estimate of the X-ray flare peak (green data point) as explained in Sect.~\ref{subsubsect:xrayflare_peakestimate} and the exponential fit to the flare decay (black curve) described in Sect.~\ref{subsubsect:xrayflare_expfit}.
} 
\label{fig:erass-flare-peak}
\end{figure}

The result of this approach is that we gain an additional data point for the X-ray light curve, representing the peak of the flare. An example of this is shown in Fig.~\ref{fig:erass-flare-peak}. 
We fixed the position of the X-ray flare peak in time to $5$ minutes after the TESS flare peak which is the typical time-delay between optical and X-ray peaks of solar and stellar flares \citep{CastellanosDuran2020, Stelzer2022}.
Evidently, the count rate of the X-ray flare peak 
estimated from the associated optical flare energy can be much higher than the highest {\it observed} eRASS flux. Similarly, the time of the X-ray flare peak 
inferred from the typical delay with respect to the associated optical flare can be significantly before the eROday with the highest count rate. It is, therefore, clear that the energetics of the X-ray flare rely strongly on the 
assumptions made in deriving this `synthesized' flare peak. This is explored in the next subsections. 

\subsubsection{Fit of the X-ray flare profiles}
\label{subsubsect:xrayflare_expfit}

Having an estimate of the peak time and count rate we can fit a decaying exponential curve to the flare count rate ($R$) as $R = R_{\rm post} + R_{\rm peak} \cdot e^{-b (t-t_{\rm x0})}$, where $R_{\rm post}$ is the post-flare count rate, $R_{\rm peak}$ is the estimated eRASS flare peak count rate,  
$b = 1 / \tau$ is the decay coefficient with $\tau$ the decay timescale, and $t_{\rm x0}$ is the start time of the X-ray flare. Note that $R_{\rm peak}$ and $t_{\rm x0}$ are not identical to the X-ray flare peak data point that we defined in Sect.~\ref{subsubsect:xrayflare_peakestimate}. $R_{\rm peak}$ is a free fit parameter, and $t_{\rm x0}$ is allowed to vary within a range of $5-10$ minutes  after the TESS flare peak.

The exponential fit is applied to the portion of the eRASS light curve starting $5$\,min after the peak of the optical flare (corresponding to the synthesized X-ray flare peak) and extending for  $16$\,h. This limit on the time length of the fitted profile appears appropriate for the following reasons: (1) Sixteen hours capture $5$ data points (the synthesized peak and four subsequent eRODays) which is the minimum needed for fitting as imposed by our 4-parameter fit function, (3) by-eye inspection revealed that typically after $16$\,h, the flare is either well into its decay and flattens out or a new flare starts. 
We can not rule out that this time-interval comprises subsequent variability not directly related to the X-ray counterpart of the optical flare. Indeed, if the flare decays back to the quiescent level, $R_{\rm post}$ should be identical to $R_{\rm qui}$ defined in Sect.\ref{sec:erass-qui-level}. However, in practice the fitted post-flare count rate is usually different from the quiescent count rate either because the corona remains in an elevated brightness after the flare (see e.g. the Great Flare on AD Leo; \citep{Stelzer2022}) or because the low number of eRASS bins does not fully constrain the decay or because subsequent flaring activity is overlaid on the first event. 

For the 58 TESS flares from which the synthetic eRASS flare peak was estimated, we identified 21 X-ray flares by eye, for which the exponential decay fit the X-ray flare light curve relatively well.
The X-ray light curves and the exponential fits of these events are presented in Appendix.\ref{fig:exp-fits}. 
The stars TIC\,150359500 and TIC\,278731274 present TESS flares with particularly strong eRASS variability that, however, could not be fit with the exponential model. These peculiar flares are discussed in Sect.~\ref{subsubsect:flares_strange}.

\subsubsection{X-ray flare energies and comparison to optical flare energies}
\label{subsubsect:xrayflare_optflare_energies}

Going one step further, we can estimate the energy of the X-ray flare ($E_{\rm x,f}$) from the profile of the exponential fit. Flare energies are commonly calculated by multiplying the quiescent luminosity with the integral over the light curve that was before normalized to quiescence, as described in Sect.~\ref{sec:rot-flare-analysis} for the TESS flares.  
Applying this to the eROSITA X-ray light curves, posits the question of the appropriate quiescent count rate. As mentioned above, for a quiet light curve with individual flares superposed this should be the count rate $R_{\rm qui}$ determined in Sect.~\ref{sec:erass-qui-level}. However, Fig.~\ref{fig:erass-flare-peak} and Appendix~\ref{fig:exp-fits} show that this low count rate level is often not reached after the X-ray flare. We, therefore performed the calculation of $E_{\rm x,f}$ for two values of the `quiescent' count rate, $R_{\rm qui}$ and $R_0$. The latter is the last count rate of our fitted curves representing the flattened portion of the decay, i.e., the count rate at 16h after the TESS flare peak. 

The calculation of $E_{\rm x,f}$ was done as follows, separately for both versions of the `quiescent' emission level. 
We first normalized the count rates of all light curve bins that have participated in the fit to the `quiescent' count rate,
and then we subtracted the normalized `quiescent' count rate, which is equal to 1.
This way we obtained a "flare-only" light curve. 
We integrated under this curve to obtain the $ED$ which has units of seconds as we are integrating over the dimensionless normalized curve. 
The X-ray flare energy, $E_{\rm x,f}$ was found by  multiplying the $ED$ with the luminosity 
obtained from the count rate $R_{\rm qui}$ or $R_{\rm 0}$ (for the two versions) after applying the `quiescent' rate-to-flux conversion factor from \cite{Magaudda2022} (see Sect.~\ref{sec:gaia-2mass-params}) and the {\it Gaia} distance of the star.


In Fig.~\ref{fig:xray-tess-flare_energy} we compare the resulting 
values of $E_{\rm x,f}$ with the energies from their optical counterparts found with TESS, $E_{\rm f,TESS}$ (see Sect.~\ref{sec:rot-flare-analysis}) for both assumptions of the quiescent level, $R_{\rm qui}$ in red and $R_{\rm 0}$ in green. 
The Pearson's correlation coefficient between $\log(E_{\rm x,f})$ and $\log(E_{\rm f,TESS})$ was found to be 0.907 with a p-value of 1.396e-8 for the exponential decays normalized to $R_{\rm qui}$, and 0.839 with a p-value of 2.03e-6 for the exponential decays normalized to $R_0$, demonstrating that
for both assumptions on the post-flare count rate the optical and X-ray flare energies are well correlated. 
\begin{figure} 
\centering
\includegraphics[width=\columnwidth]{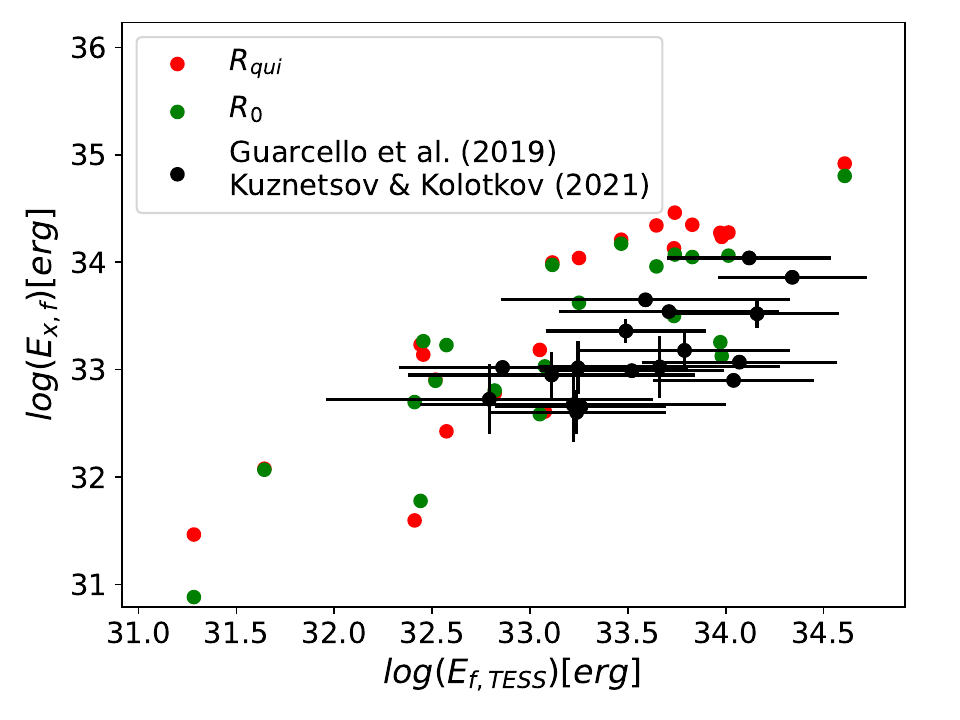}
\caption{X-ray flare energies from exponential fits normalized to $R_{\rm qui}$ (red) and $R_{0}$ (green)  respectively, vs. TESS flare energies for the 21 X-ray flares selected by eye. Energies of flares reported by \cite{Guarcello2019} and \cite{KuznetsovKolotkov2021} are marked in black.
}      
\label{fig:xray-tess-flare_energy}
\end{figure}

When compared to the small sample of joint X-ray and optical flares on M dwarfs presented in the previous literature (black in Fig.~\ref{fig:xray-tess-flare_energy}) our X-ray flares appear overluminous with respect to the corresponding  optical flares. This could be the result of the many severe assumptions made in Sect.~\ref{subsubsect:xrayflare_peakestimate} to constrain the X-ray counterpart of the TESS flares. While locating the X-ray flare peak time and flux with help of the TESS flare on the basis of earlier solar and stellar flare  observations by \cite{Namekata17} and \cite{Stelzer2022} appears solid, we observe that in many cases this synthesized flare peak exceeds the highest count rate observed in the aftermath of the TESS flare by high factors (see Fig.~\ref{fig:exp-fits}). 
This 
may lead to an overestimation of $E_{\rm x,f}$.
On the other hand, X-ray flares usually decay slower than an exponential, such that under this aspect our $E_{\rm x,f}$ values might be underestimated. As Fig.~\ref{fig:xray-tess-flare_energy} shows our 
distribution of flares to be located above that of the joint optical/X-ray flares presented in the literature, we suspect  that our X-ray peak flux estimates are too high. 

We, therefore, repeated the fitting described above after arbitrarily scaling down the inferred flare peak (green `data' point in Fig.~\ref{fig:erass-flare-peak}) by a factor two in count rate. In Fig.~\ref{fig:efx_efopt_2versions} we compare the new values for $E_{\rm x,f}$ with the optical flare energies, displaying as well the earlier result. We show only the results where $E_{\rm x,f}$ was evaluated with respect to the last fitted data point ($R_0$), because - as explained above - many flare seem not to decay all the way to $R_{\rm qui}$ and the corresponding values of $E_{\rm x,f}$ are, therefore, too high. The down-scaling of the X-ray flare peak yields on average only slightly lower values of $E_{\rm x,f}$ for a given optical flare. From visual inspection it is not evident which of the two versions of the synthetic X-ray flare peak yields better fits. Since our analysis shall serve has an explorative character we leave a more detailed  quantitative investigation of eROSITA flare light curves for future work.
%
\begin{figure}
\includegraphics[width=\columnwidth]{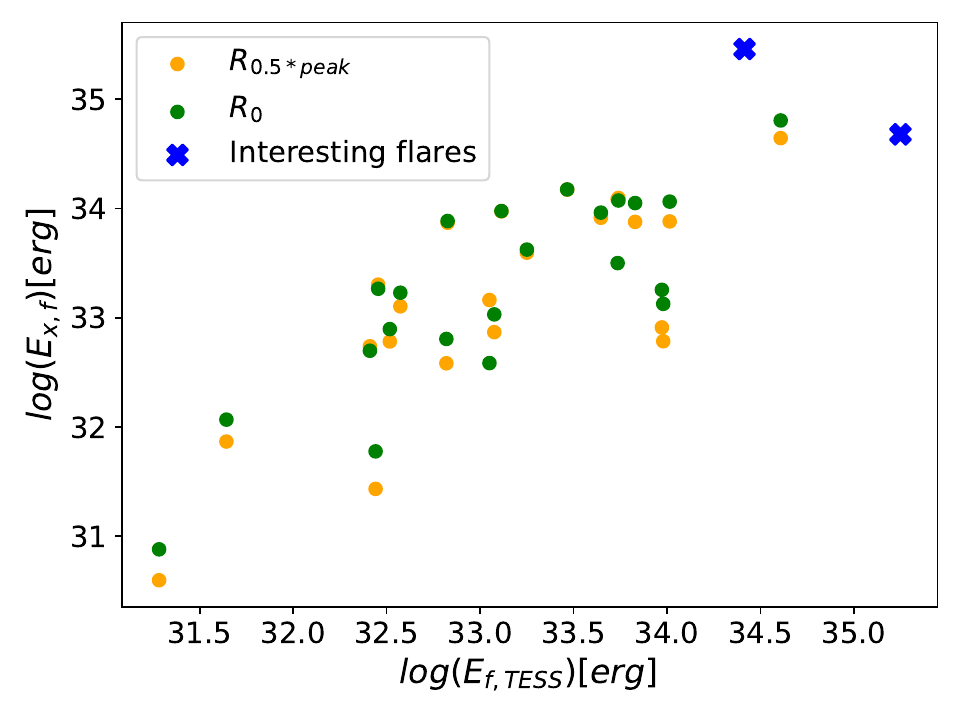}
\caption{X-ray flare energies from exponential fits normalized to $R_{0}$ vs. TESS flare energies. The X-ray flare energies are estimated from fits to the synthetic X-ray flare peak (green) and the synthetic peak scaled down by factor two  (orange). The peculiar flares described in Sect.~\ref{subsubsect:flares_strange} are marked by blue crosses.}
\label{fig:efx_efopt_2versions}
\end{figure}

\subsubsection{Peculiar flares}\label{subsubsect:flares_strange}

\begin{figure}
    \begin{center}
\includegraphics[width=0.5\textwidth]{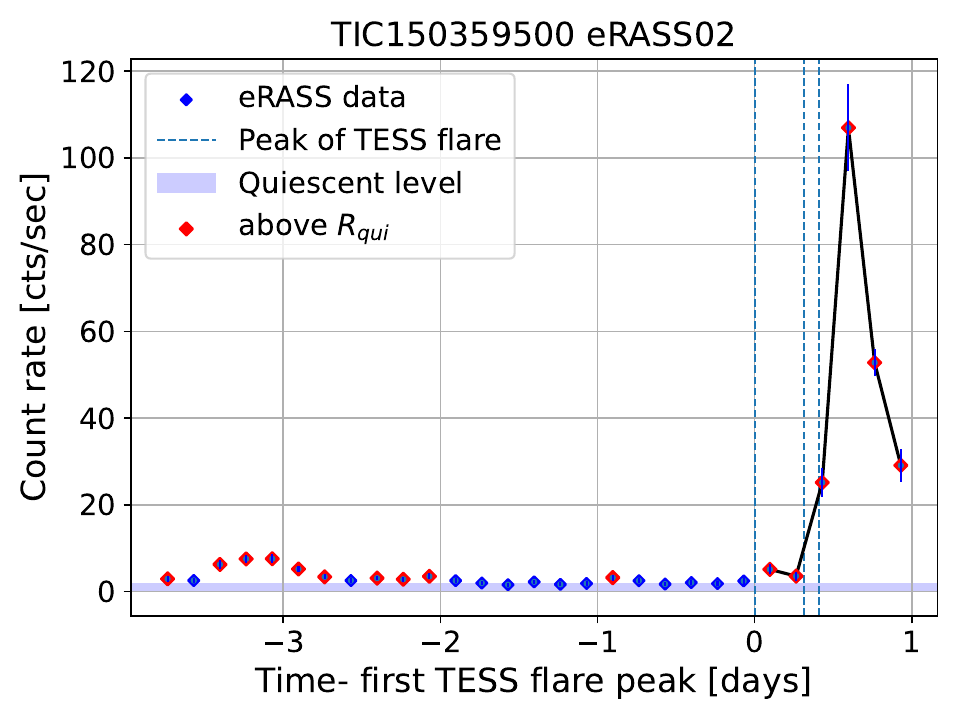}
\includegraphics[width=0.5\textwidth]{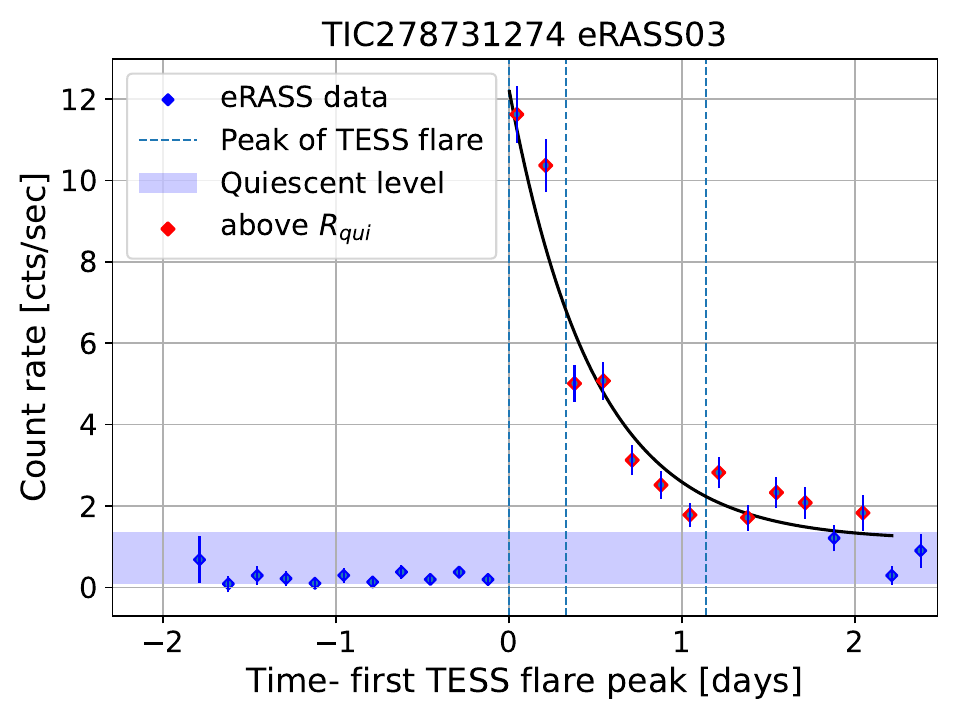} 
\caption{Two peculiar flares from the eRASS2 light curve of TIC150359500 (top) and eRASS3 light curve of TIC278731274 (bottom). The XFDPs are marked in red, The curves fit to these flares are indicated by solid black lines and the peaks of the TESS flares occurring simultaneously by the blue dashed lines.}
\label{fig:interesting-flares}
 \end{center}
\end{figure}


Two stars, TIC\,150359500 and TIC\,278731274, exhibited each an episode of remarkable flare activity in both its TESS and eRASS light curves, that could not be treated with the above method. In both cases, several TESS flares occurred close in time and contemporaneous with a longer-lasting very strong episode of X-ray brightening.
A close up of these events is shown in Fig.~\ref{fig:interesting-flares}). The times corresponding to the peaks of the TESS flares are marked with dashed vertical lines.


Since the X-ray counterparts of the individual TESS flares can not be easily distinguished we considered the episode as a whole, and we determined for each the TESS and the eRASS light curves the total energy emitted throughout the 
flaring intervals. 
For eROSITA we measured the integrated light over
all the XFDPs in the two eRASS light curves starting from the first vertical dashed line in Fig.~\ref{fig:interesting-flares} that represents the TESS peak. For the eRASS\,2 event of TIC\,150359500 we applied a linear interpolation to the XFDPs, as an exponential decay does not constrain the light curve well. For the eRASS\,3 event of TIC\,278731274, we could fit an exponential decay to the XFDPs. We then integrated under these curves normalized to $R_{\rm qui}$, to find the $ED$ and then the X-ray flare energy, as in Sect.~\ref{subsubsect:xrayflare_optflare_energies}. 
We calculated  the total optical energy emitted corresponding to the events that appear by visual inspection to be associated with the X-ray brightening by integrating over the corresponding time interval, that comprises for each of the two cases three optical flares identified by our detection algorithm but includes also some brightness enhancement above the rotational cycle that was not identified as flares by our automatic flare detection code. 
 
 The energies emitted globally in  these extraordinary events are listed in Table~\ref{tab:peculiar_flares}, and they are shown in the $E_{\rm x,f} - E_{\rm opt,f}$ plot of Fig.~\ref{fig:efx_efopt_2versions} as large cross symbols. They are on the order of $10^{35}$\,erg in both wavebands, clearly in the superflare regime.
\begin{table}
\begin{center}
\caption{Summed optical and X-ray flare energies of multiple TESS and associated  complex eRASS events.}\label{tab:peculiar_flares}
\begin{tabular}{lcc} \hline
TIC ID & $E_{\rm opt,x}[{\rm erg}]$ & $E_{\rm x,f}[{\rm erg}]$ \\ \hline
150359500 & 2.6e+34 & 2.9e+35\\
278731274 & 1.8e+35 & 4.8e+34\\ \hline
\end{tabular}   \end{center}
\end{table}

\section{Summary and conclusions}
\label{sec:conclusions}

With TESS and eROSITA for the first time in history an optical and an X-ray (nearly) all-sky survey provide partial temporal overlap, giving access to simultaneous multi-wavelength timeseries for statistical samples of  active stars. In this work we have explored the capacities of joint TESS and eRASS light curves for our understanding of flares on M dwarf stars. From a sample of $\sim 300$ M dwarfs with contemporaneous optical and X-ray light curve that are located in the ecliptic pole region, where both surveys have the highest temporal coverage, we have selected the $25$ most active ones. 

Investigations of optical flares on late-type stars have obtained a boost during the last decade with the advent of space missions like {\it Kepler} and TESS that -  thanks to their high photometric data cadence - enable a detailed  analysis of individual events
eROSITA makes a measurement of a given sky position only every $4$\,h, which is longer than the duration of many historical X-ray flares,
such that it is impossible to extract the same amount of quantitative information as for the optical flares. Also,  the strongly different data cadence
makes it non-trivial to draw a relation between the brightenings seen in both wavebands. 

To mitigate this 
we have defined a duty cycle of the fractional time the light curve deviates  from the baseline `quiescent'  emission as a simple measure for the global variability in both X-ray and optical light curves. For the TESS observations of our sample we have shown that the duty cycle is well correlated with the traditional flare rate, that is the count of flare events per day. We, thus, suggest that the duty cycle can be used as a proxy for flare activity in case of poor data sampling.
Comparing the optical and the X-ray duty cycle of our targets we find a significant correlation, meaning that 
stars that spend a higher fraction of time at enhanced X-ray emission levels also spend more time during optical flares. 
High duty cycles are mostly found among fast rotators ($P_{\rm rot} 
\lesssim 2$\,d), but a larger sample with a broader range of periods would be needed to establish a clear trend. The slowest rotator among the $25$ M dwarfs has a period of $8.6$\,d, which is due to a combination of our selection of particularly active stars and the TESS detection limit of $\lesssim 13$\,d.  

To identify in a global way if  optical and X-ray flares are associated with each other we searched each bin in the 
X-ray light curve, that is each eRODay,  for TESS flares detected in its vicinity, which means up to $2$\,h before the eRODay. This ensures that no optical  counterparts are missed, but implies that the X-ray flare - which presumably caused the enhanced X-ray  bin - must have lasted for at least $1/2$\,eROday. 
This analysis allowed us to establish that X-ray light curves with a higher number of data points above the `quiescent' emission level tend to have more optical flares preceding these  apparent X-ray events (by $\lesssim 2$\,h). 

To go beyond this coarse estimate of the association of X-ray and optical variability we 
put constraints on the X-ray flare parameters with help of the canonical (solar) flare scenario. The peak of the X-ray flare is expected to fall in most cases into the nearly $4$\,h data gap between two bins in the eRASS light curve. We constructed for each optical flare the predicted X-ray flare peak both in terms of its timing (few minutes after the optical flare) and its amplitude (constrained by the correlation with optical flare energy). 
Then we fitted the X-ray light curve starting with this reconstructed flare peak using an exponential function, and we derived the X-ray flare energy from the area under the fit function. 
A reasonable fit is obtained for $21$ events, about $1/3$ of the detected TESS flares. Our analysis has more than doubled the number of joint optical/X-ray flares observed on M dwarfs. We found larger X-ray to optical flare energy ratios than the {\it Kepler} observations of 
\cite{Guarcello2019} and \cite{KuznetsovKolotkov2021}. 
The TESS flares of our stars are well-sampled and, consequently, their energies $E_{\rm opt,f}$ are reliable. To constrain their X-ray counterparts we had to make several severe assumptions and approximations  
that involve  significant uncertainties.  As these translate into the values of $E_{\rm x,f}$, the order of magnitude agreement with the literature data can be considered to show that the solar-like flare scenario is a plausible explanation for the joint TESS and eRASS variability observed in M dwarfs. 

Alternatively, we could indeed have caught a peculiar type of flares with high X-ray output, given that our sample is biased towards highly active stars. The multi-wavelength flare energetics of stellar flares may be more complex and diverse than assumed. This is supported e.g. by comparison to a superflare on AB\,Dor observed with TESS and eROSITA: For an optical flare with an energy of $4 \cdot 10^{36}$ the X-ray flare was estimated to an energy of at most $\sim 10^{35}$\,erg \citep{schmitt21.0}. 
This event on AB\,Dor was observed from a poorly sampled eROSITA light curve, just as ours, yet - contrary to our results where $E_{\rm x,f}/E_{\rm opt,f} \gtrsim 1$ - the optical flare carried more energy than plausible estimates for the X-ray flare.

To summarize, simultaneous eROSITA and TESS data provide an entirely new  window into  statistical samples of stellar flares and many  possibilities to test the solar-stellar connection. The high uncertainty on the evolution of the X-ray variability resulting from the low ($4$\,h) cadence of the eROSITA scanning mode is the major difficulty in an assessment of the 
physical parameters of the X-ray flares.

\begin{acknowledgements}
W.M.J is supported by the Bundesministerium für Wirtschaft und Energie through the Deutsches Zentrum für Luft- und Raumfahrt e.V. (DLR) under grant number FKZ 50 OR 2208. EM is supported by Deutsche Forschungsgemeinschaft under grant STE 1068/8-1. 
This work is based on data from eROSITA, the soft X-ray instrument aboard SRG, a joint Russian-German science mission supported by the Russian Space Agency (Roskosmos), in the interests of the Russian Academy of Sciences represented by its Space Research Institute (IKI), and the Deutsches Zentrum für Luft- und Raumfahrt (DLR). The SRG spacecraft was built by Lavochkin Association (NPOL) and its subcontractors, and is operated by NPOL with support from the Max Planck Institute for Extraterrestrial Physics (MPE).

The development and construction of the eROSITA X-ray instrument was led by MPE, with contributions from the Dr. Karl Remeis Observatory Bamberg \& ECAP (FAU Erlangen-Nuernberg), the University of Hamburg Observatory, the Leibniz Institute for Astrophysics Potsdam (AIP), and the Institute for Astronomy and Astrophysics of the University of Tübingen, with the support of DLR and the Max Planck Society. The Argelander Institute for Astronomy of the University of Bonn and the Ludwig Maximilians Universität Munich also participated in the science preparation for eROSITA.

The eROSITA data shown here were processed using the eSASS/NRTA software system developed by the German eROSITA consortium. 

This paper includes data collected by the TESS mission. Funding for the TESS mission is provided by the NASA's Science Mission Directorate.

\end{acknowledgements}

%



%
\bibliographystyle{aa} 
\bibliography{biblio} 
%
\begin{appendix}
\section{Title of Appendix A}
\input{tables/gaia_2mass_tess}
\FloatBarrier

\section{Title of Appendix B}
\input{tables/erass-table}
\FloatBarrier


\section{Title of Appendix C}
\input{exp_fits_two_peaks}

\end{appendix}

\end{document}

%% file: tables/tic_rqui_erassdc.tex
\begin{table*} 
    \centering
    \caption{Table with the TIC ID,
    quiescent rate from eRASS 2 (\(R_{qui,eRASS2}\)), eRASS 3 (\(R_{qui,eRASS3}\)), the mean quiescent rate (\(R_{qui,mean}\)) and the standard deviation (\(R_{qui,std.dev}\)) from the two surveys in cts/sec, eRASS duty cycle of eRASS 2 \((\nu_{x,DC,eRASS2})\), eRASS 3 \((\nu_{x,DC,eRASS3})\), the mean eRASS duty cycle \((\nu_{x,DC,mean})\) and the standard deviation (\(\nu_{x,DC,std.dev}\)) of the two eRASS surveys}
    \label{table:tic-rqui-erassdc}
    \begin{tabular}{lllllllll}
    \hline
        TIC ID & $R_{\rm qui,eRASS2}$ & $R_{\rm qui,eRASS3}$ & $R_{\rm qui,mean}$ & $R_{\rm qui,std.dev}$ & $\nu_{x,DC,eRASS2}$ & $\nu_{x,DC,eRASS3}$ & $\nu_{x,DC,mean}$ & $\nu_{x,DC,std.dev}$ \\
        &[$cts \,s^{-1}$] &[$cts \,s^{-1}$] & [$cts \,s^{-1}$] & [$cts \,s^{-1}$] & & & & \\
        \hline
        140758953 & 0.004 & 0.027 & 0.016 & 0.012 & 0.222 & 0.278 & 0.25 & 0.028 \\ 
        141156752 & 0.026 & 0.026 & 0.026 & 0.0 & 0.333 & 0.357 & 0.345 & 0.012 \\ 
        141422340 & 0.016 & 0.006 & 0.011 & 0.005 & 0.12 & 0.059 & 0.089 & 0.031 \\ 
        142052608 & 0.004 & 0.011 & 0.008 & 0.004 & 0.051 & 0.212 & 0.132 & 0.081 \\ 
        150301109 & 0.029 & 0.029 & 0.029 & 0.0 & 0.071 & 0.143 & 0.107 & 0.036 \\ 
        150359500 & 2.517 & 1.828 & 2.172 & 0.345 & 0.5 & 0.432 & 0.466 & 0.034 \\ 
        167812509 & 0.018 & 0.005 & 0.012 & 0.007 & 0.219 & 0.31 & 0.265 & 0.046 \\ 
        167894019 & 0.024 & 0.062 & 0.043 & 0.019 & 0.172 & 0.48 & 0.326 & 0.154 \\ 
        177257889 & 0.006 & 0.017 & 0.011 & 0.006 & 0.064 & 0.057 & 0.06 & 0.004 \\ 
        220395392 & 0.325 & 0.305 & 0.315 & 0.01 & 0.413 & 0.29 & 0.352 & 0.062 \\ 
        220473309 & 3.605 & 4.603 & 4.104 & 0.499 & 0.674 & 0.438 & 0.556 & 0.118 \\ 
        260189734 & 0.024 & 0.062 & 0.043 & 0.019 & 0.333 & 0.441 & 0.387 & 0.054 \\ 
        260352702 & 0.042 & 0.015 & 0.028 & 0.014 & 0.045 & 0.034 & 0.04 & 0.006 \\ 
        277298771 & 0.641 & 0.158 & 0.4 & 0.242 & 0.706 & 0.885 & 0.795 & 0.09 \\ 
        278725797 & 0.049 & 0.195 & 0.122 & 0.073 & 0.238 & 0.222 & 0.23 & 0.008 \\ 
        278731274 & 0.706 & 1.507 & 1.107 & 0.401 & 0.333 & 0.333 & 0.333 & 0.0 \\ 
        278825715 & 0.027 & 0.224 & 0.126 & 0.099 & 0.667 & 0.75 & 0.708 & 0.042 \\ 
        300651846 & 0.092 & 0.04 & 0.066 & 0.026 & 0.297 & 0.111 & 0.204 & 0.093 \\ 
        306574570 & 0.037 & 0.027 & 0.032 & 0.005 & 0.043 & 0.121 & 0.082 & 0.039 \\ 
        31381302 & 0.028 & 0.033 & 0.031 & 0.003 & 0.143 & 0.088 & 0.116 & 0.028 \\ 
        33866201 & 0.028 & 0.018 & 0.023 & 0.005 & 0.05 & 0.148 & 0.099 & 0.049 \\ 
        350145729 & 0.064 & 0.053 & 0.059 & 0.006 & 0.556 & 0.24 & 0.398 & 0.158 \\ 
        38698751 & 0.047 & 0.07 & 0.059 & 0.012 & 0.04 & 0.122 & 0.081 & 0.041 \\ 
        391946225 & 0.07 & 0.014 & 0.042 & 0.028 & 0.13 & 0.068 & 0.099 & 0.031 \\ 
        55497266 & 0.722 & 0.995 & 0.858 & 0.137 & 0.782 & 0.73 & 0.756 & 0.026 \\ 
        \hline
    \end{tabular}
\end{table*}

%% file: tables/tic_rot_tessfr_tessdc.tex
\begin{table*} 
    \centering
    \caption{
    Rotation periods and optical variability diagnostics derived from TESS light curves. TIC ID, rotation period from eRASS 2 (\(P_{Rot,eRASS2}\))and eRASS 3 (\(P_{Rot,eRASS3}\)) in days, TESS flare rate from eRASS 2 ($\nu_{\rm o,f,eRASS2}$) and eRASS 3 ($\nu_{\rm o,f,eRASS3}$), the TESS duty cycle from eRASS 2 ($\nu_{\rm o,DC,eRASS2}$), eRASS 3 ($\nu_{\rm o,DC,eRASS3}$) and the mean ($\nu_{\rm o,DC,mean}$) and standard deviation ($\nu_{\rm o,DC,std.dev}$) between the duty cycles measured from the two surveys. The gaps indicate missing TESS light curves corresponding to the eRASS surveys.}
    \label{table:rot-tessdc-tessfr}
    \begin{tabular}{lllllllll}
    \hline
        TIC ID & $P_{\rm Rot,eRASS2}$ & $P_{\rm Rot,eRASS2}$ & $\nu_{\rm o,f,eRASS2}$ & $\nu_{\rm o,f,eRASS3}$ &  $\nu_{\rm o,DC,eRASS2}$ & $\nu_{\rm o,DC,eRASS3}$ & $\nu_{\rm o,DC,mean}$ & $\nu_{\rm o,DC,std.dev}$ \\ 
        &[d] &[d] & [${\rm d^{-1}}$] & [${\rm d^{-1}}$] & & & & \\
        \hline
        140758953 & 1.374 & 1.367 & 0.157 & 0.167 & 0.002 & 0.003 & 0.003 & 0.001 \\ 
        141156752 & \tablefootmark{a} & \tablefootmark{a} & 0.077 & 0.208 & 0.003 & 0.004 & 0.003 & 0.001 \\ 
        141422340 & \tablefootmark{a} & \tablefootmark{a} & 0.077 & 0.042 & 0.001 & 0.001 & 0.001 & 0.0 \\ 
        142052608 & 2.984 & 2.936 & 0.039 & 0.0 & 0.0 & 0.0 & 0.0 & 0.0 \\ 
        150301109 & \tablefootmark{a} & \tablefootmark{a} & 0.0 & 0.0 & 0.0 & 0.0 & 0.0 & 0.0 \\ 
        150359500 & 1.029 & ~ & 0.905 & ~ & 0.026 & ~ & 0.026 & ~ \\ 
        167812509 & \tablefootmark{a} & 0.336 & 0.079 & 0.076 & 0.001 & 0.001 & 0.001 & 0.0 \\ 
        167894019 & \tablefootmark{a} & \tablefootmark{a} & 0.118 & 0.267 & 0.003 & 0.004 & 0.004 & 0.001 \\ 
        177257889 & \tablefootmark{b} & \tablefootmark{a}\tablefootmark{b} & 0.0 & 0.0 & 0.0 & 0.0 & 0.0 & 0.0 \\ 
        220395392 & ~ & 5.066 & ~ & 0.077 & ~ & 0.001 & 0.001 & ~ \\ 
        220473309 & \tablefootmark{a} & 8.601 & 0.926 & 1.283 & 0.03 & 0.032 & 0.031 & 0.001 \\ 
        260189734 & \tablefootmark{a} & \tablefootmark{a} & 0.039 & 0.043 & 0.0 & 0.002 & 0.001 & 0.001 \\ 
        260352702 & ~ & 1.003 & ~ & 0.171 & ~ & 0.001 & 0.001 & ~ \\ 
        277298771 & 1.183 & 1.177 & 0.669 & 0.525 & 0.013 & 0.016 & 0.014 & 0.002 \\ 
        278725797 & 4.461 & \tablefootmark{a} & 0.079 & 0.294 & 0.004 & 0.005 & 0.005 & 0.001 \\ 
        278731274 & 0.498 & 0.499 & 0.472 & 0.588 & 0.013 & 0.012 & 0.013 & 0.001 \\ 
        278825715 & 6.52 & 6.438 & 0.079 & 0.294 & 0.001 & 0.004 & 0.002 & 0.002 \\ 
        300651846 & ~ & 0.344 & ~ & 0.155 & ~ & 0.0 & 0.0 & ~ \\ 
        306574570 & \tablefootmark{a} & ~ & 0.0 & ~ & 0.0 & ~ & 0.0 & ~ \\ 
        31381302 & 0.165 & 0.165 & 0.154 & 0.335 & 0.003 & 0.003 & 0.003 & 0.0 \\ 
        33866201 & \tablefootmark{c}0.258 & \tablefootmark{c}0.257 & 0.039 & 0.084 & 0.001 & 0.001 & 0.001 & 0.0 \\ 
        350145729 & \tablefootmark{a} & \tablefootmark{c}2.661 & 0.195 & 0.179 & 0.01 & 0.001 & 0.006 & 0.005 \\ 
        38698751 & 0.603 & 0.602 & 0.0 & 0.013 & 0.0 & 0.0 & 0.0 & 0.0 \\ 
        391946225 & ~ & 5.535 & ~ & 0.08 & ~ & 0.001 & 0.001 & ~ \\ 
        55497266 & 0.324 & 0.324 & 1.136 & 1.2 & 0.019 & 0.019 & 0.019 & 0.0 \\
        \hline
    \end{tabular}
    \tablefoot{
    \tablefoottext{a}{the highest peak in the periodogram was not at least twice as high as the second highest peak.}
    \tablefoottext{b}{Badly de-trended light curves.}
    \tablefoottext{c}{`double-dipped' light curves, where the longer period was chosen as the 'true' period}
    }
\end{table*}

%% file: tables/gaia_2mass_tess.tex
\begin{table*}
  \begin{center}
    \caption{Content of the 30 columns in our catalog of the 25 X-ray bright or variable stars in the TESS SCVZ which correspond to photometry, TESS detection information and stellar parameters.}
    \label{tab:gaia-2mass-tess}
    \begin{tabular}{l|l|l|p{0.6\linewidth}} 
      \textbf{\#} & \textbf{Name} & \textbf{Unit} & \textbf{Description}\\
      \hline
      1 & TIC ID & &TESS Input Catalogue Identifier.\\
      2 & GaiaDR2 & & {\it Gaia} ID from data release 2.\\
      3 & 2MASS & & 2MASS ID. \\
      4 & Distance & pc & Distance to the star from {\it Gaia} parallax.\\
      5 & PMRA & mas/year & {\it Gaia} proper motion in right ascension direction.\\
      6 & PMDEC & mas/year & {\it Gaia} proper motion in declination direction. \\
      7 & GaiaBP & mag & Magnitude in {\it Gaia} BP band. \\
      8 & GaiaRP & mag & Magnitude in {\it Gaia} RP band. \\
      9 & GaiaG & mag & Magnitude in {\it Gaia} G band. \\
      10 & V & mag & Johnson–Kron–Cousin $V$ from {\it Gaia} magnitudes $G_{\rm BP}$ and $G_{\rm RP}$.\\
      11 & J & mag & Magnitude in 2MASS J band.\\
      12 & H & mag & Magnitude in 2MASS H band.\\
      13 & $K_{s}$ & mag & Magnitude in 2MASS $K_{s}$ band.\\
      14 & Radius & Solar radius & Stellar radius based on empirical relations from \citep{Mann2015} using $M_{\rm Ks}$.\\
      15 & Mass & Solar mass & Stellar mass based on empirical relations from \citep{Mann2015} using $M_{\rm Ks}$.\\
      16 & Teff\_v\_j\_j\_h & Kelvin & Stellar effective temperatures based on empirical relations from \citep{Mann2015} using J, H and V.\\
      17 & Teff\_bp\_rp\_j\_h & Kelvin & Stellar effective temperatures based on empirical relations from \citep{Mann2015} using GaiaBP, GaiaRP, J and H.\\
      18 & L\_Bol & ergs/sec & Bolometric luminosities calculated from Bolometric corrections to {\it Gaia} magnitudes.\\
      19 & TESS\_sector\_XX & & Boolean value to indicate whether the star was observed in TESS sector XX. XX=27,...,39.
            
    \end{tabular}
  \end{center}
\end{table*}

%% file: tables/erass-table.tex
\begin{table*}
  \begin{center}
    \caption{Content of the 8 columns in our catalog of the 25 X-ray bright or variable stars in the TESS SCVZ which correspond to eRASS data and X-ray stellar parameters.}
    \label{tab:erass}
    \begin{tabular}{l|l|l|p{0.6\linewidth}} 
      \textbf{\#} & \textbf{Name} & \textbf{Unit} & \textbf{Description}\\
      \hline
      1 & GaiaDR2 & & {\it Gaia} ID from data release 2.\\
      2 & RA CORR <survey> & deg & Corrected X-ray right ascension.\\
      3 & DEC CORR <survey> & deg & Corrected X-ray declination.\\
      4 & Separation <survey> & arcsec & separation along a great circle between the proper motion corrected {\it Gaia}-DR2 coordinates and the matched eRASS coordinates RA CORR <survey> and DEC CORR <survey>\\
      5 & DET\_LIKE <survey> & & Detection likelihood in energy band 0.2-5 KeV\\
      6 & ML\_RATE\_0 <survey> & cts/sec & Count rate in energy band 0.2-5 KeV\\
      7 & $L_{X}$ & ergs/sec & X-ray luminosity from the average ML\_RATE\_0 from eRASS 2 and eRASS 3 and {\it Gaia} distances\\
      8 & $F_{X,surf}$ & ${ergs\,sec^{-1}\,cm^{-2}}$ & X-ray surface flux from the average ML\_RATE\_0 from eRASS 2 and eRASS 3\\
            
    \end{tabular}
  \end{center}
\end{table*}

%% file: exp_fits_two_peaks.tex
\begin{figure*}[h]
\centering
\parbox{\textwidth}{
	\parbox{0.5\textwidth}{\includegraphics[width=0.5\textwidth]{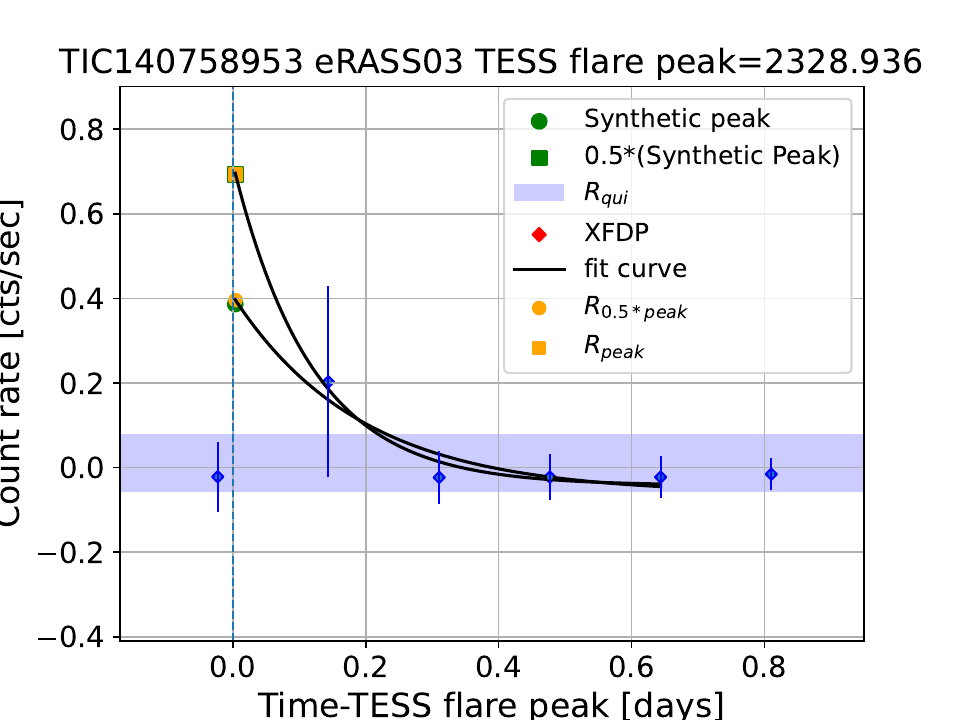}}
	\parbox{0.5\textwidth}{\includegraphics[width=0.5\textwidth]{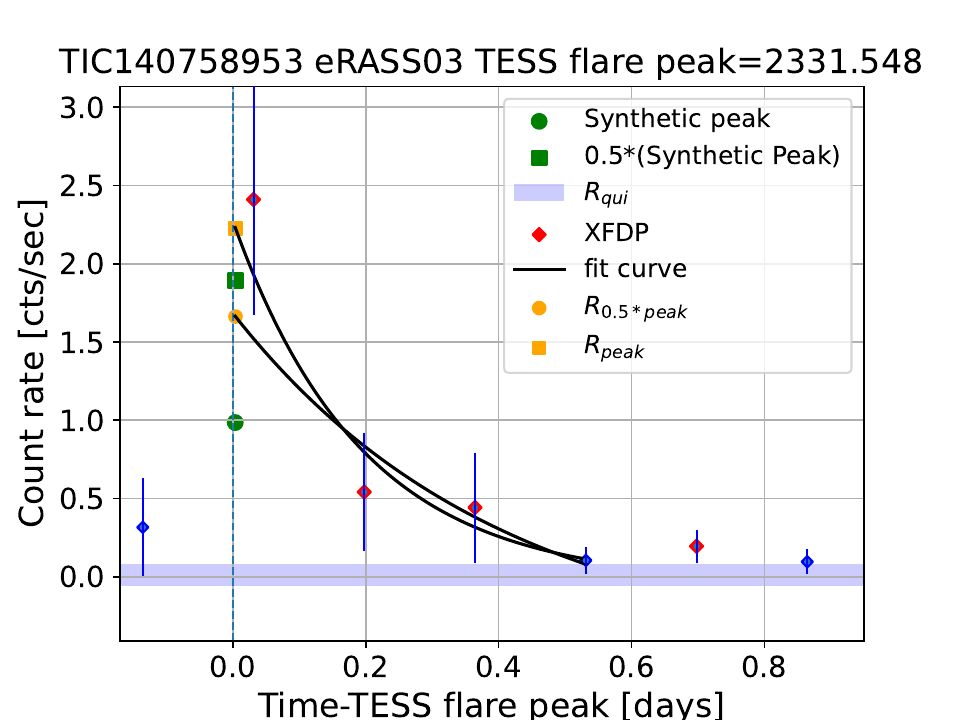}}
}
\parbox{\textwidth}{
	\parbox{0.5\textwidth}{\includegraphics[width=0.5\textwidth]{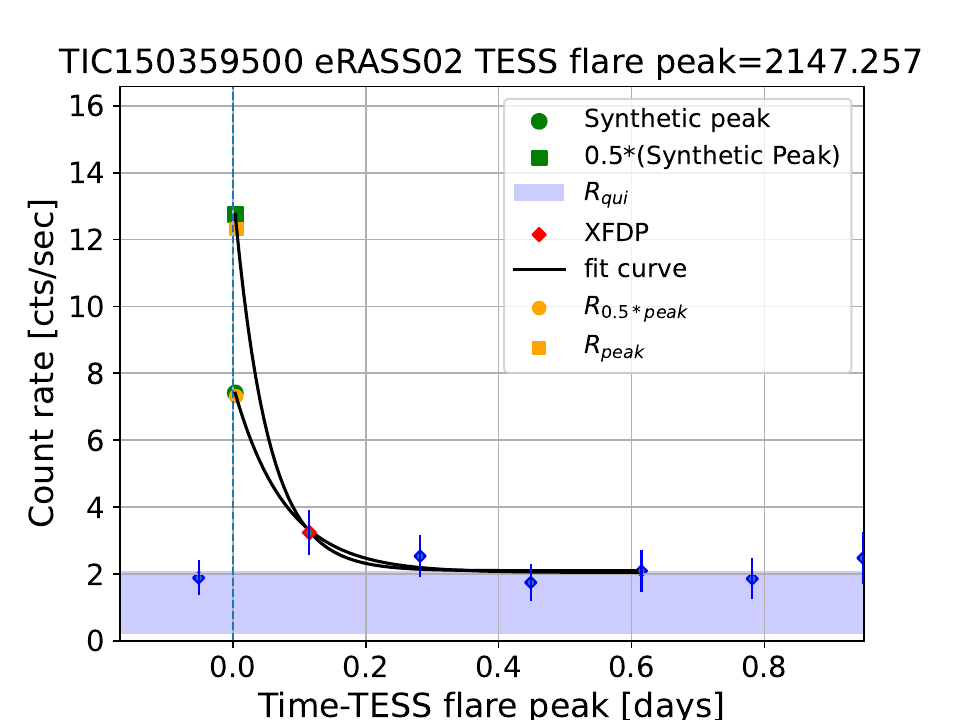}}
	\parbox{0.5\textwidth}{\includegraphics[width=0.5\textwidth]{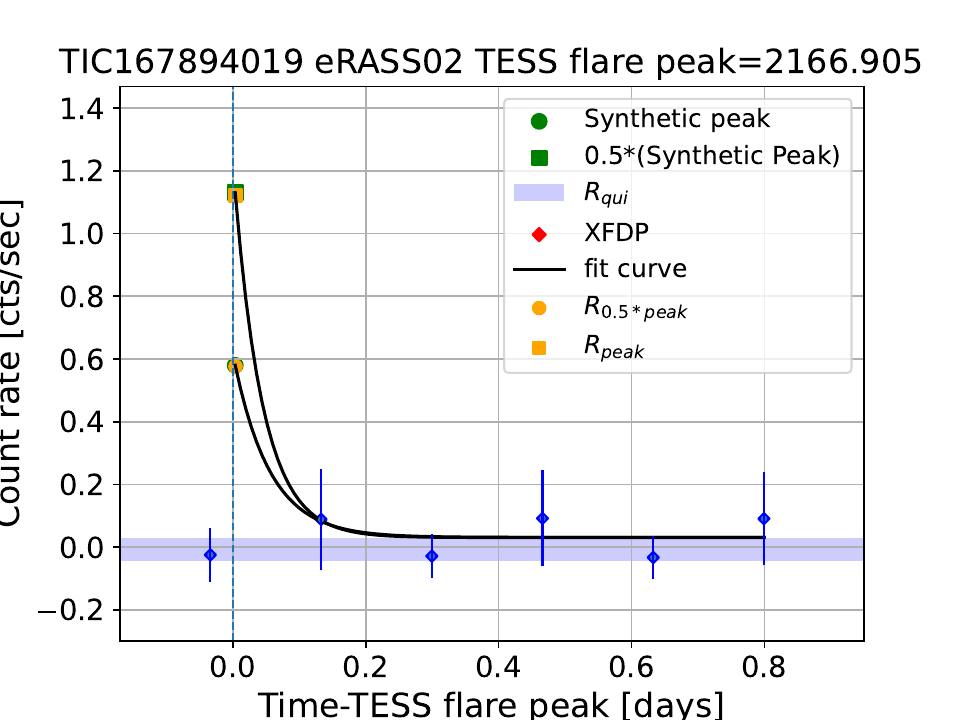}}
}
\parbox{\textwidth}{
	\parbox{0.5\textwidth}{\includegraphics[width=0.5\textwidth]{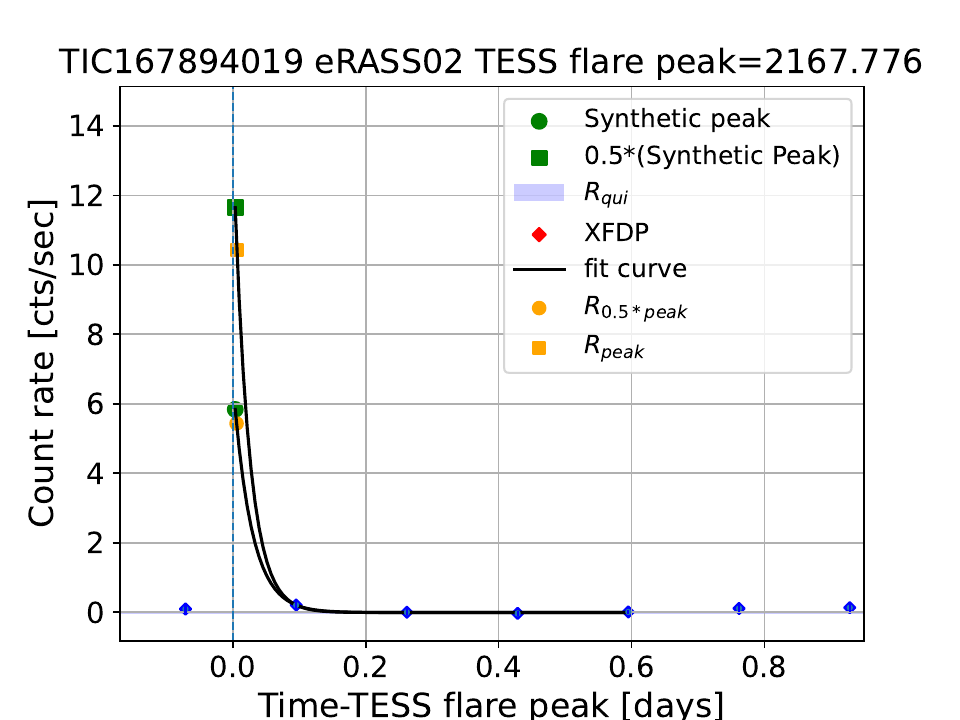}}
	\parbox{0.5\textwidth}{\includegraphics[width=0.5\textwidth]{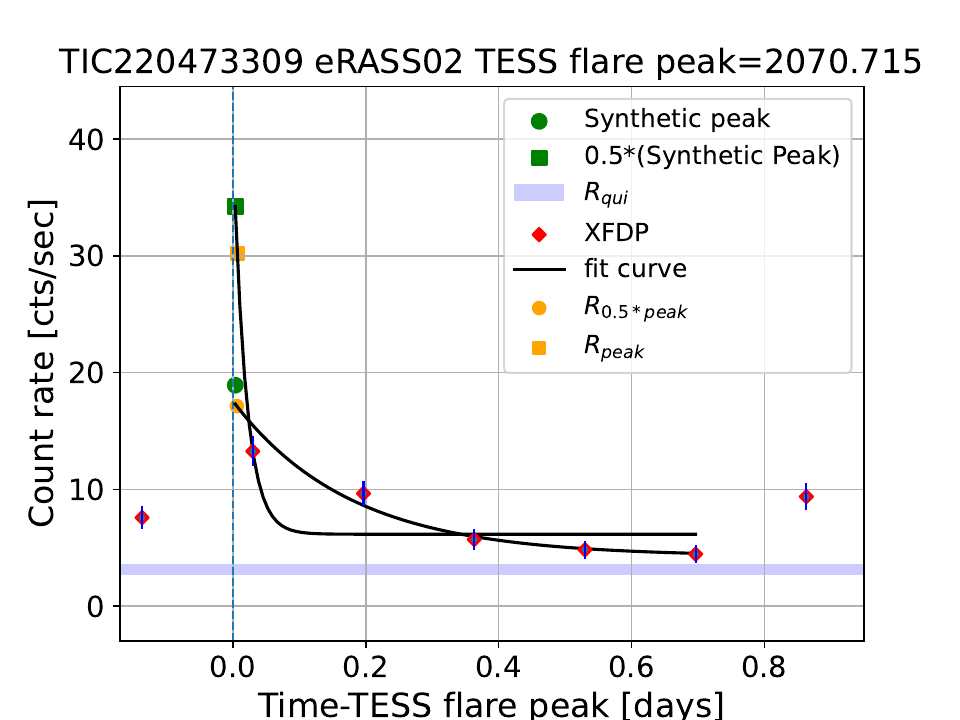}}
}
\caption{eRASS light curves illustrating our estimate of the X-ray flare peak (red data point) as explained in Sect.~\ref{subsubsect:xrayflare_peakestimate} and the exponential fit to the flare decay (black curve) described in Sect.~\ref{subsubsect:xrayflare_expfit}.}
\label{fig:exp-fits}
\end{figure*}

\begin{figure*}[h]
\centering
\parbox{\textwidth}{
	\parbox{0.5\textwidth}{\includegraphics[width=0.5\textwidth]{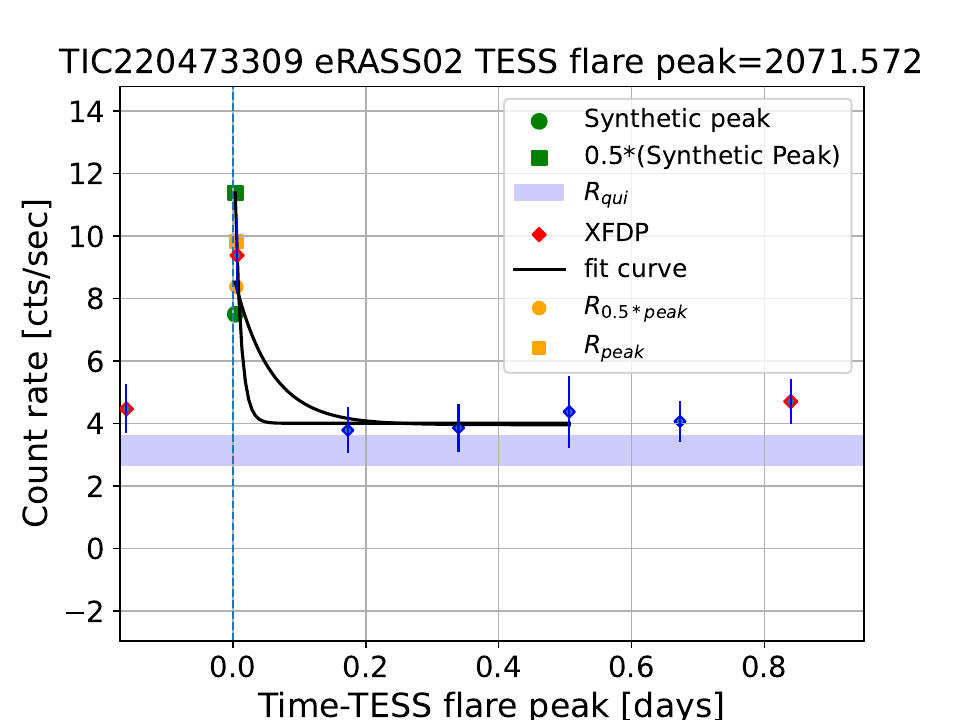}}
	\parbox{0.5\textwidth}{\includegraphics[width=0.5\textwidth]{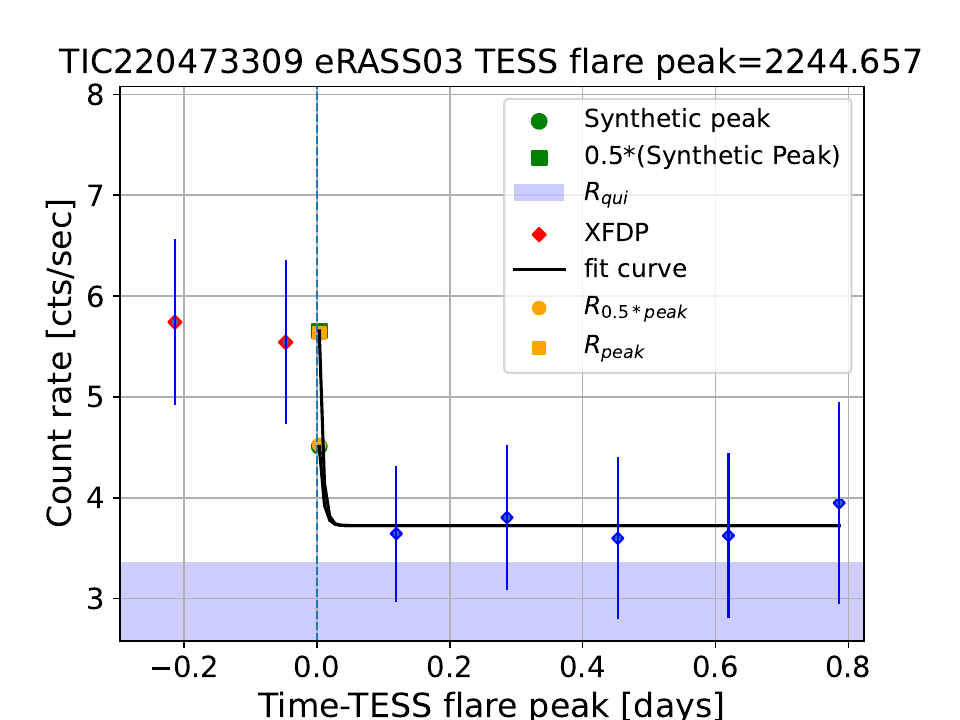}}
}
\parbox{\textwidth}{
	\parbox{0.5\textwidth}{\includegraphics[width=0.5\textwidth]{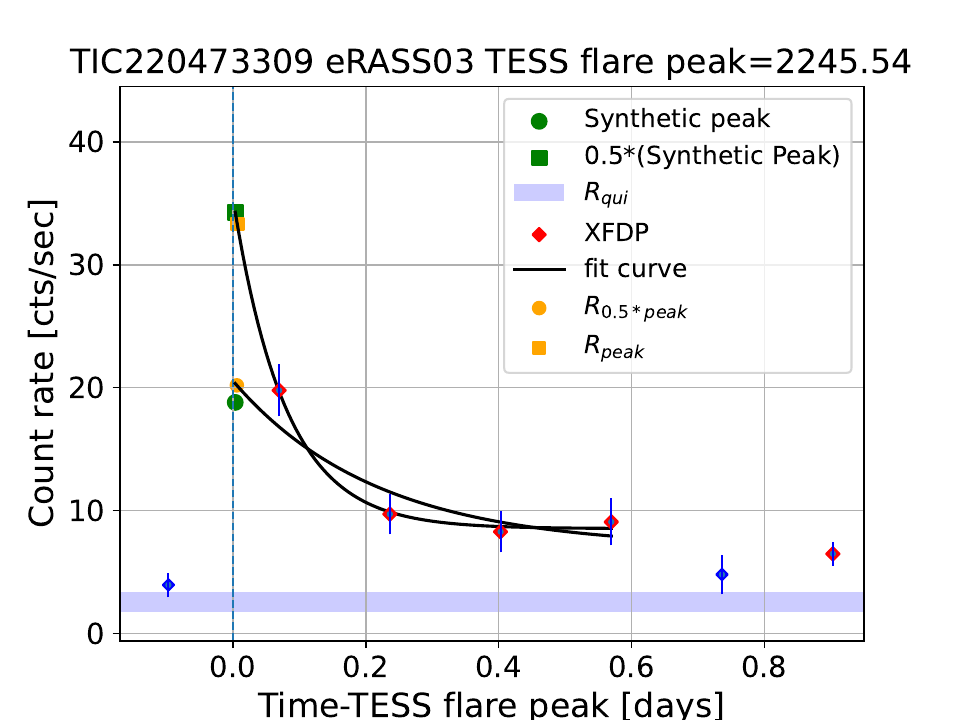}}
	\parbox{0.5\textwidth}{\includegraphics[width=0.5\textwidth]{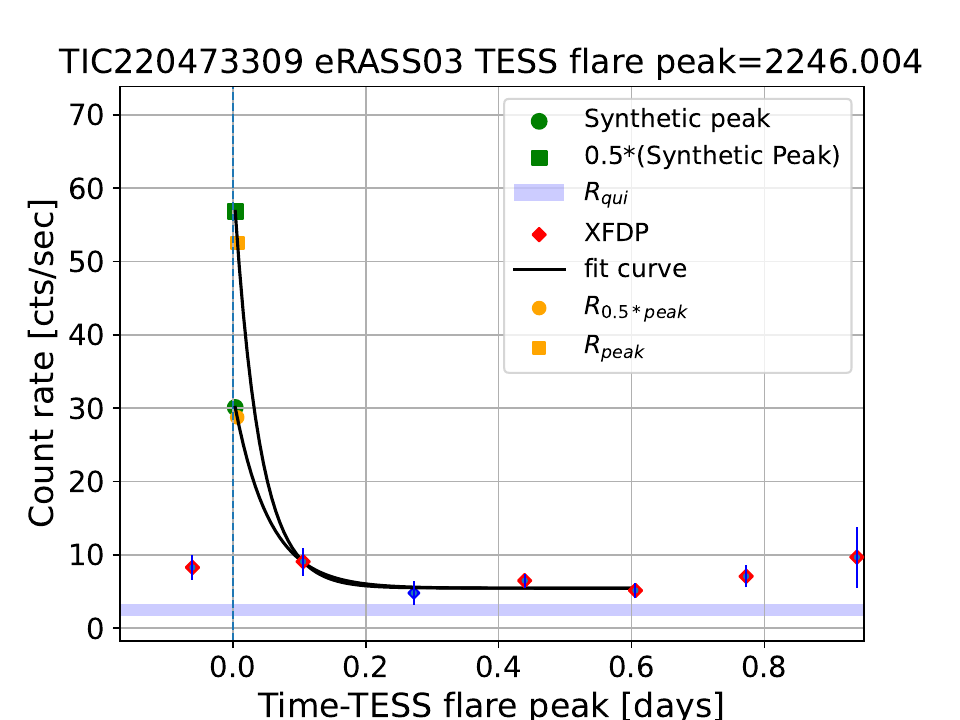}}
}
\parbox{\textwidth}{
	\parbox{0.5\textwidth}{\includegraphics[width=0.5\textwidth]{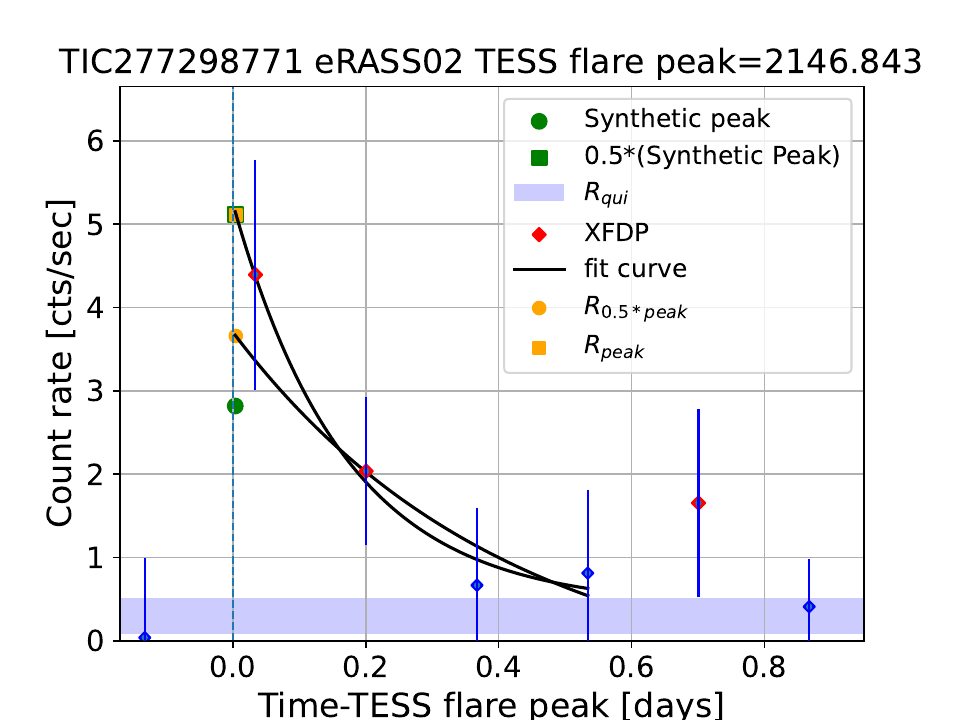}}
	\parbox{0.5\textwidth}{\includegraphics[width=0.5\textwidth]{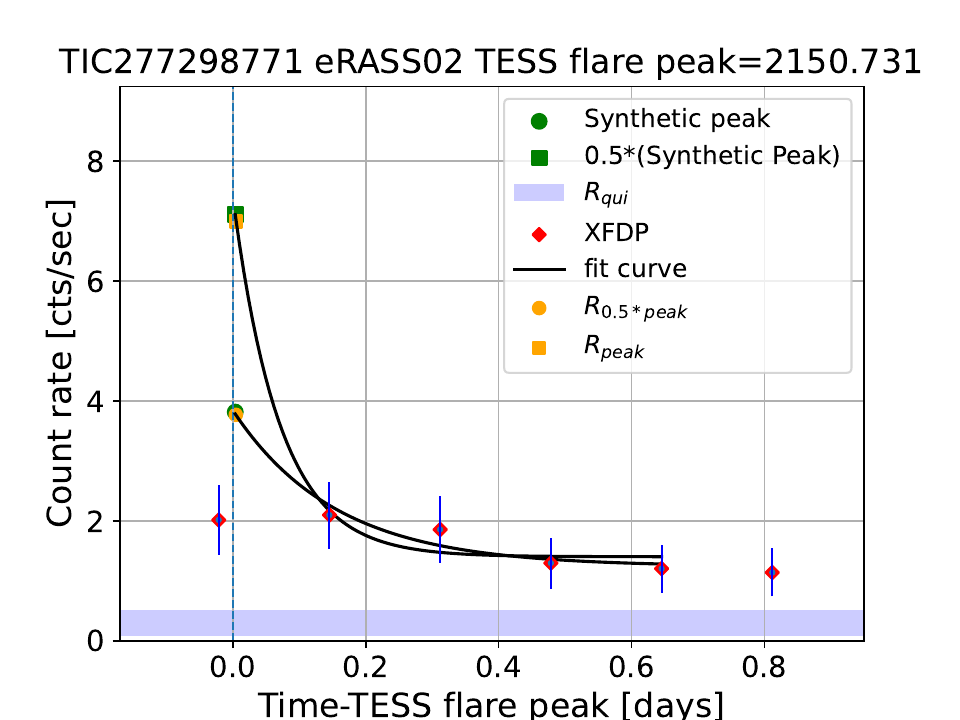}}
}
\addtocounter{figure}{-1}
\caption{Continued}
\label{fig:exp-fits-2}
\end{figure*}

\begin{figure*}[h]
\centering
\parbox{\textwidth}{
	\parbox{0.5\textwidth}{\includegraphics[width=0.5\textwidth]{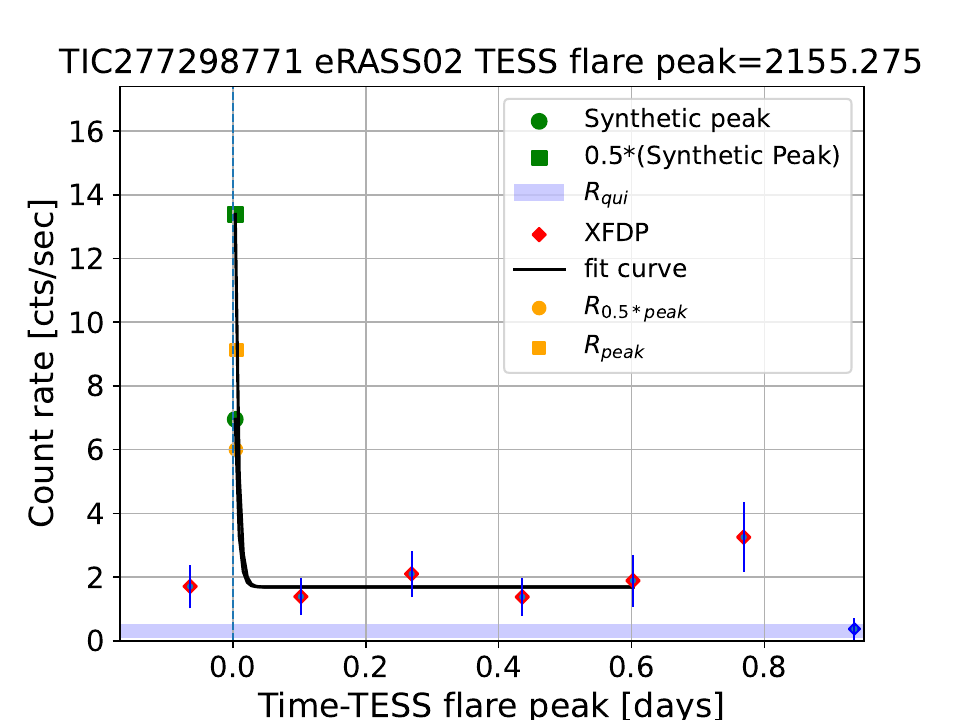}}
	\parbox{0.5\textwidth}{\includegraphics[width=0.5\textwidth]{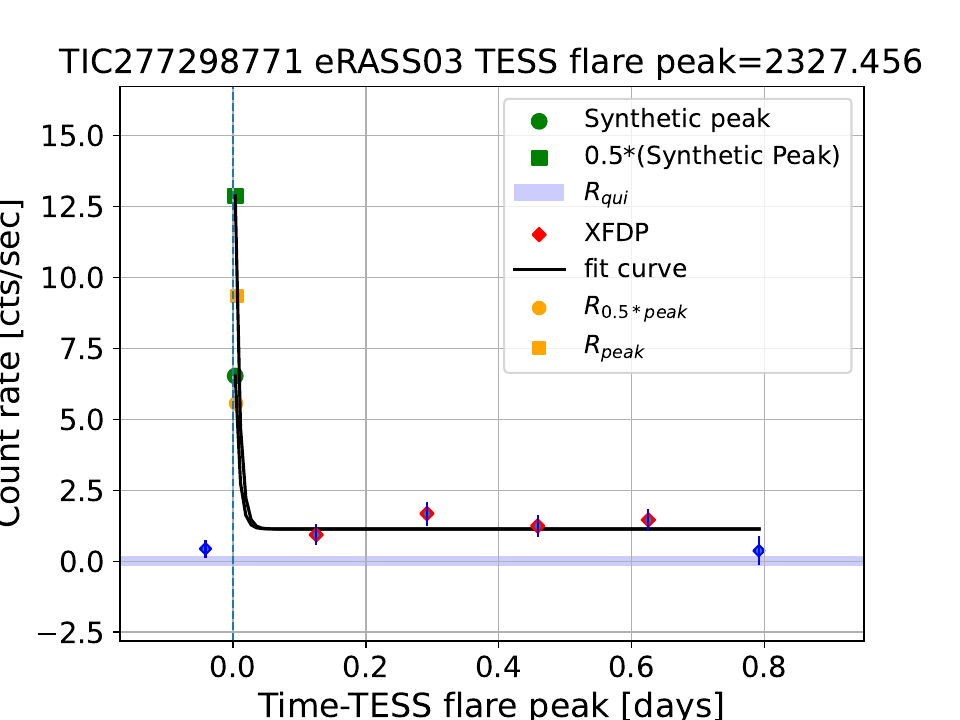}}
}
\parbox{\textwidth}{
	\parbox{0.5\textwidth}{\includegraphics[width=0.5\textwidth]{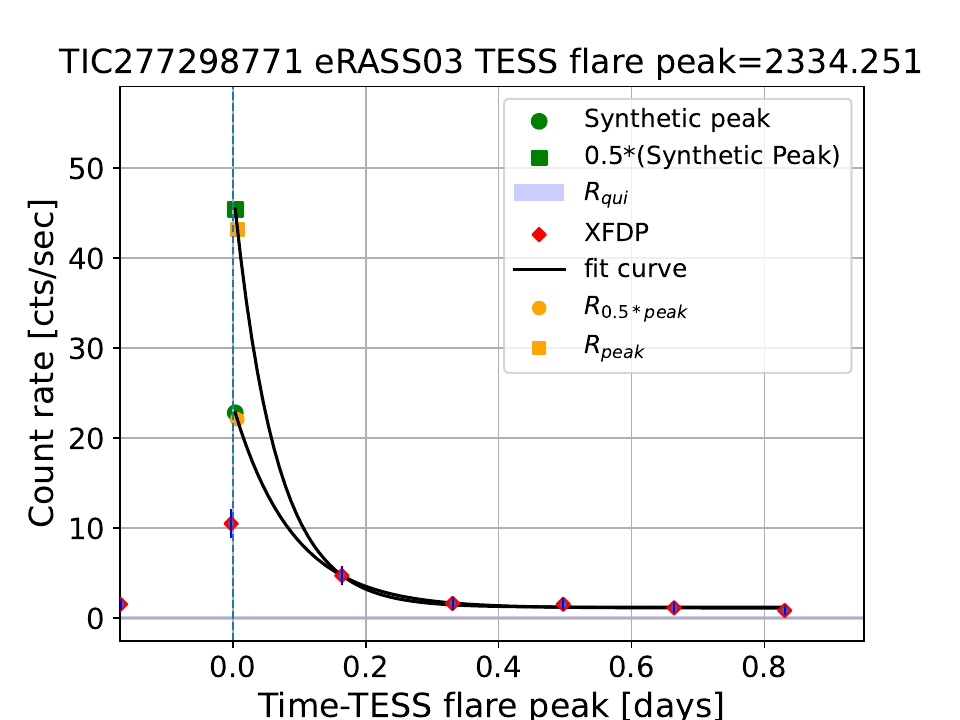}}
	\parbox{0.5\textwidth}{\includegraphics[width=0.5\textwidth]{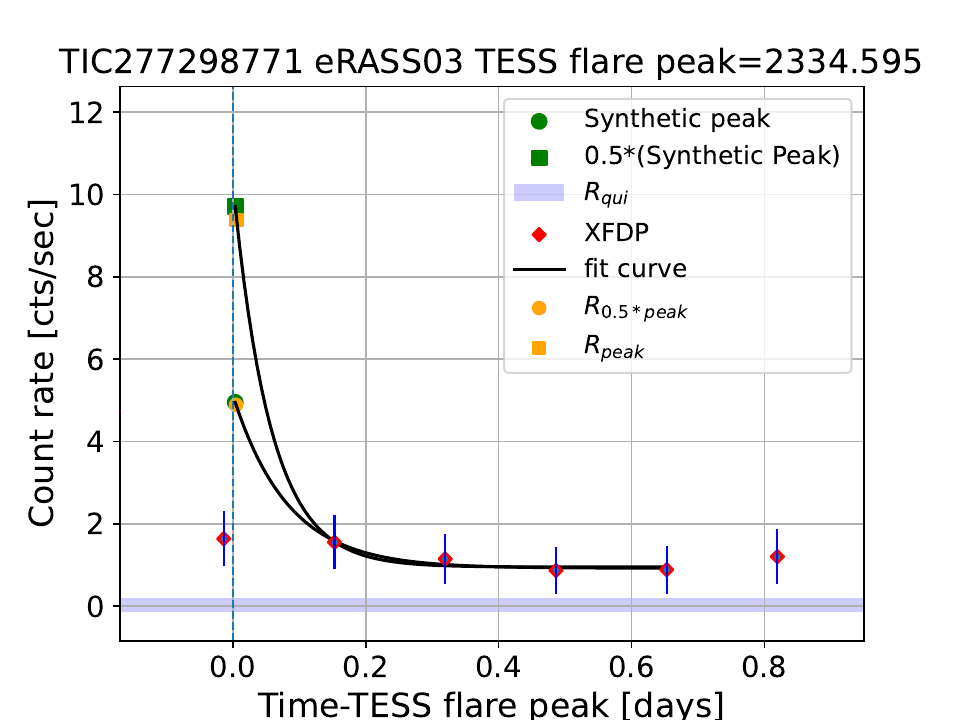}}
}
\parbox{\textwidth}{
	\parbox{0.5\textwidth}{\includegraphics[width=0.5\textwidth]{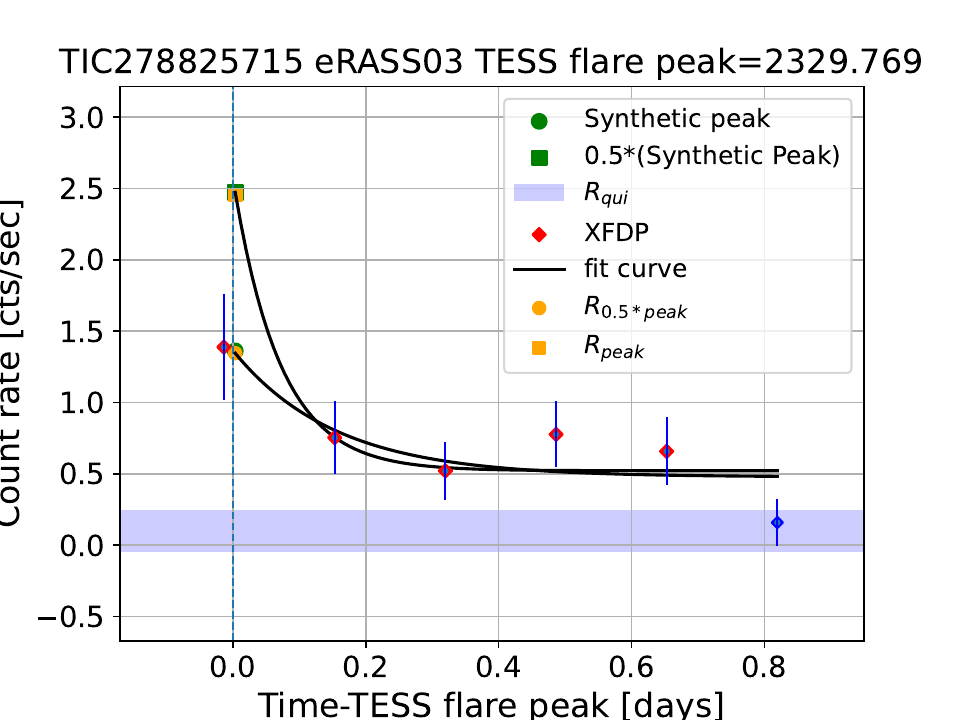}}
	\parbox{0.5\textwidth}{\includegraphics[width=0.5\textwidth]{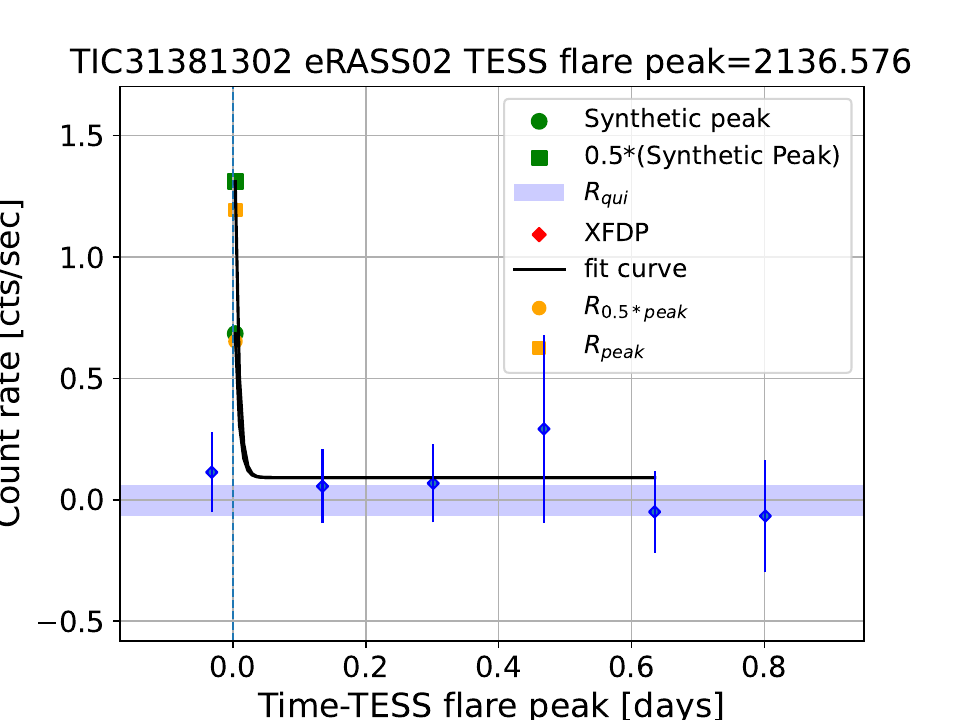}}
}
\addtocounter{figure}{-1}
\caption{Continued}
\label{fig:exp-fits-3}
\end{figure*}

\begin{figure*}[h]
\centering
\parbox{\textwidth}{
	\parbox{0.5\textwidth}{\includegraphics[width=0.5\textwidth]{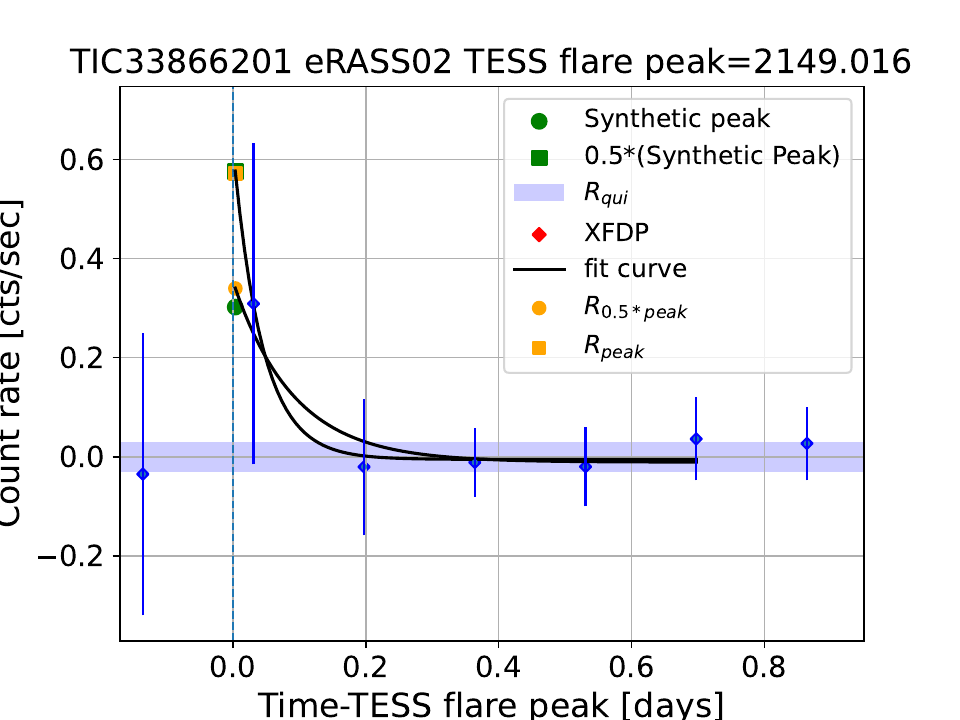}}
	\parbox{0.5\textwidth}{\includegraphics[width=0.5\textwidth]{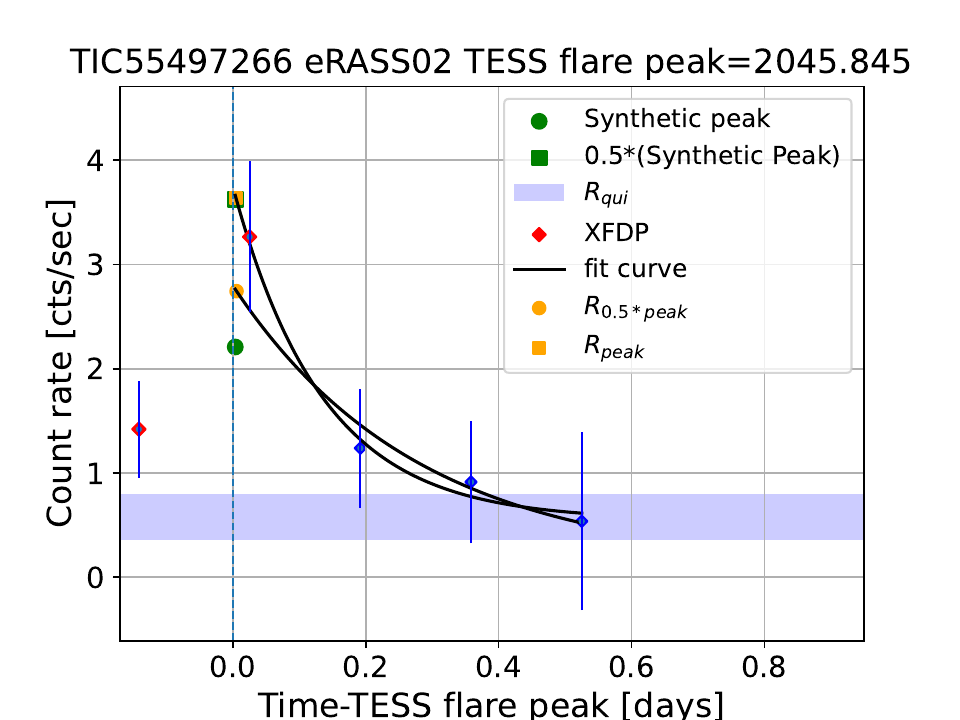}}
}
\parbox{\textwidth}{
	\parbox{0.5\textwidth}{\includegraphics[width=0.5\textwidth]{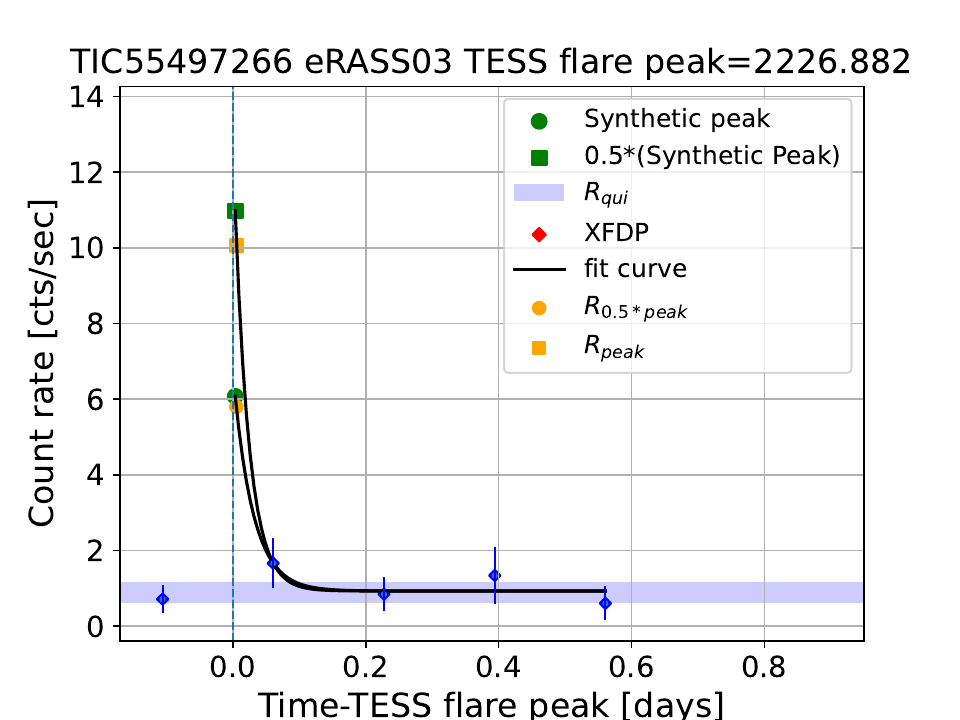}}
 }
\addtocounter{figure}{-1}
\caption{Continued}
\label{fig:exp-fits-4}
\end{figure*}

%% file: main.bbl
\begin{thebibliography}{59}
\expandafter\ifx\csname natexlab\endcsname\relax\def\natexlab#1{#1}\fi

\bibitem[{{Astropy Collaboration} {et~al.}(2022){Astropy Collaboration},
  {Price-Whelan}, {Lim}, {Earl}, {Starkman}, {Bradley}, {Shupe}, {Patil},
  {Corrales}, {Brasseur}, {N{"o}the}, {Donath}, {Tollerud}, {Morris},
  {Ginsburg}, {Vaher}, {Weaver}, {Tocknell}, {Jamieson}, {van Kerkwijk},
  {Robitaille}, {Merry}, {Bachetti}, {G{"u}nther}, {Aldcroft},
  {Alvarado-Montes}, {Archibald}, {B{'o}di}, {Bapat}, {Barentsen}, {Baz{'a}n},
  {Biswas}, {Boquien}, {Burke}, {Cara}, {Cara}, {Conroy}, {Conseil}, {Craig},
  {Cross}, {Cruz}, {D'Eugenio}, {Dencheva}, {Devillepoix}, {Dietrich},
  {Eigenbrot}, {Erben}, {Ferreira}, {Foreman-Mackey}, {Fox}, {Freij}, {Garg},
  {Geda}, {Glattly}, {Gondhalekar}, {Gordon}, {Grant}, {Greenfield}, {Groener},
  {Guest}, {Gurovich}, {Handberg}, {Hart}, {Hatfield-Dodds}, {Homeier},
  {Hosseinzadeh}, {Jenness}, {Jones}, {Joseph}, {Kalmbach}, {Karamehmetoglu},
  {Ka{l}uszy{'n}ski}, {Kelley}, {Kern}, {Kerzendorf}, {Koch}, {Kulumani},
  {Lee}, {Ly}, {Ma}, {MacBride}, {Maljaars}, {Muna}, {Murphy}, {Norman},
  {O'Steen}, {Oman}, {Pacifici}, {Pascual}, {Pascual-Granado}, {Patil},
  {Perren}, {Pickering}, {Rastogi}, {Roulston}, {Ryan}, {Rykoff}, {Sabater},
  {Sakurikar}, {Salgado}, {Sanghi}, {Saunders}, {Savchenko}, {Schwardt},
  {Seifert-Eckert}, {Shih}, {Jain}, {Shukla}, {Sick}, {Simpson},
  {Singanamalla}, {Singer}, {Singhal}, {Sinha}, {Sip{H{o}}cz}, {Spitler},
  {Stansby}, {Streicher}, {{{S}}umak}, {Swinbank}, {Taranu}, {Tewary},
  {Tremblay}, {Val-Borro}, {Van Kooten}, {Vasovi{'c}}, {Verma}, {de Miranda
  Cardoso}, {Williams}, {Wilson}, {Winkel}, {Wood-Vasey}, {Xue}, {Yoachim},
  {Zhang}, {Zonca}, \& {Astropy Project Contributors}}]{astropy:2022}
{Astropy Collaboration}, {Price-Whelan}, A.~M., {Lim}, P.~L., {et~al.} 2022,
  \apj, 935, 167

\bibitem[{{Astropy Collaboration} {et~al.}(2018){Astropy Collaboration},
  {Price-Whelan}, {Sip{\H{o}}cz}, {G{\"u}nther}, {Lim}, {Crawford}, {Conseil},
  {Shupe}, {Craig}, {Dencheva}, {Ginsburg}, {Vand erPlas}, {Bradley},
  {P{\'e}rez-Su{\'a}rez}, {de Val-Borro}, {Aldcroft}, {Cruz}, {Robitaille},
  {Tollerud}, {Ardelean}, {Babej}, {Bach}, {Bachetti}, {Bakanov}, {Bamford},
  {Barentsen}, {Barmby}, {Baumbach}, {Berry}, {Biscani}, {Boquien}, {Bostroem},
  {Bouma}, {Brammer}, {Bray}, {Breytenbach}, {Buddelmeijer}, {Burke},
  {Calderone}, {Cano Rodr{\'\i}guez}, {Cara}, {Cardoso}, {Cheedella}, {Copin},
  {Corrales}, {Crichton}, {D'Avella}, {Deil}, {Depagne}, {Dietrich}, {Donath},
  {Droettboom}, {Earl}, {Erben}, {Fabbro}, {Ferreira}, {Finethy}, {Fox},
  {Garrison}, {Gibbons}, {Goldstein}, {Gommers}, {Greco}, {Greenfield},
  {Groener}, {Grollier}, {Hagen}, {Hirst}, {Homeier}, {Horton}, {Hosseinzadeh},
  {Hu}, {Hunkeler}, {Ivezi{\'c}}, {Jain}, {Jenness}, {Kanarek}, {Kendrew},
  {Kern}, {Kerzendorf}, {Khvalko}, {King}, {Kirkby}, {Kulkarni}, {Kumar},
  {Lee}, {Lenz}, {Littlefair}, {Ma}, {Macleod}, {Mastropietro}, {McCully},
  {Montagnac}, {Morris}, {Mueller}, {Mumford}, {Muna}, {Murphy}, {Nelson},
  {Nguyen}, {Ninan}, {N{\"o}the}, {Ogaz}, {Oh}, {Parejko}, {Parley}, {Pascual},
  {Patil}, {Patil}, {Plunkett}, {Prochaska}, {Rastogi}, {Reddy Janga},
  {Sabater}, {Sakurikar}, {Seifert}, {Sherbert}, {Sherwood-Taylor}, {Shih},
  {Sick}, {Silbiger}, {Singanamalla}, {Singer}, {Sladen}, {Sooley},
  {Sornarajah}, {Streicher}, {Teuben}, {Thomas}, {Tremblay}, {Turner},
  {Terr{\'o}n}, {van Kerkwijk}, {de la Vega}, {Watkins}, {Weaver}, {Whitmore},
  {Woillez}, {Zabalza}, \& {Astropy Contributors}}]{astropy:2018}
{Astropy Collaboration}, {Price-Whelan}, A.~M., {Sip{\H{o}}cz}, B.~M., {et~al.}
  2018, \aj, 156, 123

\bibitem[{{Astropy Collaboration} {et~al.}(2013){Astropy Collaboration},
  {Robitaille}, {Tollerud}, {Greenfield}, {Droettboom}, {Bray}, {Aldcroft},
  {Davis}, {Ginsburg}, {Price-Whelan}, {Kerzendorf}, {Conley}, {Crighton},
  {Barbary}, {Muna}, {Ferguson}, {Grollier}, {Parikh}, {Nair}, {Unther},
  {Deil}, {Woillez}, {Conseil}, {Kramer}, {Turner}, {Singer}, {Fox}, {Weaver},
  {Zabalza}, {Edwards}, {Azalee Bostroem}, {Burke}, {Casey}, {Crawford},
  {Dencheva}, {Ely}, {Jenness}, {Labrie}, {Lim}, {Pierfederici}, {Pontzen},
  {Ptak}, {Refsdal}, {Servillat}, \& {Streicher}}]{astropy:2013}
{Astropy Collaboration}, {Robitaille}, T.~P., {Tollerud}, E.~J., {et~al.} 2013,
  \aap, 558, A33

\bibitem[{{Benz}(2017)}]{2017LRSP...14....2B}
{Benz}, A.~O. 2017, Living Reviews in Solar Physics, 14, 2

\bibitem[{{Brunner} {et~al.}(2022){Brunner}, {Liu}, {Lamer}, {Georgakakis},
  {Merloni}, {Brusa}, {Bulbul}, {Dennerl}, {Friedrich}, {Liu}, {Maitra},
  {Nandra}, {Ramos-Ceja}, {Sanders}, {Stewart}, {Boller}, {Buchner}, {Clerc},
  {Comparat}, {Dwelly}, {Eckert}, {Finoguenov}, {Freyberg}, {Ghirardini},
  {Gueguen}, {Haberl}, {Kreykenbohm}, {Krumpe}, {Osterhage}, {Pacaud},
  {Predehl}, {Reiprich}, {Robrade}, {Salvato}, {Santangelo}, {Schrabback},
  {Schwope}, \& {Wilms}}]{Brunner22.0}
{Brunner}, H., {Liu}, T., {Lamer}, G., {et~al.} 2022, \aap, 661, A1

\bibitem[{{Cargill} \& {Priest}(1983)}]{cargill83.0}
{Cargill}, P.~J. \& {Priest}, E.~R. 1983, \apj, 266, 383

\bibitem[{{Castellanos Dur{\'a}n} \& {Kleint}(2020)}]{CastellanosDuran2020}
{Castellanos Dur{\'a}n}, J.~S. \& {Kleint}, L. 2020, \apj, 904, 96

\bibitem[{{Chabrier}(2001)}]{Chabrier2001}
{Chabrier}, G. 2001, \apj, 554, 1274

\bibitem[{{Davenport} {et~al.}(2014){Davenport}, {Hawley}, {Hebb},
  {Wisniewski}, {Kowalski}, {Johnson}, {Malatesta}, {Peraza}, {Keil},
  {Silverberg}, {Jansen}, {Scheffler}, {Berdis}, {Larsen}, \&
  {Hilton}}]{Davenport2014}
{Davenport}, J. R.~A., {Hawley}, S.~L., {Hebb}, L., {et~al.} 2014, \apj, 797,
  122

\bibitem[{Freedman \& Diaconis(1981)}]{Freedman1981}
Freedman, D. \& Diaconis, P. 1981, Zeitschrift f{\"u}r
  Wahrscheinlichkeitstheorie und Verwandte Gebiete, 57, 453

\bibitem[{{Fuhrmeister} {et~al.}(2011){Fuhrmeister}, {Lalitha}, {Poppenhaeger},
  {Rudolf}, {Liefke}, {Reiners}, {Schmitt}, \& {Ness}}]{Fuhrmeister2011}
{Fuhrmeister}, B., {Lalitha}, S., {Poppenhaeger}, K., {et~al.} 2011, \aap, 534,
  A133

\bibitem[{{Gaia Collaboration} {et~al.}(2018){Gaia Collaboration}, {Brown},
  {Vallenari}, {Prusti}, {de Bruijne}, {Babusiaux}, {Bailer-Jones}, {Biermann},
  {Evans}, {Eyer}, {Jansen}, {Jordi}, {Klioner}, {Lammers}, {Lindegren},
  {Luri}, {Mignard}, {Panem}, {Pourbaix}, {Randich}, {Sartoretti}, {Siddiqui},
  {Soubiran}, {van Leeuwen}, {Walton}, {Arenou}, {Bastian}, {Cropper},
  {Drimmel}, {Katz}, {Lattanzi}, {Bakker}, {Cacciari}, {Casta{\~n}eda},
  {Chaoul}, {Cheek}, {De Angeli}, {Fabricius}, {Guerra}, {Holl}, {Masana},
  {Messineo}, {Mowlavi}, {Nienartowicz}, {Panuzzo}, {Portell}, {Riello},
  {Seabroke}, {Tanga}, {Th{\'e}venin}, {Gracia-Abril}, {Comoretto},
  {Garcia-Reinaldos}, {Teyssier}, {Altmann}, {Andrae}, {Audard},
  {Bellas-Velidis}, {Benson}, {Berthier}, {Blomme}, {Burgess}, {Busso},
  {Carry}, {Cellino}, {Clementini}, {Clotet}, {Creevey}, {Davidson}, {De
  Ridder}, {Delchambre}, {Dell'Oro}, {Ducourant},
  {Fern{\'a}ndez-Hern{\'a}ndez}, {Fouesneau}, {Fr{\'e}mat}, {Galluccio},
  {Garc{\'\i}a-Torres}, {Gonz{\'a}lez-N{\'u}{\~n}ez}, {Gonz{\'a}lez-Vidal},
  {Gosset}, {Guy}, {Halbwachs}, {Hambly}, {Harrison}, {Hern{\'a}ndez},
  {Hestroffer}, {Hodgkin}, {Hutton}, {Jasniewicz}, {Jean-Antoine-Piccolo},
  {Jordan}, {Korn}, {Krone-Martins}, {Lanzafame}, {Lebzelter}, {L{\"o}ffler},
  {Manteiga}, {Marrese}, {Mart{\'\i}n-Fleitas}, {Moitinho}, {Mora}, {Muinonen},
  {Osinde}, {Pancino}, {Pauwels}, {Petit}, {Recio-Blanco}, {Richards},
  {Rimoldini}, {Robin}, {Sarro}, {Siopis}, {Smith}, {Sozzetti}, {S{\"u}veges},
  {Torra}, {van Reeven}, {Abbas}, {Abreu Aramburu}, {Accart}, {Aerts},
  {Altavilla}, {{\'A}lvarez}, {Alvarez}, {Alves}, {Anderson}, {Andrei},
  {Anglada Varela}, {Antiche}, {Antoja}, {Arcay}, {Astraatmadja}, {Bach},
  {Baker}, {Balaguer-N{\'u}{\~n}ez}, {Balm}, {Barache}, {Barata}, {Barbato},
  {Barblan}, {Barklem}, {Barrado}, {Barros}, {Barstow}, {Bartholom{\'e}
  Mu{\~n}oz}, {Bassilana}, {Becciani}, {Bellazzini}, {Berihuete}, {Bertone},
  {Bianchi}, {Bienaym{\'e}}, {Blanco-Cuaresma}, {Boch}, {Boeche}, {Bombrun},
  {Borrachero}, {Bossini}, {Bouquillon}, {Bourda}, {Bragaglia}, {Bramante},
  {Breddels}, {Bressan}, {Brouillet}, {Br{\"u}semeister}, {Brugaletta},
  {Bucciarelli}, {Burlacu}, {Busonero}, {Butkevich}, {Buzzi}, {Caffau},
  {Cancelliere}, {Cannizzaro}, {Cantat-Gaudin}, {Carballo}, {Carlucci},
  {Carrasco}, {Casamiquela}, {Castellani}, {Castro-Ginard}, {Charlot},
  {Chemin}, {Chiavassa}, {Cocozza}, {Costigan}, {Cowell}, {Crifo}, {Crosta},
  {Crowley}, {Cuypers}, {Dafonte}, {Damerdji}, {Dapergolas}, {David}, {David},
  {de Laverny}, {De Luise}, {De March}, {de Martino}, {de Souza}, {de Torres},
  {Debosscher}, {del Pozo}, {Delbo}, {Delgado}, {Delgado}, {Di Matteo},
  {Diakite}, {Diener}, {Distefano}, {Dolding}, {Drazinos}, {Dur{\'a}n},
  {Edvardsson}, {Enke}, {Eriksson}, {Esquej}, {Eynard Bontemps}, {Fabre},
  {Fabrizio}, {Faigler}, {Falc{\~a}o}, {Farr{\`a}s Casas}, {Federici},
  {Fedorets}, {Fernique}, {Figueras}, {Filippi}, {Findeisen}, {Fonti},
  {Fraile}, {Fraser}, {Fr{\'e}zouls}, {Gai}, {Galleti}, {Garabato},
  {Garc{\'\i}a-Sedano}, {Garofalo}, {Garralda}, {Gavel}, {Gavras}, {Gerssen},
  {Geyer}, {Giacobbe}, {Gilmore}, {Girona}, {Giuffrida}, {Glass}, {Gomes},
  {Granvik}, {Gueguen}, {Guerrier}, {Guiraud}, {Guti{\'e}rrez-S{\'a}nchez},
  {Haigron}, {Hatzidimitriou}, {Hauser}, {Haywood}, {Heiter}, {Helmi}, {Heu},
  {Hilger}, {Hobbs}, {Hofmann}, {Holland}, {Huckle}, {Hypki}, {Icardi},
  {Jan{\ss}en}, {Jevardat de Fombelle}, {Jonker}, {Juh{\'a}sz}, {Julbe},
  {Karampelas}, {Kewley}, {Klar}, {Kochoska}, {Kohley}, {Kolenberg},
  {Kontizas}, {Kontizas}, {Koposov}, {Kordopatis}, {Kostrzewa-Rutkowska},
  {Koubsky}, {Lambert}, {Lanza}, {Lasne}, {Lavigne}, {Le Fustec}, {Le
  Poncin-Lafitte}, {Lebreton}, {Leccia}, {Leclerc}, {Lecoeur-Taibi},
  {Lenhardt}, {Leroux}, {Liao}, {Licata}, {Lindstr{\o}m}, {Lister}, {Livanou},
  {Lobel}, {L{\'o}pez}, {Managau}, {Mann}, {Mantelet}, {Marchal}, {Marchant},
  {Marconi}, {Marinoni}, {Marschalk{\'o}}, {Marshall}, {Martino}, {Marton},
  {Mary}, {Massari}, {Matijevi{\v{c}}}, {Mazeh}, {McMillan}, {Messina},
  {Michalik}, {Millar}, {Molina}, {Molinaro}, {Moln{\'a}r}, {Montegriffo},
  {Mor}, {Morbidelli}, {Morel}, {Morris}, {Mulone}, {Muraveva}, {Musella},
  {Nelemans}, {Nicastro}, {Noval}, {O'Mullane}, {Ord{\'e}novic},
  {Ord{\'o}{\~n}ez-Blanco}, {Osborne}, {Pagani}, {Pagano}, {Pailler},
  {Palacin}, {Palaversa}, {Panahi}, {Pawlak}, {Piersimoni}, {Pineau}, {Plachy},
  {Plum}, {Poggio}, {Poujoulet}, {Pr{\v{s}}a}, {Pulone}, {Racero}, {Ragaini},
  {Rambaux}, {Ramos-Lerate}, {Regibo}, {Reyl{\'e}}, {Riclet}, {Ripepi}, {Riva},
  {Rivard}, {Rixon}, {Roegiers}, {Roelens}, {Romero-G{\'o}mez}, {Rowell},
  {Royer}, {Ruiz-Dern}, {Sadowski}, {Sagrist{\`a} Sell{\'e}s}, {Sahlmann},
  {Salgado}, {Salguero}, {Sanna}, {Santana-Ros}, {Sarasso}, {Savietto},
  {Schultheis}, {Sciacca}, {Segol}, {Segovia}, {S{\'e}gransan}, {Shih},
  {Siltala}, {Silva}, {Smart}, {Smith}, {Solano}, {Solitro}, {Sordo}, {Soria
  Nieto}, {Souchay}, {Spagna}, {Spoto}, {Stampa}, {Steele},
  {Steidelm{\"u}ller}, {Stephenson}, {Stoev}, {Suess}, {Surdej}, {Szabados},
  {Szegedi-Elek}, {Tapiador}, {Taris}, {Tauran}, {Taylor}, {Teixeira},
  {Terrett}, {Teyssandier}, {Thuillot}, {Titarenko}, {Torra Clotet}, {Turon},
  {Ulla}, {Utrilla}, {Uzzi}, {Vaillant}, {Valentini}, {Valette}, {van Elteren},
  {Van Hemelryck}, {van Leeuwen}, {Vaschetto}, {Vecchiato}, {Veljanoski},
  {Viala}, {Vicente}, {Vogt}, {von Essen}, {Voss}, {Votruba}, {Voutsinas},
  {Walmsley}, {Weiler}, {Wertz}, {Wevers}, {Wyrzykowski}, {Yoldas},
  {{\v{Z}}erjal}, {Ziaeepour}, {Zorec}, {Zschocke}, {Zucker}, {Zurbach}, \&
  {Zwitter}}]{Gaia2018}
{Gaia Collaboration}, {Brown}, A.~G.~A., {Vallenari}, A., {et~al.} 2018, \aap,
  616, A1

\bibitem[{{Gaia Collaboration} {et~al.}(2016){Gaia Collaboration}, {Prusti},
  {de Bruijne}, {Brown}, {Vallenari}, {Babusiaux}, {Bailer-Jones}, {Bastian},
  {Biermann}, {Evans}, {Eyer}, {Jansen}, {Jordi}, {Klioner}, {Lammers},
  {Lindegren}, {Luri}, {Mignard}, {Milligan}, {Panem}, {Poinsignon},
  {Pourbaix}, {Randich}, {Sarri}, {Sartoretti}, {Siddiqui}, {Soubiran},
  {Valette}, {van Leeuwen}, {Walton}, {Aerts}, {Arenou}, {Cropper}, {Drimmel},
  {H{\o}g}, {Katz}, {Lattanzi}, {O'Mullane}, {Grebel}, {Holland}, {Huc},
  {Passot}, {Bramante}, {Cacciari}, {Casta{\~n}eda}, {Chaoul}, {Cheek}, {De
  Angeli}, {Fabricius}, {Guerra}, {Hern{\'a}ndez}, {Jean-Antoine-Piccolo},
  {Masana}, {Messineo}, {Mowlavi}, {Nienartowicz}, {Ord{\'o}{\~n}ez-Blanco},
  {Panuzzo}, {Portell}, {Richards}, {Riello}, {Seabroke}, {Tanga},
  {Th{\'e}venin}, {Torra}, {Els}, {Gracia-Abril}, {Comoretto},
  {Garcia-Reinaldos}, {Lock}, {Mercier}, {Altmann}, {Andrae}, {Astraatmadja},
  {Bellas-Velidis}, {Benson}, {Berthier}, {Blomme}, {Busso}, {Carry},
  {Cellino}, {Clementini}, {Cowell}, {Creevey}, {Cuypers}, {Davidson}, {De
  Ridder}, {de Torres}, {Delchambre}, {Dell'Oro}, {Ducourant}, {Fr{\'e}mat},
  {Garc{\'\i}a-Torres}, {Gosset}, {Halbwachs}, {Hambly}, {Harrison}, {Hauser},
  {Hestroffer}, {Hodgkin}, {Huckle}, {Hutton}, {Jasniewicz}, {Jordan},
  {Kontizas}, {Korn}, {Lanzafame}, {Manteiga}, {Moitinho}, {Muinonen},
  {Osinde}, {Pancino}, {Pauwels}, {Petit}, {Recio-Blanco}, {Robin}, {Sarro},
  {Siopis}, {Smith}, {Smith}, {Sozzetti}, {Thuillot}, {van Reeven}, {Viala},
  {Abbas}, {Abreu Aramburu}, {Accart}, {Aguado}, {Allan}, {Allasia},
  {Altavilla}, {{\'A}lvarez}, {Alves}, {Anderson}, {Andrei}, {Anglada Varela},
  {Antiche}, {Antoja}, {Ant{\'o}n}, {Arcay}, {Atzei}, {Ayache}, {Bach},
  {Baker}, {Balaguer-N{\'u}{\~n}ez}, {Barache}, {Barata}, {Barbier}, {Barblan},
  {Baroni}, {Barrado y Navascu{\'e}s}, {Barros}, {Barstow}, {Becciani},
  {Bellazzini}, {Bellei}, {Bello Garc{\'\i}a}, {Belokurov}, {Bendjoya},
  {Berihuete}, {Bianchi}, {Bienaym{\'e}}, {Billebaud}, {Blagorodnova},
  {Blanco-Cuaresma}, {Boch}, {Bombrun}, {Borrachero}, {Bouquillon}, {Bourda},
  {Bouy}, {Bragaglia}, {Breddels}, {Brouillet}, {Br{\"u}semeister},
  {Bucciarelli}, {Budnik}, {Burgess}, {Burgon}, {Burlacu}, {Busonero}, {Buzzi},
  {Caffau}, {Cambras}, {Campbell}, {Cancelliere}, {Cantat-Gaudin}, {Carlucci},
  {Carrasco}, {Castellani}, {Charlot}, {Charnas}, {Charvet}, {Chassat},
  {Chiavassa}, {Clotet}, {Cocozza}, {Collins}, {Collins}, {Costigan}, {Crifo},
  {Cross}, {Crosta}, {Crowley}, {Dafonte}, {Damerdji}, {Dapergolas}, {David},
  {David}, {De Cat}, {de Felice}, {de Laverny}, {De Luise}, {De March}, {de
  Martino}, {de Souza}, {Debosscher}, {del Pozo}, {Delbo}, {Delgado},
  {Delgado}, {di Marco}, {Di Matteo}, {Diakite}, {Distefano}, {Dolding}, {Dos
  Anjos}, {Drazinos}, {Dur{\'a}n}, {Dzigan}, {Ecale}, {Edvardsson}, {Enke},
  {Erdmann}, {Escolar}, {Espina}, {Evans}, {Eynard Bontemps}, {Fabre},
  {Fabrizio}, {Faigler}, {Falc{\~a}o}, {Farr{\`a}s Casas}, {Faye}, {Federici},
  {Fedorets}, {Fern{\'a}ndez-Hern{\'a}ndez}, {Fernique}, {Fienga}, {Figueras},
  {Filippi}, {Findeisen}, {Fonti}, {Fouesneau}, {Fraile}, {Fraser}, {Fuchs},
  {Furnell}, {Gai}, {Galleti}, {Galluccio}, {Garabato}, {Garc{\'\i}a-Sedano},
  {Gar{\'e}}, {Garofalo}, {Garralda}, {Gavras}, {Gerssen}, {Geyer}, {Gilmore},
  {Girona}, {Giuffrida}, {Gomes}, {Gonz{\'a}lez-Marcos},
  {Gonz{\'a}lez-N{\'u}{\~n}ez}, {Gonz{\'a}lez-Vidal}, {Granvik}, {Guerrier},
  {Guillout}, {Guiraud}, {G{\'u}rpide}, {Guti{\'e}rrez-S{\'a}nchez}, {Guy},
  {Haigron}, {Hatzidimitriou}, {Haywood}, {Heiter}, {Helmi}, {Hobbs},
  {Hofmann}, {Holl}, {Holland}, {Hunt}, {Hypki}, {Icardi}, {Irwin}, {Jevardat
  de Fombelle}, {Jofr{\'e}}, {Jonker}, {Jorissen}, {Julbe}, {Karampelas},
  {Kochoska}, {Kohley}, {Kolenberg}, {Kontizas}, {Koposov}, {Kordopatis},
  {Koubsky}, {Kowalczyk}, {Krone-Martins}, {Kudryashova}, {Kull}, {Bachchan},
  {Lacoste-Seris}, {Lanza}, {Lavigne}, {Le Poncin-Lafitte}, {Lebreton},
  {Lebzelter}, {Leccia}, {Leclerc}, {Lecoeur-Taibi}, {Lemaitre}, {Lenhardt},
  {Leroux}, {Liao}, {Licata}, {Lindstr{\o}m}, {Lister}, {Livanou}, {Lobel},
  {L{\"o}ffler}, {L{\'o}pez}, {Lopez-Lozano}, {Lorenz}, {Loureiro},
  {MacDonald}, {Magalh{\~a}es Fernandes}, {Managau}, {Mann}, {Mantelet},
  {Marchal}, {Marchant}, {Marconi}, {Marie}, {Marinoni}, {Marrese},
  {Marschalk{\'o}}, {Marshall}, {Mart{\'\i}n-Fleitas}, {Martino}, {Mary},
  {Matijevi{\v{c}}}, {Mazeh}, {McMillan}, {Messina}, {Mestre}, {Michalik},
  {Millar}, {Miranda}, {Molina}, {Molinaro}, {Molinaro}, {Moln{\'a}r},
  {Moniez}, {Montegriffo}, {Monteiro}, {Mor}, {Mora}, {Morbidelli}, {Morel},
  {Morgenthaler}, {Morley}, {Morris}, {Mulone}, {Muraveva}, {Musella},
  {Narbonne}, {Nelemans}, {Nicastro}, {Noval}, {Ord{\'e}novic},
  {Ordieres-Mer{\'e}}, {Osborne}, {Pagani}, {Pagano}, {Pailler}, {Palacin},
  {Palaversa}, {Parsons}, {Paulsen}, {Pecoraro}, {Pedrosa}, {Pentik{\"a}inen},
  {Pereira}, {Pichon}, {Piersimoni}, {Pineau}, {Plachy}, {Plum}, {Poujoulet},
  {Pr{\v{s}}a}, {Pulone}, {Ragaini}, {Rago}, {Rambaux}, {Ramos-Lerate},
  {Ranalli}, {Rauw}, {Read}, {Regibo}, {Renk}, {Reyl{\'e}}, {Ribeiro},
  {Rimoldini}, {Ripepi}, {Riva}, {Rixon}, {Roelens}, {Romero-G{\'o}mez},
  {Rowell}, {Royer}, {Rudolph}, {Ruiz-Dern}, {Sadowski}, {Sagrist{\`a}
  Sell{\'e}s}, {Sahlmann}, {Salgado}, {Salguero}, {Sarasso}, {Savietto},
  {Schnorhk}, {Schultheis}, {Sciacca}, {Segol}, {Segovia}, {Segransan},
  {Serpell}, {Shih}, {Smareglia}, {Smart}, {Smith}, {Solano}, {Solitro},
  {Sordo}, {Soria Nieto}, {Souchay}, {Spagna}, {Spoto}, {Stampa}, {Steele},
  {Steidelm{\"u}ller}, {Stephenson}, {Stoev}, {Suess}, {S{\"u}veges}, {Surdej},
  {Szabados}, {Szegedi-Elek}, {Tapiador}, {Taris}, {Tauran}, {Taylor},
  {Teixeira}, {Terrett}, {Tingley}, {Trager}, {Turon}, {Ulla}, {Utrilla},
  {Valentini}, {van Elteren}, {Van Hemelryck}, {van Leeuwen}, {Varadi},
  {Vecchiato}, {Veljanoski}, {Via}, {Vicente}, {Vogt}, {Voss}, {Votruba},
  {Voutsinas}, {Walmsley}, {Weiler}, {Weingrill}, {Werner}, {Wevers},
  {Whitehead}, {Wyrzykowski}, {Yoldas}, {{\v{Z}}erjal}, {Zucker}, {Zurbach},
  {Zwitter}, {Alecu}, {Allen}, {Allende Prieto}, {Amorim},
  {Anglada-Escud{\'e}}, {Arsenijevic}, {Azaz}, {Balm}, {Beck}, {Bernstein},
  {Bigot}, {Bijaoui}, {Blasco}, {Bonfigli}, {Bono}, {Boudreault}, {Bressan},
  {Brown}, {Brunet}, {Bunclark}, {Buonanno}, {Butkevich}, {Carret}, {Carrion},
  {Chemin}, {Ch{\'e}reau}, {Corcione}, {Darmigny}, {de Boer}, {de Teodoro}, {de
  Zeeuw}, {Delle Luche}, {Domingues}, {Dubath}, {Fodor}, {Fr{\'e}zouls},
  {Fries}, {Fustes}, {Fyfe}, {Gallardo}, {Gallegos}, {Gardiol}, {Gebran},
  {Gomboc}, {G{\'o}mez}, {Grux}, {Gueguen}, {Heyrovsky}, {Hoar}, {Iannicola},
  {Isasi Parache}, {Janotto}, {Joliet}, {Jonckheere}, {Keil}, {Kim},
  {Klagyivik}, {Klar}, {Knude}, {Kochukhov}, {Kolka}, {Kos}, {Kutka}, {Lainey},
  {LeBouquin}, {Liu}, {Loreggia}, {Makarov}, {Marseille}, {Martayan},
  {Martinez-Rubi}, {Massart}, {Meynadier}, {Mignot}, {Munari}, {Nguyen},
  {Nordlander}, {Ocvirk}, {O'Flaherty}, {Olias Sanz}, {Ortiz}, {Osorio},
  {Oszkiewicz}, {Ouzounis}, {Palmer}, {Park}, {Pasquato}, {Peltzer}, {Peralta},
  {P{\'e}turaud}, {Pieniluoma}, {Pigozzi}, {Poels}, {Prat}, {Prod'homme},
  {Raison}, {Rebordao}, {Risquez}, {Rocca-Volmerange}, {Rosen}, {Ruiz-Fuertes},
  {Russo}, {Sembay}, {Serraller Vizcaino}, {Short}, {Siebert}, {Silva},
  {Sinachopoulos}, {Slezak}, {Soffel}, {Sosnowska}, {Strai{\v{z}}ys}, {ter
  Linden}, {Terrell}, {Theil}, {Tiede}, {Troisi}, {Tsalmantza}, {Tur},
  {Vaccari}, {Vachier}, {Valles}, {Van Hamme}, {Veltz}, {Virtanen}, {Wallut},
  {Wichmann}, {Wilkinson}, {Ziaeepour}, \& {Zschocke}}]{Gaia2016}
{Gaia Collaboration}, {Prusti}, T., {de Bruijne}, J.~H.~J., {et~al.} 2016,
  \aap, 595, A1

\bibitem[{{Guarcello} {et~al.}(2019){Guarcello}, {Micela}, {Sciortino},
  {L{\'o}pez-Santiago}, {Argiroffi}, {Reale}, {Flaccomio},
  {Alvarado-G{\'o}mez}, {Antoniou}, {Drake}, {Pillitteri}, {Rebull}, \&
  {Stauffer}}]{Guarcello2019}
{Guarcello}, M.~G., {Micela}, G., {Sciortino}, S., {et~al.} 2019, \aap, 622,
  A210

\bibitem[{{G{\"u}nther} {et~al.}(2020){G{\"u}nther}, {Zhan}, {Seager},
  {Rimmer}, {Ranjan}, {Stassun}, {Oelkers}, {Daylan}, {Newton}, {Kristiansen},
  {Olah}, {Gillen}, {Rappaport}, {Ricker}, {Vanderspek}, {Latham}, {Winn},
  {Jenkins}, {Glidden}, {Fausnaugh}, {Levine}, {Dittmann}, {Quinn},
  {Krishnamurthy}, \& {Ting}}]{2020AJ....159...60G}
{G{\"u}nther}, M.~N., {Zhan}, Z., {Seager}, S., {et~al.} 2020, \aj, 159, 60

\bibitem[{Harris {et~al.}(2020)Harris, Millman, van~der Walt, Gommers,
  Virtanen, Cournapeau, Wieser, Taylor, Berg, Smith, Kern, Picus, Hoyer, van
  Kerkwijk, Brett, Haldane, del R{\'{i}}o, Wiebe, Peterson,
  G{\'{e}}rard-Marchant, Sheppard, Reddy, Weckesser, Abbasi, Gohlke, \&
  Oliphant}]{harris2020array}
Harris, C.~R., Millman, K.~J., van~der Walt, S.~J., {et~al.} 2020, Nature, 585,
  357

\bibitem[{{Hawley} {et~al.}(2014){Hawley}, {Davenport}, {Kowalski},
  {Wisniewski}, {Hebb}, {Deitrick}, \& {Hilton}}]{Hawley2014}
{Hawley}, S.~L., {Davenport}, J. R.~A., {Kowalski}, A.~F., {et~al.} 2014, \apj,
  797, 121

\bibitem[{{Howard} {et~al.}(2012){Howard}, {Marcy}, {Bryson}, {Jenkins},
  {Rowe}, {Batalha}, {Borucki}, {Koch}, {Dunham}, {Gautier}, {Van Cleve},
  {Cochran}, {Latham}, {Lissauer}, {Torres}, {Brown}, {Gilliland}, {Buchhave},
  {Caldwell}, {Christensen-Dalsgaard}, {Ciardi}, {Fressin}, {Haas}, {Howell},
  {Kjeldsen}, {Seager}, {Rogers}, {Sasselov}, {Steffen}, {Basri},
  {Charbonneau}, {Christiansen}, {Clarke}, {Dupree}, {Fabrycky}, {Fischer},
  {Ford}, {Fortney}, {Tarter}, {Girouard}, {Holman}, {Johnson}, {Klaus},
  {Machalek}, {Moorhead}, {Morehead}, {Ragozzine}, {Tenenbaum}, {Twicken},
  {Quinn}, {Isaacson}, {Shporer}, {Lucas}, {Walkowicz}, {Welsh}, {Boss},
  {Devore}, {Gould}, {Smith}, {Morris}, {Prsa}, {Morton}, {Still}, {Thompson},
  {Mullally}, {Endl}, \& {MacQueen}}]{Howard2012}
{Howard}, A.~W., {Marcy}, G.~W., {Bryson}, S.~T., {et~al.} 2012, \apjs, 201, 15

\bibitem[{{Hudson}(1991)}]{1991SoPh..133..357H}
{Hudson}, H.~S. 1991, \solphys, 133, 357

\bibitem[{{Hudson}(2011)}]{hudson11.0}
{Hudson}, H.~S. 2011, \ssr, 158, 5

\bibitem[{{Hunt-Walker} {et~al.}(2012){Hunt-Walker}, {Hilton}, {Kowalski},
  {Hawley}, \& {Matthews}}]{Hunt-Walker2012}
{Hunt-Walker}, N.~M., {Hilton}, E.~J., {Kowalski}, A.~F., {Hawley}, S.~L., \&
  {Matthews}, J.~M. 2012, \pasp, 124, 545

\bibitem[{{Jao} {et~al.}(2018){Jao}, {Henry}, {Gies}, \& {Hambly}}]{Jao2018}
{Jao}, W.-C., {Henry}, T.~J., {Gies}, D.~R., \& {Hambly}, N.~C. 2018, \apjl,
  861, L11

\bibitem[{{Joy} \& {Humason}(1949)}]{JoyHumason1949}
{Joy}, A.~H. \& {Humason}, M.~L. 1949, \pasp, 61, 133

\bibitem[{{Kaltenegger} {et~al.}(2019){Kaltenegger}, {Pepper}, {Stassun}, \&
  {Oelkers}}]{2019ApJ...874L...8K}
{Kaltenegger}, L., {Pepper}, J., {Stassun}, K., \& {Oelkers}, R. 2019, \apjl,
  874, L8

\bibitem[{{Katsova} {et~al.}(2002){Katsova}, {Livshits}, \&
  {Schmitt}}]{Katsova2002}
{Katsova}, M.~M., {Livshits}, M.~A., \& {Schmitt}, J.~H.~M.~M. 2002, in
  Astronomical Society of the Pacific Conference Series, Vol. 277, Stellar
  Coronae in the Chandra and XMM-NEWTON Era, ed. F.~{Favata} \& J.~J. {Drake},
  515

\bibitem[{{Kretzschmar}(2011)}]{kretzschmar11.0}
{Kretzschmar}, M. 2011, \aap, 530, A84

\bibitem[{{Kuznetsov} \& {Kolotkov}(2021)}]{KuznetsovKolotkov2021}
{Kuznetsov}, A.~A. \& {Kolotkov}, D.~Y. 2021, \apj, 912, 81

\bibitem[{{Lomb}(1976)}]{1976Ap&SS..39..447L}
{Lomb}, N.~R. 1976, \apss, 39, 447

\bibitem[{{Magaudda} {et~al.}(2020){Magaudda}, {Stelzer}, {Covey}, {Raetz},
  {Matt}, \& {Scholz}}]{Magaudda2020}
{Magaudda}, E., {Stelzer}, B., {Covey}, K.~R., {et~al.} 2020, \aap, 638, A20

\bibitem[{{Magaudda} {et~al.}(2022{\natexlab{a}}){Magaudda}, {Stelzer}, \&
  {Raetz}}]{Magaudda2022AN}
{Magaudda}, E., {Stelzer}, B., \& {Raetz}, S. 2022{\natexlab{a}}, Astronomische
  Nachrichten, 343, e20220049

\bibitem[{{Magaudda} {et~al.}(2022{\natexlab{b}}){Magaudda}, {Stelzer},
  {Raetz}, {Klutsch}, {Salvato}, \& {Wolf}}]{Magaudda2022}
{Magaudda}, E., {Stelzer}, B., {Raetz}, S., {et~al.} 2022{\natexlab{b}}, \aap,
  661, A29

\bibitem[{{Mann} {et~al.}(2015){Mann}, {Feiden}, {Gaidos}, {Boyajian}, \& {von
  Braun}}]{Mann2015}
{Mann}, A.~W., {Feiden}, G.~A., {Gaidos}, E., {Boyajian}, T., \& {von Braun},
  K. 2015, \apj, 804, 64

\bibitem[{{McQuillan} {et~al.}(2013){McQuillan}, {Aigrain}, \&
  {Mazeh}}]{McQuillan2013}
{McQuillan}, A., {Aigrain}, S., \& {Mazeh}, T. 2013, \mnras, 432, 1203

\bibitem[{{Merloni} {et~al.}(2024){Merloni}, {Larmer}, {Liu}, {Ramos-Ceja},
  {Brunner}, {Bubul}, {Dennerli}, {Doroshenko}, {Freyberg}, {Friedrich},
  {Gatuzz}, {Georgakakis}, {Haberl}, {Igo}, {Kreykenbohm}, {Liu}, {Maitra},
  {Malyali}, {Mayer}, {Nandra}, {Predehl}, {Robrade}, {Salvato}, {Sanders},
  {Stewart}, {Tubín-Arenas}, {Weber}, {Wilms}, {Arcodia}, {Artis},
  {Aschersleben}, {Avakyan}, {Aydar}, {Bahar}, {Balzer}, {Becker}, {Berger},
  {Boller}, {Bornemann}, {Brüggen}, {Brusa}, {Buchner}, {Burwitz},
  {Camilloni}, {Clerc}, {Comparat}, {Coutinho}, {Czesla}, {Dannhauer},
  {Dauner}, {Dauser}, {Dietl}, {Dolag}, {Dwelly}, {Egg}, {Ehl}, {Freund},
  {Friedrich}, {Gaida}, {Garrel}, {Ghirardini}, {Gokus}, {Grünwald},
  {Grandis}, {Grotova}, {Gruen}, {Gueguen}, {Hämmerich}, {Hamaus}, {Hasinger},
  {Haubner}, {Homan}, {IderChitham}, {Joseph}, {Joyce}, {König},
  {Kaltenbrunner}, {Khokhriakova}, {Kink}, {Kirsch}, {Kluge}, {Knies},
  {Krippendorf}, {Krumpe}, {Kurpas}, {Li}, {Liu}, {Locatelli}, {Lorenz},
  {Müller}, {Magaudda}, {Mannes}, {McCall}, {Meidinger}, {Michailidis},
  {Migkas}, {Muñoz-Giraldo}, {Musiimenta}, {Nguyen-Dang}, {Ni}, {Olechowska},
  {Ota}, {Pacaud}, {Pasini}, {Perinati}, {Pires}, {Pommranz}, {Ponti},
  {Poppenhaeger}, {Pühlhofer}, {Rau}, {Reh}, {Reiprich}, {Roster}, {Saeedi},
  {Santangelo}, {Sasaki}, {Schmitt}, {Schneider}, {Schrabback}, {Schuster},
  {Schwope}, {Seppi}, {Serim}, {Shreeram}, {Sokolova-Lapa}, {Starck},
  {Stelzer}, {Stierhof}, {Suleimanov}, {Tenzer}, {Traulsen}, {Trümper},
  {Tsuge}, {Urrutia}, {Veronica}, {Waddell}, {Willer}, {Wolf}, {Yeung},
  {Zainab}, {Zangrandi}, {Zhang}, {Zhang}, \& {Zheng}}]{Merloni23.0}
{Merloni}, A., {Larmer}, G., {Liu}, T., {et~al.} 2024, \aap, 682, A34

\bibitem[{Namekata {et~al.}(2020)Namekata, Maehara, Sasaki, Kawai, Notsu,
  Kowalski, Allred, Iwakiri, Tsuboi, Murata, Niwano, Shiraishi, Adachi, Iida,
  Oeda, Honda, Tozuka, Katoh, Onozato, Okamoto, Isogai, Kimura, Kojiguchi,
  Wakamatsu, Tampo, Nogami, \& Shibata}]{Namekata_2020}
Namekata, K., Maehara, H., Sasaki, R., {et~al.} 2020, Publications of the
  Astronomical Society of Japan, 72

\bibitem[{{Namekata} {et~al.}(2017){Namekata}, {Sakaue}, {Watanabe}, {Asai},
  {Maehara}, {Notsu}, {Notsu}, {Honda}, {Ishii}, {Ikuta}, {Nogami}, \&
  {Shibata}}]{Namekata17}
{Namekata}, K., {Sakaue}, T., {Watanabe}, K., {et~al.} 2017, \apj, 851, 91

\bibitem[{{Predehl} {et~al.}(2021){Predehl}, {Andritschke}, {Arefiev},
  {Babyshkin}, {Batanov}, {Becker}, {B{\"o}hringer}, {Bogomolov}, {Boller},
  {Borm}, {Bornemann}, {Br{\"a}uninger}, {Br{\"u}ggen}, {Brunner}, {Brusa},
  {Bulbul}, {Buntov}, {Burwitz}, {Burkert}, {Clerc}, {Churazov}, {Coutinho},
  {Dauser}, {Dennerl}, {Doroshenko}, {Eder}, {Emberger}, {Eraerds},
  {Finoguenov}, {Freyberg}, {Friedrich}, {Friedrich}, {F{\"u}rmetz},
  {Georgakakis}, {Gilfanov}, {Granato}, {Grossberger}, {Gueguen}, {Gureev},
  {Haberl}, {H{\"a}lker}, {Hartner}, {Hasinger}, {Huber}, {Ji}, {Kienlin},
  {Kink}, {Korotkov}, {Kreykenbohm}, {Lamer}, {Lomakin}, {Lapshov}, {Liu},
  {Maitra}, {Meidinger}, {Menz}, {Merloni}, {Mernik}, {Mican}, {Mohr},
  {M{\"u}ller}, {Nandra}, {Nazarov}, {Pacaud}, {Pavlinsky}, {Perinati},
  {Pfeffermann}, {Pietschner}, {Ramos-Ceja}, {Rau}, {Reiffers}, {Reiprich},
  {Robrade}, {Salvato}, {Sanders}, {Santangelo}, {Sasaki}, {Scheuerle},
  {Schmid}, {Schmitt}, {Schwope}, {Shirshakov}, {Steinmetz}, {Stewart},
  {Str{\"u}der}, {Sunyaev}, {Tenzer}, {Tiedemann}, {Tr{\"u}mper}, {Voron},
  {Weber}, {Wilms}, \& {Yaroshenko}}]{Predehl2021}
{Predehl}, P., {Andritschke}, R., {Arefiev}, V., {et~al.} 2021, \aap, 647, A1

\bibitem[{{Raetz} {et~al.}(2020){Raetz}, {Stelzer}, {Damasso}, \&
  {Scholz}}]{Raetz2020}
{Raetz}, S., {Stelzer}, B., {Damasso}, M., \& {Scholz}, A. 2020, \aap, 637, A22

\bibitem[{{Reiners} {et~al.}(2014){Reiners}, {Sch{\"u}ssler}, \&
  {Passegger}}]{Reiners2014}
{Reiners}, A., {Sch{\"u}ssler}, M., \& {Passegger}, V.~M. 2014, \apj, 794, 144

\bibitem[{{Ricker} {et~al.}(2014){Ricker}, {Winn}, {Vanderspek}, {Latham},
  {Bakos}, {Bean}, {Berta-Thompson}, {Brown}, {Buchhave}, {Butler}, {Butler},
  {Chaplin}, {Charbonneau}, {Christensen-Dalsgaard}, {Clampin}, {Deming},
  {Doty}, {De Lee}, {Dressing}, {Dunham}, {Endl}, {Fressin}, {Ge}, {Henning},
  {Holman}, {Howard}, {Ida}, {Jenkins}, {Jernigan}, {Johnson}, {Kaltenegger},
  {Kawai}, {Kjeldsen}, {Laughlin}, {Levine}, {Lin}, {Lissauer}, {MacQueen},
  {Marcy}, {McCullough}, {Morton}, {Narita}, {Paegert}, {Palle}, {Pepe},
  {Pepper}, {Quirrenbach}, {Rinehart}, {Sasselov}, {Sato}, {Seager},
  {Sozzetti}, {Stassun}, {Sullivan}, {Szentgyorgyi}, {Torres}, {Udry}, \&
  {Villasenor}}]{Ricker2014}
{Ricker}, G.~R., {Winn}, J.~N., {Vanderspek}, R., {et~al.} 2014, in Society of
  Photo-Optical Instrumentation Engineers (SPIE) Conference Series, Vol. 9143,
  Space Telescopes and Instrumentation 2014: Optical, Infrared, and Millimeter
  Wave, ed. J.~{Oschmann}, Jacobus~M., M.~{Clampin}, G.~G. {Fazio}, \& H.~A.
  {MacEwen}, 914320

\bibitem[{{Riello} {et~al.}(2018){Riello}, {De Angeli}, {Evans}, {Busso},
  {Hambly}, {Davidson}, {Burgess}, {Montegriffo}, {Osborne}, {Kewley},
  {Carrasco}, {Fabricius}, {Jordi}, {Cacciari}, {van Leeuwen}, \&
  {Holland}}]{Riello2018}
{Riello}, M., {De Angeli}, F., {Evans}, D.~W., {et~al.} 2018, \aap, 616, A3

\bibitem[{{Scargle}(1982)}]{1982ApJ...263..835S}
{Scargle}, J.~D. 1982, \apj, 263, 835

\bibitem[{{Schmitt} {et~al.}(2021){Schmitt}, {Ioannidis}, {Robrade}, {Predehl},
  {Czesla}, \& {Schneider}}]{schmitt21.0}
{Schmitt}, J.~H.~M.~M., {Ioannidis}, P., {Robrade}, J., {et~al.} 2021, \aap,
  652, A135

\bibitem[{Scott(2009)}]{SturgesEstimator}
Scott, D.~W. 2009, WIREs Computational Statistics, 1, 303

\bibitem[{{Shibayama} {et~al.}(2013){Shibayama}, {Maehara}, {Notsu}, {Notsu},
  {Nagao}, {Honda}, {Ishii}, {Nogami}, \& {Shibata}}]{Shibayama2013}
{Shibayama}, T., {Maehara}, H., {Notsu}, S., {et~al.} 2013, \apjs, 209, 5

\bibitem[{{Stassun}(2019)}]{Stassun2019}
{Stassun}, K.~G. 2019, VizieR Online Data Catalog, IV/38

\bibitem[{{Stassun} {et~al.}(2018){Stassun}, {Oelkers}, {Pepper}, {Paegert},
  {De Lee}, {Torres}, {Latham}, {Charpinet}, {Dressing}, {Huber}, {Kane},
  {L{\'e}pine}, {Mann}, {Muirhead}, {Rojas-Ayala}, {Silvotti}, {Fleming},
  {Levine}, \& {Plavchan}}]{Stassun2018}
{Stassun}, K.~G., {Oelkers}, R.~J., {Pepper}, J., {et~al.} 2018, \aj, 156, 102

\bibitem[{{Stelzer} {et~al.}(2022{\natexlab{a}}){Stelzer}, {Bogner},
  {Magaudda}, \& {Raetz}}]{Stelzer2022HabitableZones}
{Stelzer}, B., {Bogner}, M., {Magaudda}, E., \& {Raetz}, S. 2022{\natexlab{a}},
  \aap, 665, A30

\bibitem[{{Stelzer} {et~al.}(2022{\natexlab{b}}){Stelzer}, {Caramazza},
  {Raetz}, {Argiroffi}, \& {Coffaro}}]{Stelzer2022}
{Stelzer}, B., {Caramazza}, M., {Raetz}, S., {Argiroffi}, C., \& {Coffaro}, M.
  2022{\natexlab{b}}, \aap, 667, L9

\bibitem[{{Stelzer} {et~al.}(2016){Stelzer}, {Damasso}, {Scholz}, \&
  {Matt}}]{Stelzer2016}
{Stelzer}, B., {Damasso}, M., {Scholz}, A., \& {Matt}, S.~P. 2016, \mnras, 463,
  1844

\bibitem[{{Stelzer} {et~al.}(2006){Stelzer}, {Schmitt}, {Micela}, \&
  {Liefke}}]{Stelzer2006}
{Stelzer}, B., {Schmitt}, J.~H.~M.~M., {Micela}, G., \& {Liefke}, C. 2006,
  \aap, 460, L35

\bibitem[{{Sunyaev} {et~al.}(2021){Sunyaev}, {Arefiev}, {Babyshkin},
  {Bogomolov}, {Borisov}, {Buntov}, {Brunner}, {Burenin}, {Churazov},
  {Coutinho}, {Eder}, {Eismont}, {Freyberg}, {Gilfanov}, {Gureyev}, {Hasinger},
  {Khabibullin}, {Kolmykov}, {Komovkin}, {Krivonos}, {Lapshov}, {Levin},
  {Lomakin}, {Lutovinov}, {Medvedev}, {Merloni}, {Mernik}, {Mikhailov},
  {Molodtsov}, {Mzhelsky}, {M{\"u}ller}, {Nandra}, {Nazarov}, {Pavlinsky},
  {Poghodin}, {Predehl}, {Robrade}, {Sazonov}, {Scheuerle}, {Shirshakov},
  {Tkachenko}, \& {Voron}}]{Sunyaev2021}
{Sunyaev}, R., {Arefiev}, V., {Babyshkin}, V., {et~al.} 2021, \aap, 656, A132

\bibitem[{{Tarter} {et~al.}(2007){Tarter}, {Backus}, {Mancinelli}, {Aurnou},
  {Backman}, {Basri}, {Boss}, {Clarke}, {Deming}, {Doyle}, {Feigelson},
  {Freund}, {Grinspoon}, {Haberle}, {Hauck}, {Heath}, {Henry}, {Hollingsworth},
  {Joshi}, {Kilston}, {Liu}, {Meikle}, {Reid}, {Rothschild}, {Scalo}, {Segura},
  {Tang}, {Tiedje}, {Turnbull}, {Walkowicz}, {Weber}, \& {Young}}]{Tarter2007}
{Tarter}, J.~C., {Backus}, P.~R., {Mancinelli}, R.~L., {et~al.} 2007,
  Astrobiology, 7, 30

\bibitem[{{Tilley} {et~al.}(2019){Tilley}, {Segura}, {Meadows}, {Hawley}, \&
  {Davenport}}]{Tilley2019}
{Tilley}, M.~A., {Segura}, A., {Meadows}, V., {Hawley}, S., \& {Davenport}, J.
  2019, Astrobiology, 19, 64

\bibitem[{{Truemper}(1982)}]{Truemper1982}
{Truemper}, J. 1982, Advances in Space Research, 2, 241

\bibitem[{{VanderPlas}(2018)}]{2018ApJS..236...16V}
{VanderPlas}, J.~T. 2018, \apjs, 236, 16

\bibitem[{{Vanderspek} {et~al.}(2018){Vanderspek}, {Doty}, {Fausnaugh},
  {Villase}, {Jenkins}, {Berta-Thompson}, {Burke}, \&
  {Ricker}}]{tess-inst-handbook}
{Vanderspek}, R., {Doty}, J.~P., {Fausnaugh}, M., {et~al.} 2018, 73

\bibitem[{{Veronig} {et~al.}(2002){Veronig}, {Temmer}, {Hanslmeier}, {Otruba},
  \& {Messerotti}}]{2002A&A...382.1070V}
{Veronig}, A., {Temmer}, M., {Hanslmeier}, A., {Otruba}, W., \& {Messerotti},
  M. 2002, \aap, 382, 1070

\bibitem[{{Wright} {et~al.}(2011){Wright}, {Drake}, {Mamajek}, \&
  {Henry}}]{2011ApJ...743...48W}
{Wright}, N.~J., {Drake}, J.~J., {Mamajek}, E.~E., \& {Henry}, G.~W. 2011,
  \apj, 743, 48

\end{thebibliography}
